\documentclass{aa}  

\usepackage{appendix}
\usepackage{graphicx}
\usepackage{txfonts}
\usepackage{overpic}
\usepackage{lipsum,float}
\usepackage{subfig,caption}
\usepackage{color}
\begin{document}

   \title{First Discovery and Confirmation of PN Candidates Found from AI and Deep Learning Techniques Applied to VPHAS+ Survey Data}


   \author{Yushan Li\inst{1,2,3}
          \and
          Quentin Parker\inst{2}
          \thanks{quentinp@hku.hk}
          \and
          Peng Jia\inst{3,4}
          }

   \institute{The Department of Physics, The University of Hong Kong, Hong Kong SAR, China
        \and  The Laboratory for Space Research, The University of Hong Kong, Hong Kong, 999077, China
        \and  College of Electronic Information and Optical Engineering, Taiyuan University of Technology, Taiyuan, 030024, China
        \and  Peng Cheng Lab, Shenzhen, 518066, China
             }

   \titlerunning{First Discovery and Confirmation of PN Candidates Found by DL}
   \authorrunning{Yushan Li et al.}

   \date{Received -; accepted -}

 
  \abstract
   {We have developed deep learning (DL) and AI based tools to search extant narrow-band wide-field H$\alpha$ surveys of the Galactic Plane for elusive planetary nebulae (PNe) which are hidden in dense star fields towards the Galactic centre. They are faint, low-surface brightness, usually resolved sources, which are not discovered by previous automatic searches that depend on photometric data for point-like sources. These sources are very challenging to find by traditional visual inspection in such crowded fields and many have been missed.  We have successfully adopted a novel 'Swin-Transformer' AI algorithm, which we described in detail in the preceding Techniques paper (Paper~I).}
   {Here, we present preliminary results from our first spectroscopic follow-up run for 31 top-quality PN candidates found by the algorithm from the high-resolution H$\alpha$ survey VPHAS+. This survey has not yet undergone extensive manual, systematic searching.}
   {Our candidate PNe were observed with the SpUpNIC spectrograph on the 1.9 m telescope at the South African Astronomical Observatory (SAAO) in June 2023. We performed standard IRAF spectroscopic reduction and then followed our normal HASH PN identification and classification procedures.}
   {Our reduced spectra confirmed that these candidates include 22 true, likely and possible PNe (70.97\%), 3 emission-line galaxies, 2 emission-line stars, 2 late-type star contaminants and 2 other H$\alpha$ sources including a newly identified detached fragment of SNR RCW 84. We present the imaging and spectral data of these candidates and a preliminary analysis of their properties. These data provide strong input to help to evaluate and refine the behaviour of the AI algorithm when searching for PNe in wide-field H$\alpha$ surveys.}
   {}


   \maketitle
%

\section{Introduction}

A planetary nebula (PN) is a brief but crucial stage in the late stage evolution of 
low- to intermediate-mass stars. Their properties can reveal the physics of stellar 
nucleosynthesis and mass loss processes occurring during this phase, e.g. 
\citet{1995PhR...250....2I}, Galactic chemical enrichment and any Galactic abundance 
gradients, see \citet{2003A&A...397..667M, 
2010ApJ...724..748H, 2015RMxAA..51..165M, 2018ApJ...862...45S}. As one of the most 
complex and beautiful astrophysical object classes to study, a PN is the final, ejected 
gaseous envelope from its central star (CSPN) after the asymptotic giant branch (post-AGB) phase. These envelopes are ionized by the high-energy radiation emitted by the 
CSPN as it evolves towards becoming a white dwarf (WD). They experience dramatic 
evolution over cosmological short periods of time from typically $\sim21,000 \pm5,000$ 
years \citep{2013A&A...558A..78J, 2015ApJ...804L..25B},
although an even older example of a PN in an open cluster is estimated at an age of 
$\sim50,000$ years \citep{2022ApJ...935L..35F}. This brevity makes them rare. They 
exhibit a diverse range of properties, including morphology, size, surface brightness 
distribution, ionization characteristics, elemental abundances, densities and expansion 
rates related to the mass, temperature and properties of the CSPN. 
PNe can be easily confused with other objects, such as H~II regions, 
young stellar objects (YSOs), supernova remnants (SNRs), Wolf-Rayet shells, nova shells, stellar-ejecta and symbiotic systems as described in 
\citep{2010PASA...27..129F} and \citep{2022FrASS...9.5287P}. Such shared 
characteristics among various types of nebulae can cause difficulties in hunting for 
them with automatic software which generally performs better on more regular targets 
like stars and galaxies. With the advent of deep, wide-field, H$\alpha$ 
surveys of the Galactic plane in the past 20 years, such as the SHS 
\citep{2005MNRAS.362..689P} in the South and IPHAS \citep{2005MNRAS.362..753D} in  
North, a remarkable contribution of ~50\% to the $\sim3800$ currently 
known Galactic PNe \citep{2016JPhCS.728c2008P} was made by visual examination from 
H$\alpha$ narrow-band images contrasted with corresponding red broad-band images in a 
major series of papers led by and involving the second author
\citep{2006MNRAS.373...79P, 2008MNRAS.384..525M, 2014MNRAS.443.3388S}.

The most recent, high-resolution H$\alpha$ survey VPHAS+ \citep{2014MNRAS.440.2036D} of 
the Southern Galactic plane, using the VST telescope in Chile, finished 91.6\% of its planned footprint up to
August 2018 with a coverage of $\sim 2000$ sq.deg. So far it has not been effectively 
scanned by human eyes. In the Galactic bulge the survey was extended from its standard 
$\pm5$~degrees Galactic latitude covered to $\rm |b|=10^\circ$. The survey has an 
angular resolution of $\sim0.21$~arcsec/pixel well matched to the natural seeing that
often approaches 0.5~arcsec. This higher resolution Southern survey, which includes the 
rich Galactic bulge region, provides a great opportunity to hunt for small resolved PNe 
hidden behind the dense star field towards the Galactic centre. To utilize the data in 
a more efficient, objective and consistent way, we constructed a deep learning object 
detection algorithm for PNe based on the novel Swin-Transformer model 
\citep{2021arXiv210314030L}. For details  see \citet{2024MNRAS.528.4733S}. This 
data-driven method was adjusted according to the diverse properties of PNe, trained 
with 1137 individual images from IPHAS and corresponding catalogue from HASH database 
\footnote{HASH: online at \url{http://www.hashpn.space}. HASH federates available multi-
wavelength imaging, 
spectroscopic and other data for all known Galactic and Magellanic Cloud PNe.} 
\citep{2016JPhCS.728c2008P, 2022FrASS...9.5287P}, and then validated with a further 454 
PNe images. The validation results are the successful detection of 97.8\% of 
all known PNe. Of the detected targets 96.5\% are real in HASH, which showed 
very decent performance for our model. Previous machine learning and deep learning work 
in PNe research, \citet{1996A&AS..116..395F} used hierarchical cluster 
analysis and supervised ANN (artificial neural network) to classify PNe according to 
their chemical composition (so based on spectroscopy), while 
\citet{2019MNRAS.488.3238A} used a decision tree model to 
find compact PNe in infrared photometric data. Most recently
\citet{2020Galax...8...88A} applied multiple DNN (deep neural network) models to 
several infrared and optical image datasets to distinguish PNe from all the types of 
non-PNe, and classify their morphologies with some success.

This paper is organised as follows: the spectral observations and reduction process are 
described in Section \ref{sec:spec}, the results are presented in Section \ref{sec:res} 
in several categories, and the conclusions are given in Section \ref{sec:con}.

\section{Spectrum Observation and Reduction}
\label{sec:spec}

The spectra of PNe have many strong, characteristic emission-lines across the 
electromagnetic spectrum though most available spectra are optical. We conducted 
spectral confirmation for a high quality, AI/ML discovered sample of PNe candidates. 
Our observations were performed between Jun 13$^{th}$-21$^{st}$, 2023, on the 1.9m SAAO 
telescope using the SpUpNIC spectrograph \citep{2016SPIE.9908E..27C} under the RUN 
\#484. We used grating gr7 (300 lines per mm, blaze of 4600 \r{A}, range of 5550 \r{A}, 
dispersion of 2.72 \r{A}/pixel), and typically used a slit width  17 
($\sim$2 arcseconds).  We adjusted exposure times according to the brightness of the 
targets, typically 300s for brighter, more compact PNe, 600s for intermediate surface 
brightness candidates and 900s for our fainter targets including some repeated 900s for 
such very faint ones. For the specific target YP0900-4457 (see later), we observed both 
its compact and apparent nebula jet components parts, with exposure times of 300s and 
900s.

We reduced the spectra with IRAF V2.17, 2021 \citep{1986SPIE..627..733T, 
1993ASPC...52..173T}. We created master bias with {\it zerocombine} task, and then 
smoothed it with {\it imsurfit} task. For dome and sky flats, the master flats were 
created with the {\it flatcombine} task. The master dome flat was fitted with the {\it 
response} task, then the new sky flat was generated by the {\it ccdproc} task, finally 
the residual slope was removed with 
{\it illum} task from the new sky flat, and a 'perfect flat' was created by the {\it 
imarith} task, multiplying the fitted master dome flat by the final new sky flat. We 
can then correct for both bias and flat and then trim the spectra with the {\it 
ccdproc} task on the spectra of all object, ARC and flux calibrator spectra.

After this data reduction process, we combined any repeated observation with the {\it 
imcombine} task, and cleaned cosmic ray events (CRE) with the {\it imedit} task.  
The 1-D spectra of targets and ARCs were then extracted from the 2-D spectra using the 
{\it apall} task after identifying the nebula region and sky regions (for sky 
subtraction). The Wavelength calibration was conducted with the {\it identify}, {\it 
reidentify}, {\it refspec} and {\it dispcor} tasks, to determinate the dispersion 
solutions from the ARC calibration spectra taken for each target observation and 
assigned to the target spectra as appropriate. Finally,
flux calibration was conducted with the {\it standard}, {\it sensfunc} and {\it 
calibrate} tasks, for our targets using selected standard 
stars LTT4364 and LTT9239 observed during each night.

\begin{table*}
\centering
\caption{Names, HASH IDs, positions, PN Status, angular size (in arcseconds) and (where possible) the PNe morphologies (R-Round, E-Elliptical, B-Bipolar, I-Irregular, S-quasi-Stellar, A-Asymmetric) of the observed PN candidates and any associated nebulosity or outflows.}
\label{tab:prop}
\renewcommand\arraystretch{1.35}
\resizebox{\textwidth}{!}{
\begin{tabular}{lcccccccc} 
 \hline
 Common Name & HASH & RA (J2000) & Dec (J2000) & l & b & Object & Diam. & ERBIAS \\ 
  & ID & hh:mm:ss.s & $\pm$dd:mm:ss & deg & deg & Status & arcsec & Morph. \\
 \hline
 YP0821-4253 & 33724 & 08:21:56.5 & -42:53:05 & 260.28 & -3.41 & True PN & 7.0 & R \\
 YP0821-4253b & 33798 & 08:22:06.0 & -42:53:30 & 260.30 & -3.39 & True PN & 226.9 & B \\
 YP0827-3033 & 33725 & 08:27:57.4 & -30:33:06 & 250.88 & 4.68 & EM galaxy & 3.1 & S \\
 YP0900-4457 & 33726 & 09:00:57.4 & -44:57:05 & 266.22 & 0.87 & EM star & 7.0 & \\
 YP0900-4457b & 33726 & 09:00:57.4 & -44:56:56 & 266.22 & 0.87 & Outflow & 6.7 & I \\
 YP1028-5714 & 33727 & 10:28:41.2 & -57:14:28 & 284.53 & 0.44 & Likely PN & 3.4 & R \\
 YP1048-6127 & 33728 & 10:48:57.0 & -61:27:10 & 288.83 & -1.98 & True PN & 4.7 & I \\ 
 YP1151-6247 & 33729 & 11:51:51.5 & -62:47:10 & 296.18 & -0.70 & EM galaxy & 6.3 & B \\ 
 YP1209-6332 & 33730 & 12:09:16.4 & -63:32:03 & 298.25 & -1.05 & Possible PN & 2.4 & S \\ 
 YP1214-6226 & 33731 & 12:14:34.6 & -62:26:07 & 298.68 & 0.13 & Possible PN & 2.9 & S \\ 
 YP1317-6513 & 33732 & 13:17:52.3 & -65:13:27 & 305.70 & -2.49 & True PN & 8.5 & E \\ 
 YP1327-6328 & 33733 & 13:27:40.1 & -63:28:05 & 306.96 & -0.88 & True PN & 9.2 & E \\ 
 YP1429-6214 & 33734 & 14:29:34.2 & -62:14:01 & 314.09 & -1.51 & LT star+True PN & 4.1 & B \\ 
 YP1441-6239 & 33735 & 14:41:05.5 & -62:39:44 & 315.15 & -2.42 & Possible SNR & 5.6 & I \\ 
 YP1502-6040 & 33736 & 15:02:47.1 & -60:40:47 & 318.34 & -1.80 & True PN & 3.0 & R \\ 
 YP1612-5209 & 33737 & 16:12:36.0 & -52:09:20 & 331.10 & -0.64 & True PN & 9.2 & E \\ 
 YP1647-4550 & 33738 & 16:47:05.9 & -45:50:28 & 339.56 & -0.40 & Neb in Cluster & 3.7 & I \\ 
 YP1650-4328 & 33739 & 16:50:40.5 & -43:28:12 & 341.79 & 0.64 & True PN & 2.8 & S \\ 
 YP1701-2519 & 33740 & 17:01:30.0 & -25:19:48 & 357.48 & 10.15 & EM galaxy & 3.4 & R \\ 
 YP1705-2524 & 33741 & 17:05:11.0 & -25:24:40 & 357.91 & 9.43 & LT star & 8.4 & \\ 
 YP1716-2705 & 33742 & 17:16:42.2 & -27:05:19 & 358.05 & 6.35 & Possible PN & 2.5 & R \\ 
 YP1731-3011 & 33743 & 17:31:17.8 & -30:11:36 & 357.27 & 1.96 & WD+True PN & 2.6 & S \\ 
 YP1731-3011b & 33743 & 17:31:17.8 & -30:11:36 & 357.27 & 1.96 & Jet \& Bub & 33.6 & B \\
 YP1754-1815 & 33744 & 17:54:32.4 & -18:15:48 & 10.22 & 3.71 & True PN & 3.3 & R \\ 
 YP1759-2036 & 33745 & 17:59:41.1 & -20:36:28 & 8.79 & 1.49 & True PN & 7.6 & E \\ 
 YP1802-1440 & 33746 & 18:02:57.9 & -14:40:19 & 14.35 & 3.74 & True PN & 3.4 & R \\ 
 YP1810-2450 & 33747 & 18:10:39.6 & -24:50:28 & 6.34 & -2.79 & True PN & 7.3 & A \\ 
 YP1813-2146 & 33748 & 18:13:22.1 & -21:46:50 & 9.33 & -1.86 & EM star & 2.2 & \\ 
 YP1818-1211 & 33749 & 18:18:29.2 & -12:11:11 & 18.35 & 1.63 & True PN & 8.4 & B \\ 
 YP1819-2405 & 33750 & 18:19:17.6 & -24:05:27 & 7.93 & -4.16 & LT Star & 7.5 & \\ 
 YP1826-1029 & 33751 & 18:26:23.0 & -10:29:47 & 20.75 & 0.71 & True PN & 6.4 & I \\ 
 YP1837-0354 & 33752 & 18:37:54.3 & -03:54:31 & 27.91 & 1.22 & Likely PN & 3.0 & R \\ 
 YP1845-0511 & 33753 & 18:45:42.7 & -05:11:50 & 27.65 & -1.10 & True PN & 3.5 & R \\ 
 YP1856-0424 & 33754 & 18:56:46.1 & -04:24:46 & 29.61 & -3.19 & Possible PN & 2.4 & R \\ 
 \hline
\end{tabular}}
\end{table*}

\section{Results}
\label{sec:res}

We determine whether a target is a PN or another object type by holistically assessing key features, following the precepts summarised in \cite{2022FrASS...9.5287P} - refer Figure 11 and Section 12.2 of that paper. This includes whether the target appears as an emission object in narrow-band imaging (our ML/AI algorithm detects as many unknown emission objects as possible), whether the target exhibits conforms to typical PN morphological classifications. We adopt the HASH ERBIAS/sparm morphological classification scheme \citep{2006MNRAS.373...79P}. We also assess whether nebula emission lines of reasonable S/N in the candidate follow-up spectra align with characteristics specific of typical PN types and evolutionary stages. However, typical emission line ratios of certain nebulae can still be mistaken for PNe. It may sometimes be necessary to consider the equivalent width of the H$\alpha$ line for making judgments \citep{2022FrASS...9.3485S}. In the subsequent subsections, we will provide examples of how to determine the type of target in specific cases.

In summary, of the 31 top-quality PNe candidates we were able to observe, we 
spectroscopically confirmed 16 true, 2 likely and 4 possible PNe. We further 
identified 3 emission-line galaxies (here their redshifts are low enough that the [N~II] 
and H-alpha lines were not shifted out of the VPHAS+ H$\alpha$ narrow-band filter), 2 
emission-line stars, 2 late-type stars, 1 possible SNR fragment and 1 known emission 
blob in a star cluster. It is clear that all but the two late-type stars had real 
emission lines. The properties of these candidates are summarised in Table 
\ref{tab:prop}, including common names, RA (J2000) \& Dec (J2000), Galactic l \& b, 
object status, angular diameters (arcseconds) and morphology (if applicable). Note that 
the diameters of candidates were measured with the DS9 {\it region} function on VPHAS+ 
H$\alpha$ images whose  scale is {\it zscale}. We provided examples for each category and included all of our images and spectra in Appendix.

\subsection{True PNe}
\label{sec:res:subsec:true}

We spectroscopically confirmed 16 candidates as true PNe (accounting for 51.61\% of the total 
sample). The assignment of the true 'T' class followed standard HASH procedures as outlined in 
\citep{2016JPhCS.728c2008P, 2022FrASS...9.5287P} based on holistic assessment of all available data 
for each object. Of course here the spectroscopic confirmation is vital and based on the nature of 
the target's emission lines. This includes presence of H$\alpha$, [N~II], [O~III], [S~II] and [O~I] 
in the red and the detection of He~II (if present this is strong proof of PNe nature) [O~III] and 
H$\beta$ in the blue and in ratio's typical for PNe. See Figure \ref{fig:true_pn} for their VPHAS+ detected PNG images labelled by red boxes, SHS H$\alpha$-Rband quotient images, VPHAS+ H$\alpha$-Rband quotient images, and our confirmatory SAAO spectra with spectral lines labelled.

\begin{figure*}
\includegraphics[width=0.2\textwidth]{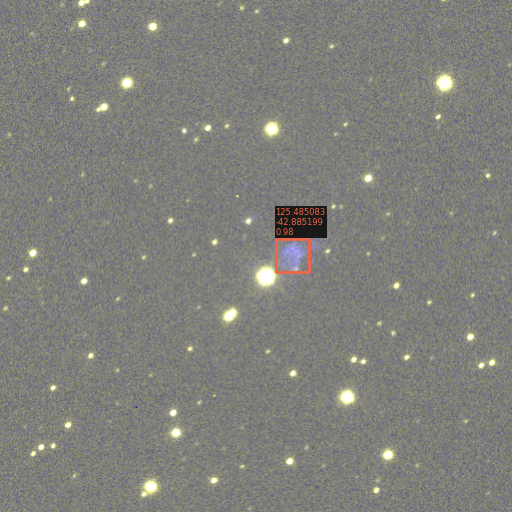}
\includegraphics[width=0.205\textwidth]{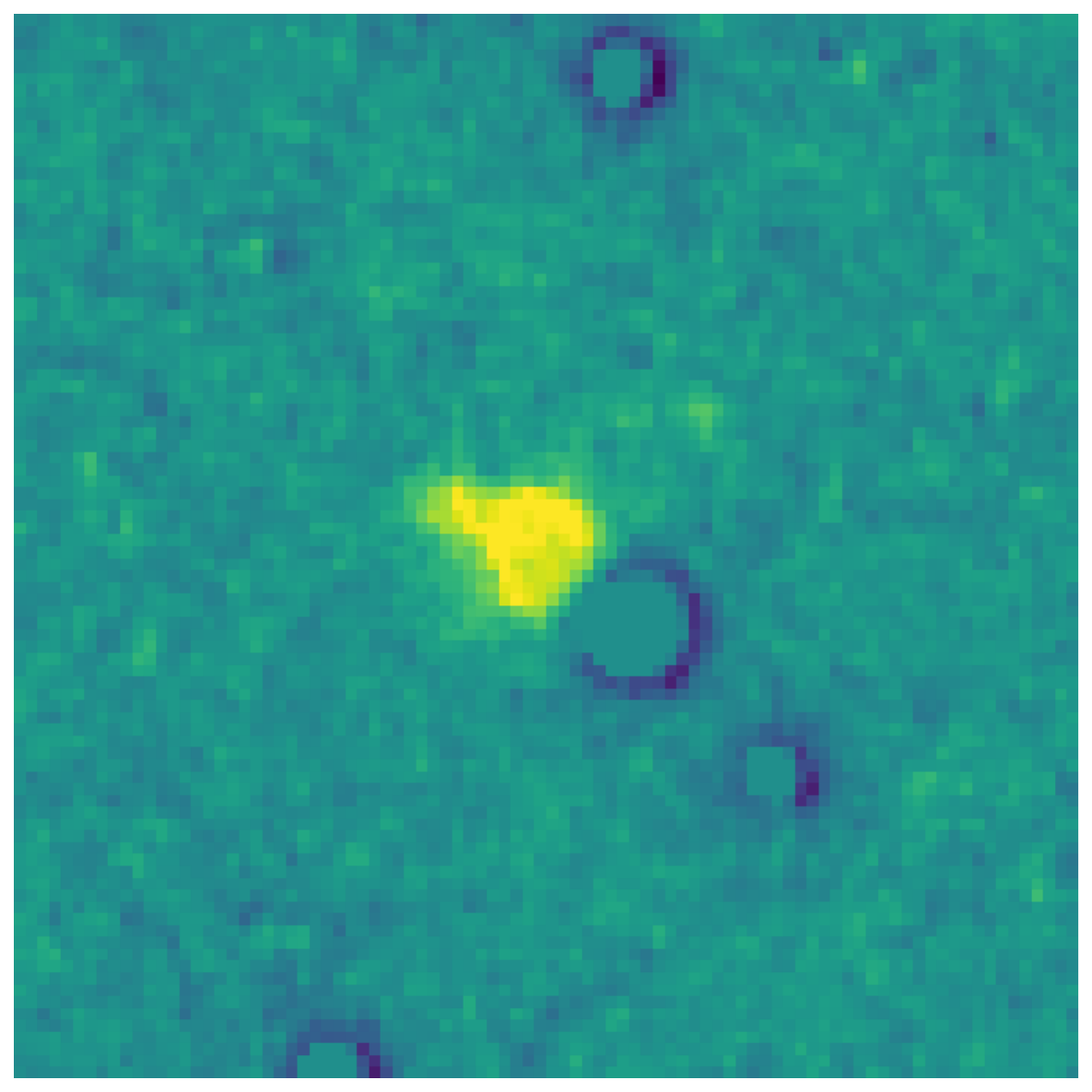}
\includegraphics[width=0.205\textwidth]{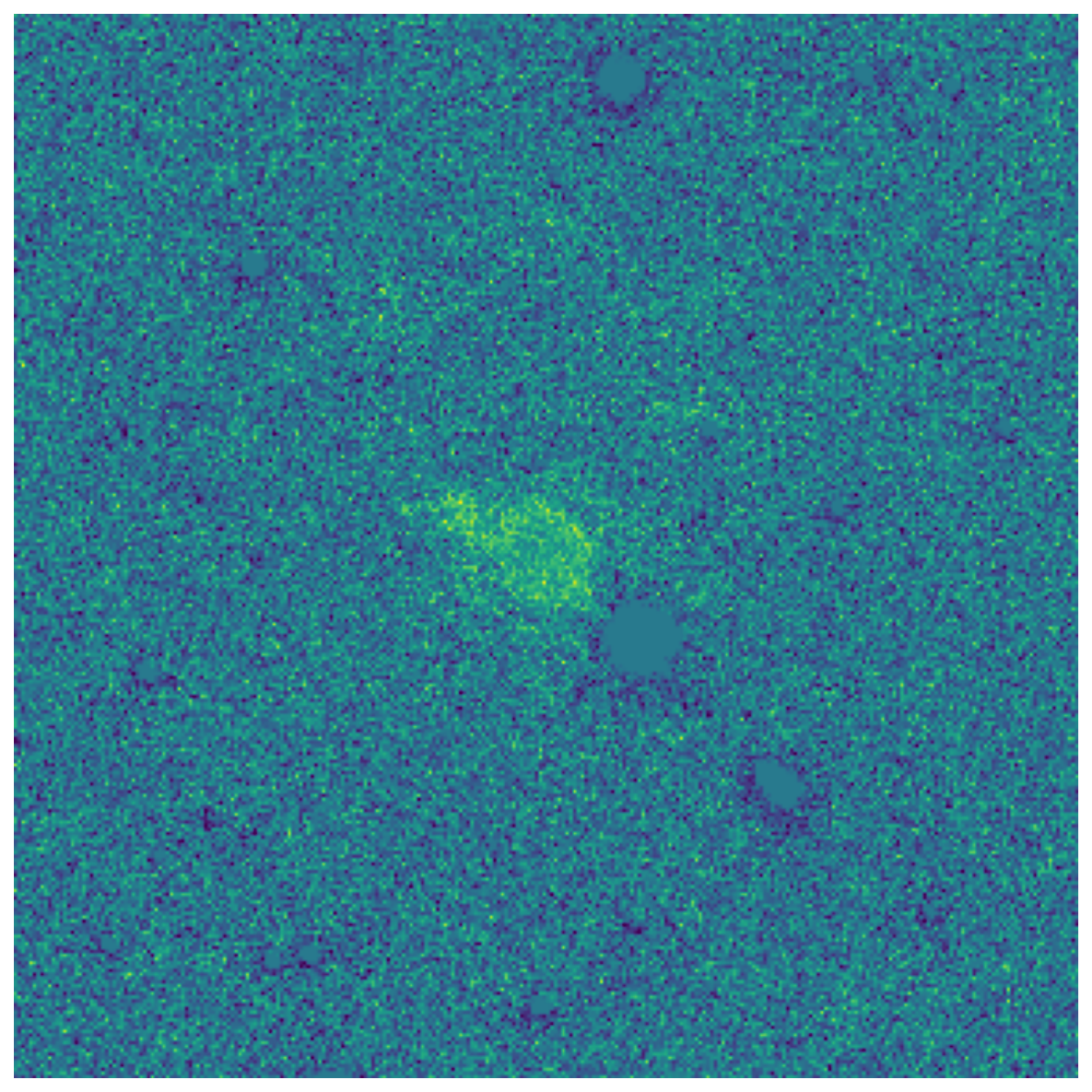}
\begin{overpic}[width=0.4\textwidth]{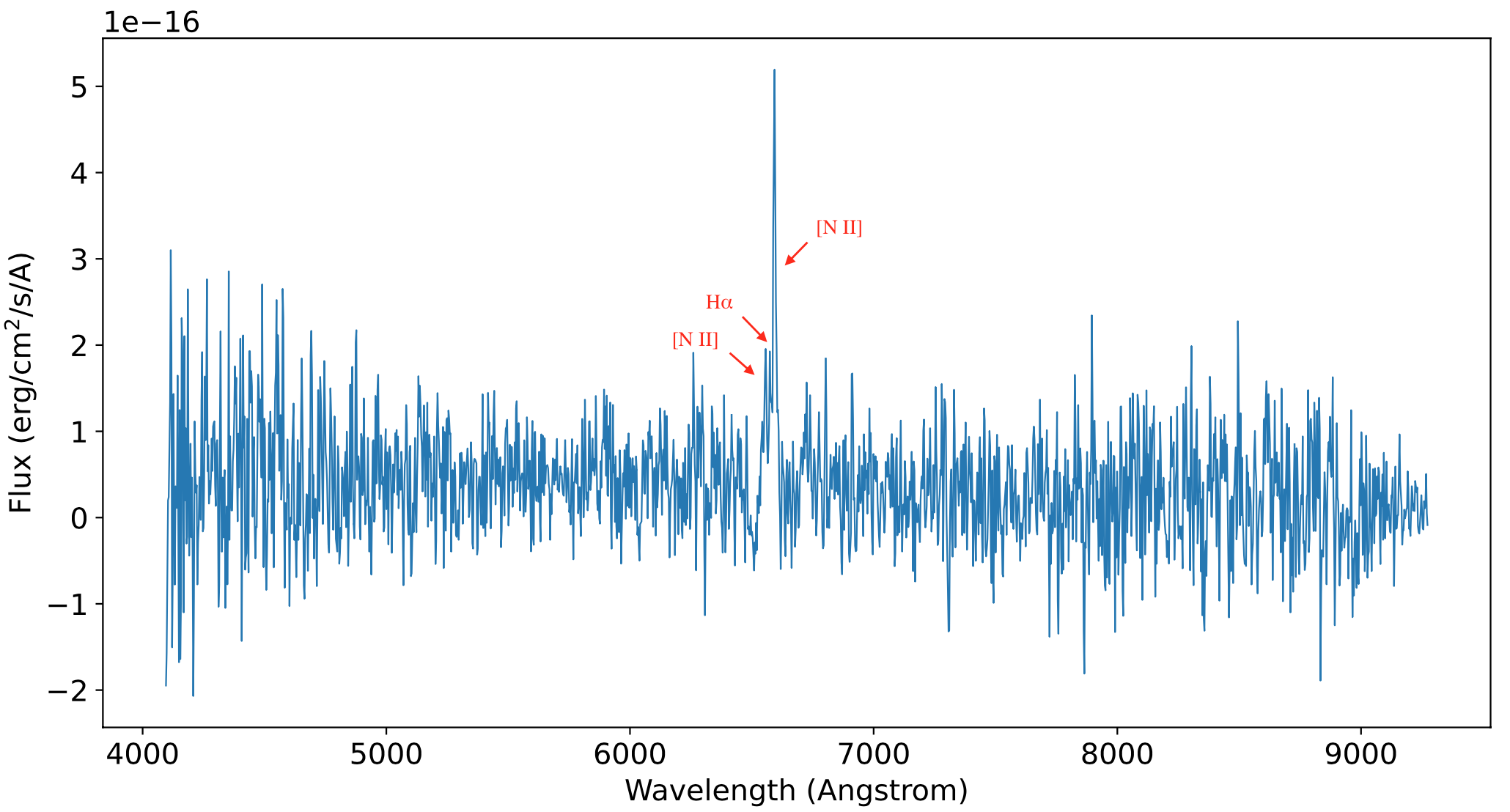}
\put(70,45){\small{YP0821-4253}}
\put(70,38){\small{True PN}}
\put(70,31){\small{Large \& Fragmented}}
\end{overpic}
\hspace*{1.44in}
\includegraphics[width=0.205\textwidth]{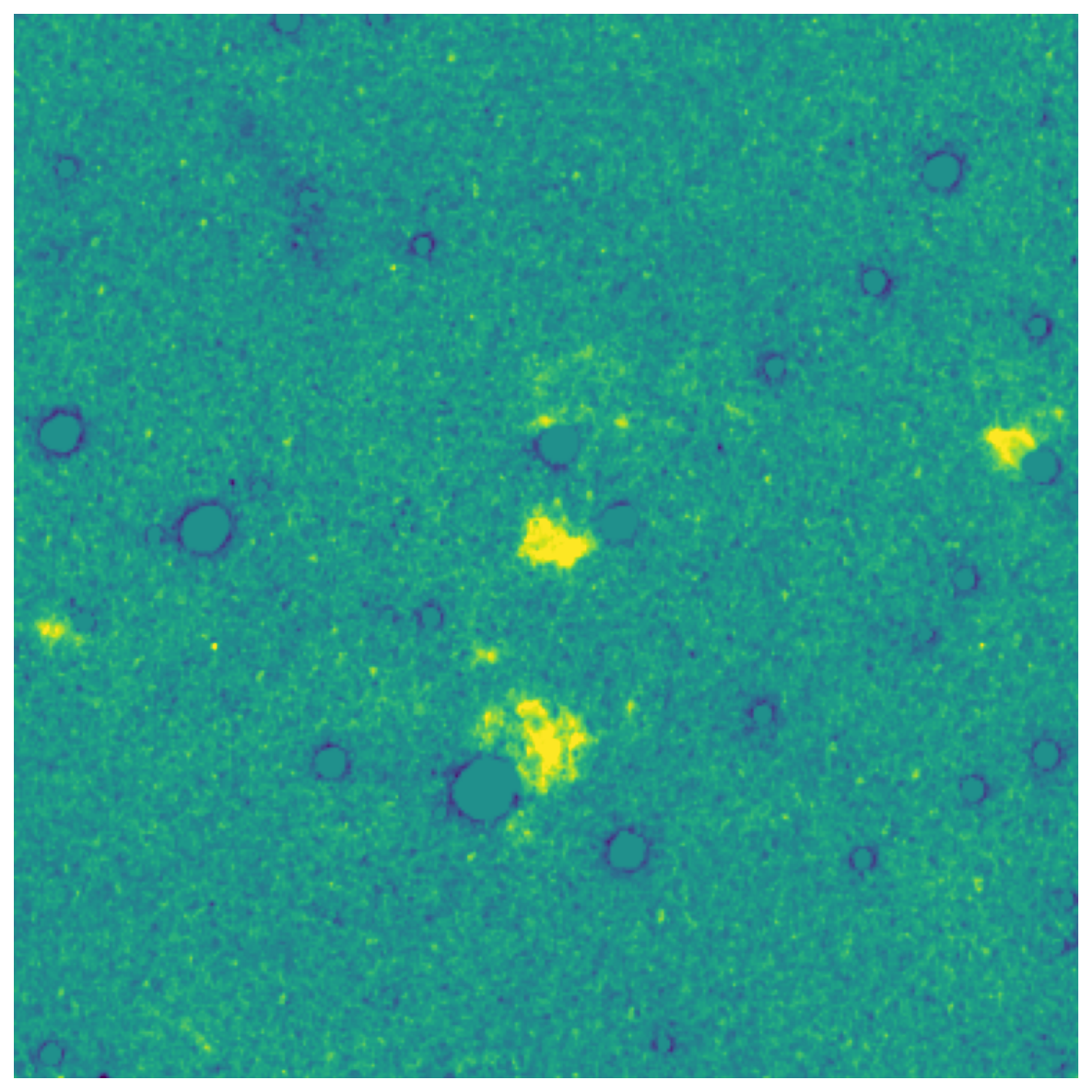}
\includegraphics[width=0.205\textwidth]{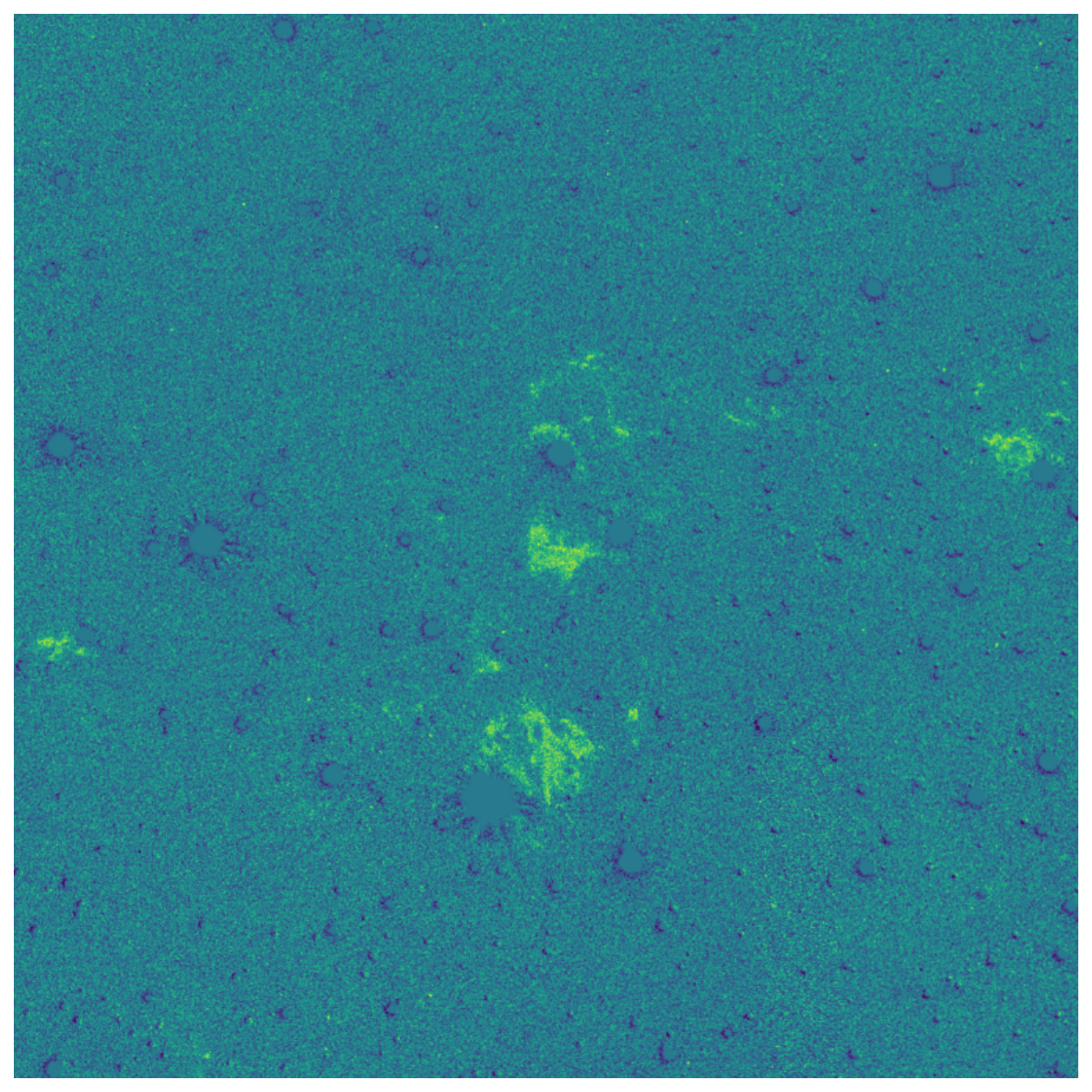}\\
\includegraphics[width=0.2\textwidth]{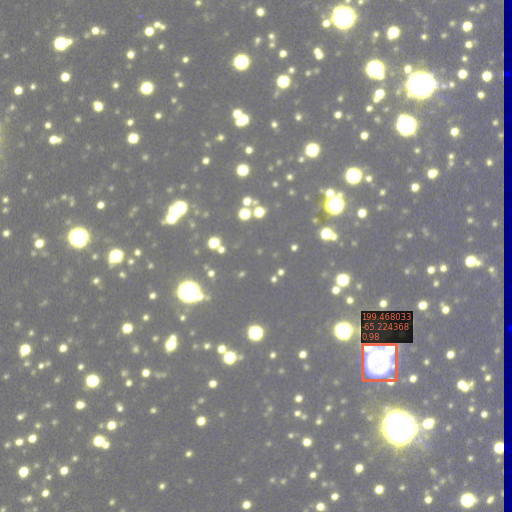}
\includegraphics[width=0.205\textwidth]{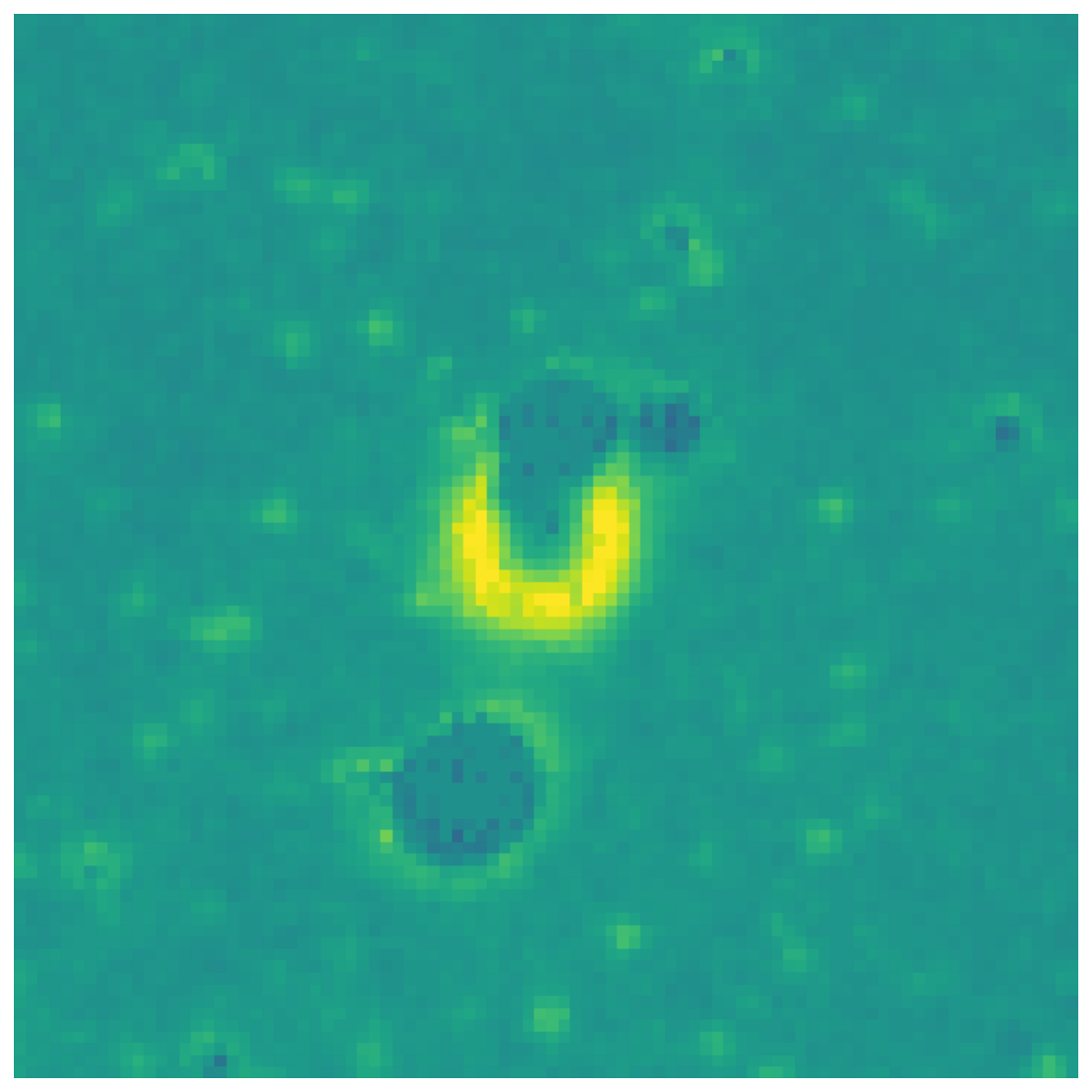}
\includegraphics[width=0.205\textwidth]{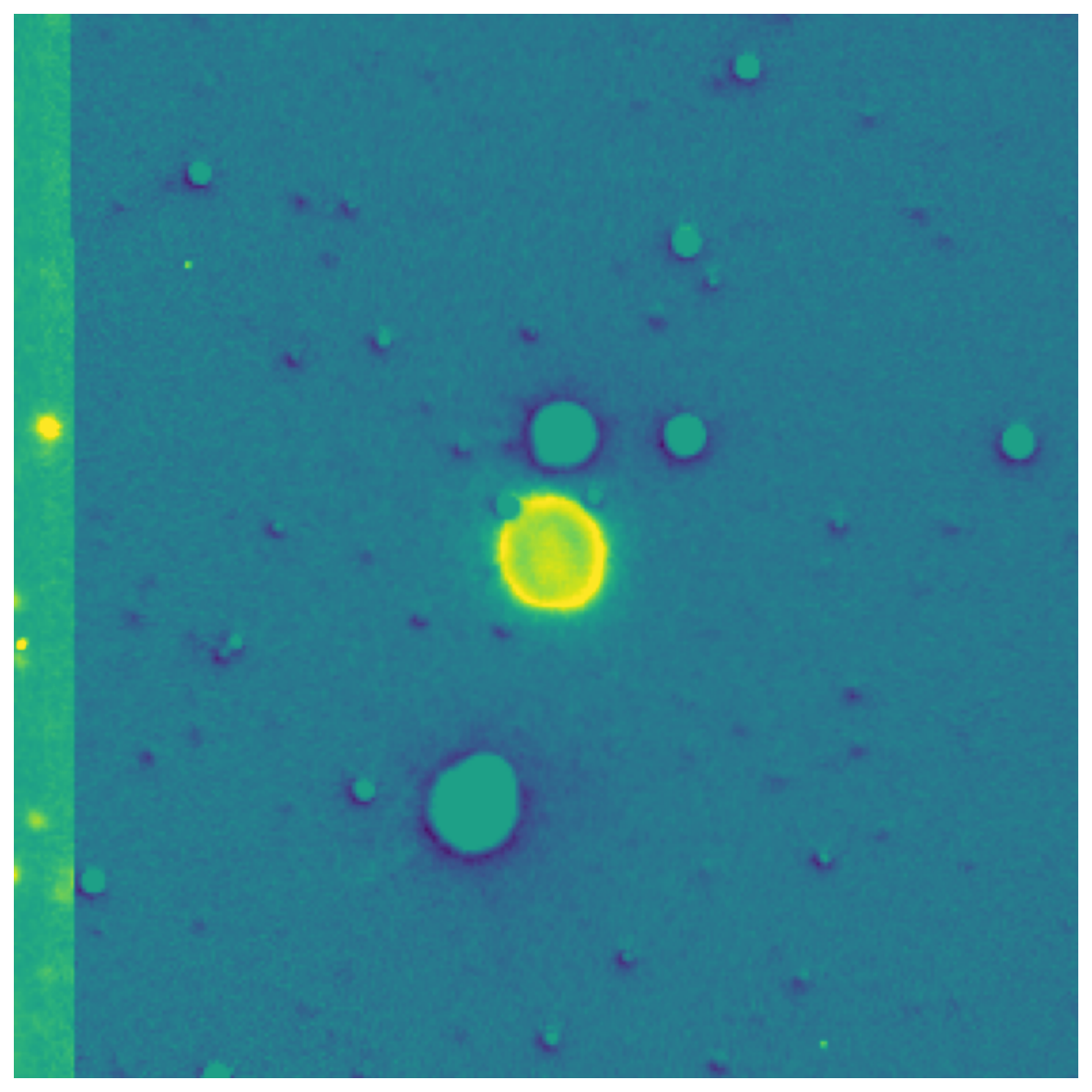}
\begin{overpic}[width=0.4\textwidth]{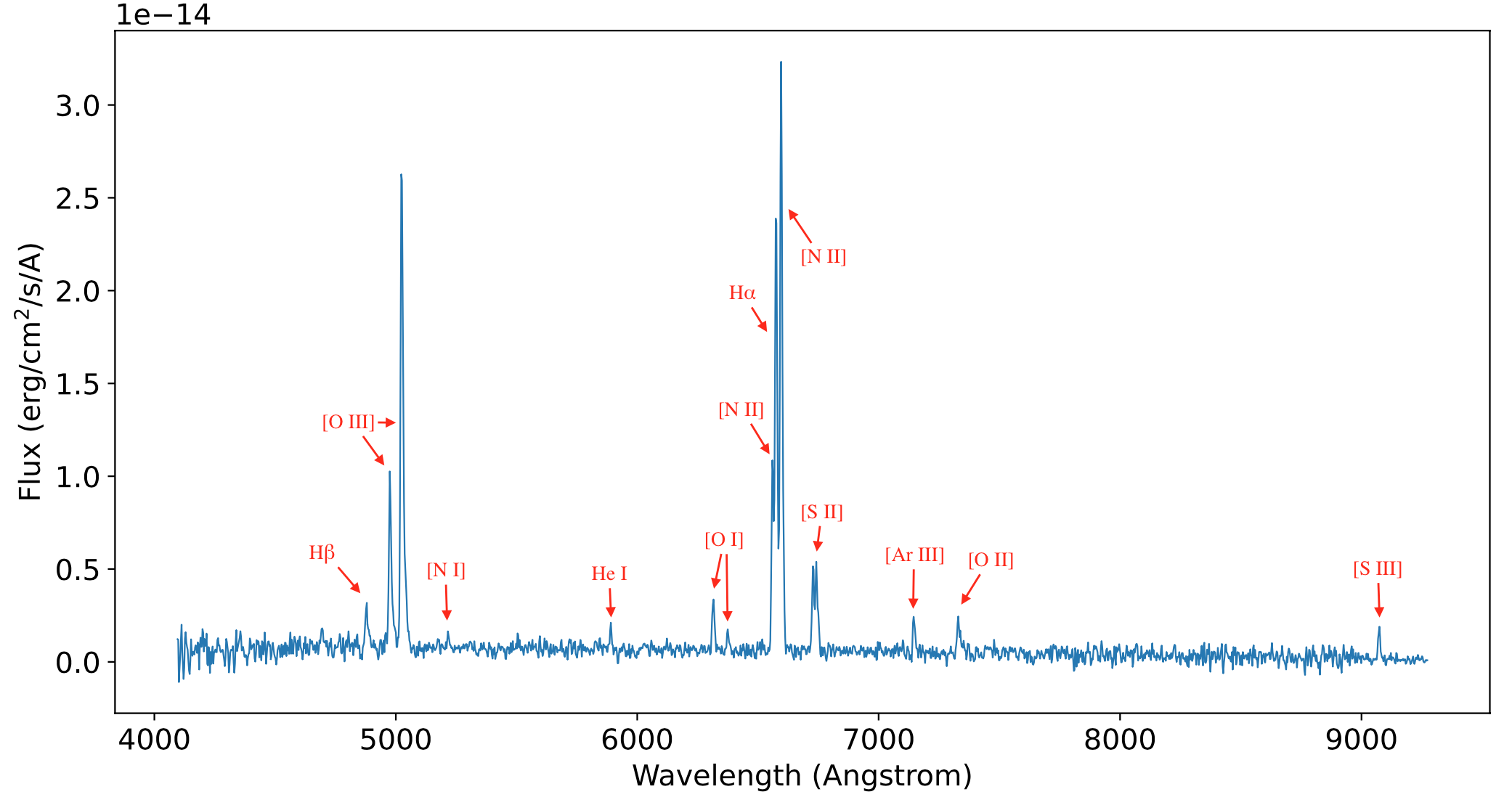}
\put(70,45){\small{YP1317-6513}}
\put(70,38){\small{True PN}}
\put(70,31){\small{Bright \& Resolved}}
\end{overpic}
\includegraphics[width=0.2\textwidth]{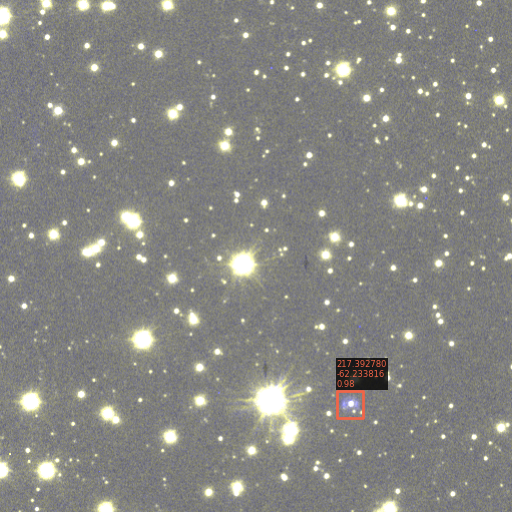}
\includegraphics[width=0.205\textwidth]{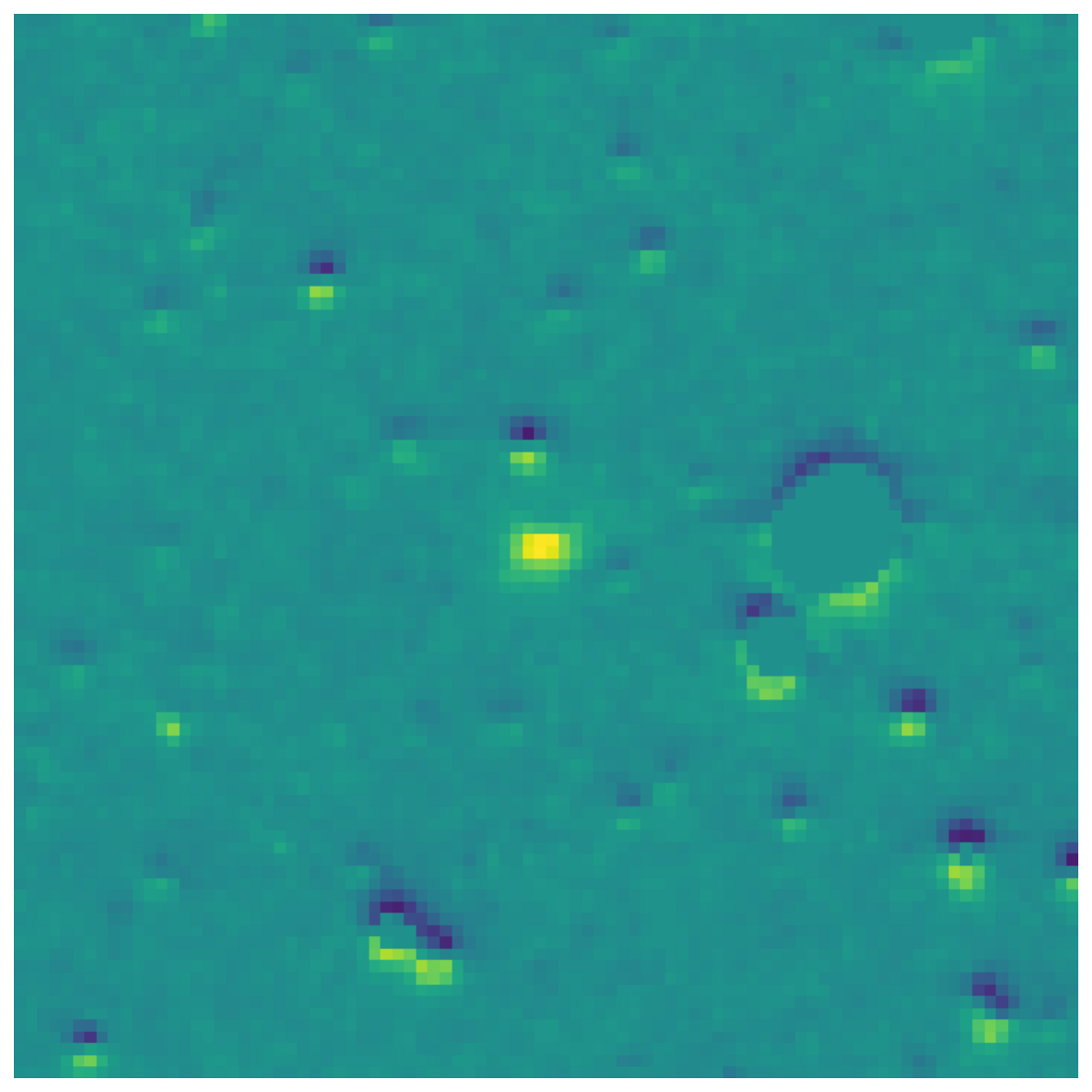}
\includegraphics[width=0.205\textwidth]{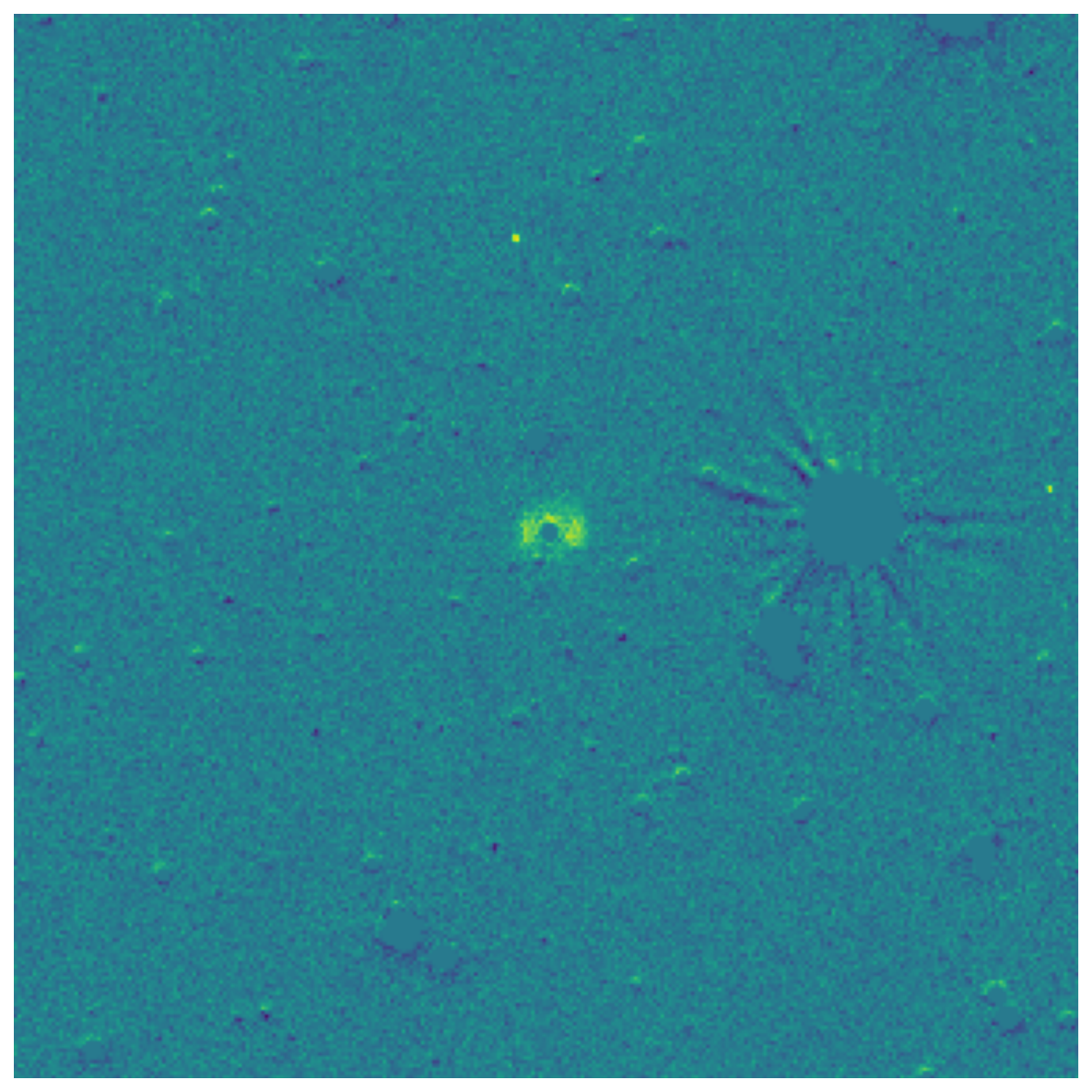}
\begin{overpic}[width=0.4\textwidth]{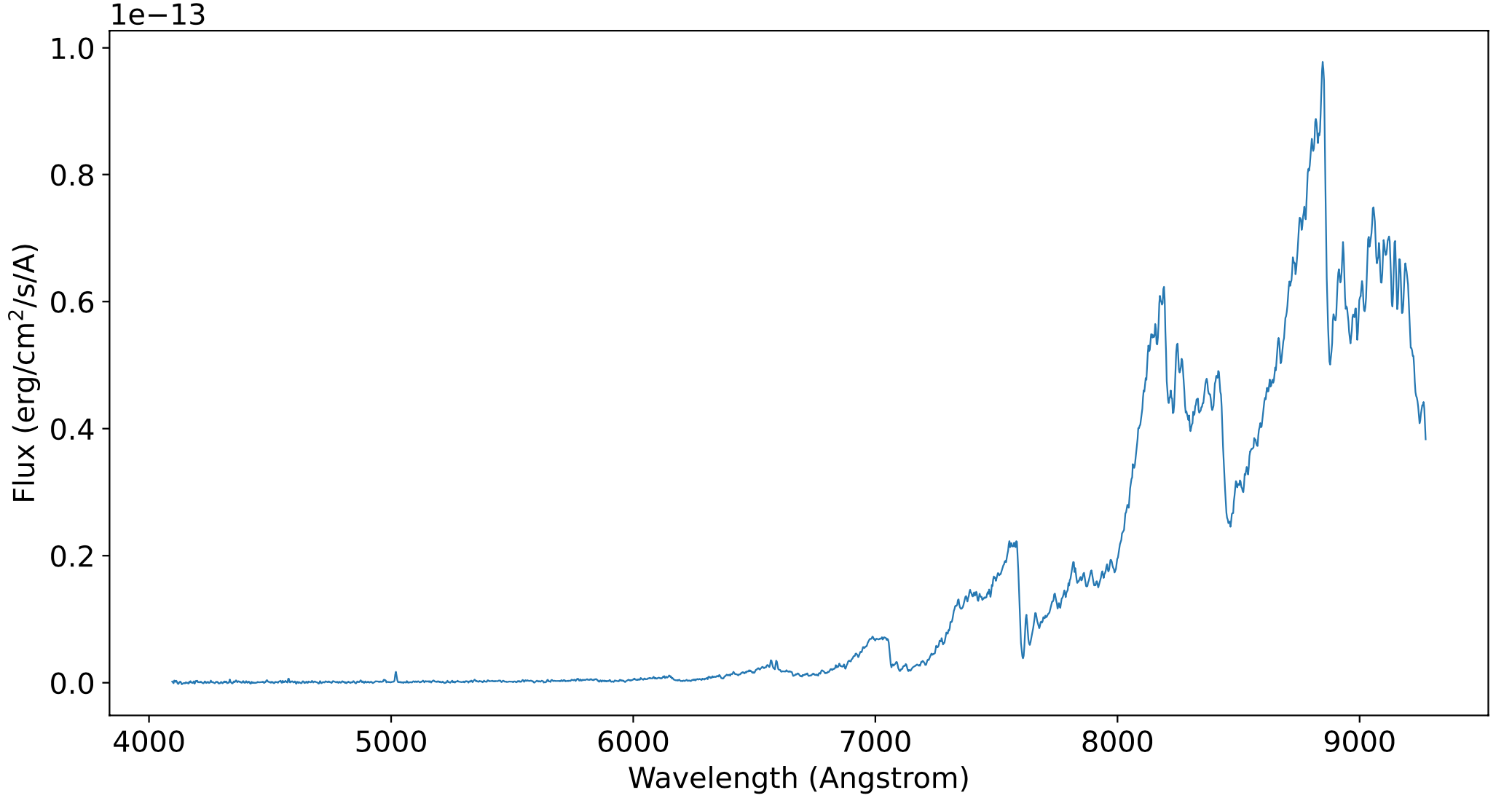}
\put(70,45){\small{YP1429-6214}}
\put(70,38){\small{True PN}}
\put(70,31){\small{Behind a Star}}
\put(8,18){\includegraphics[scale=0.12]{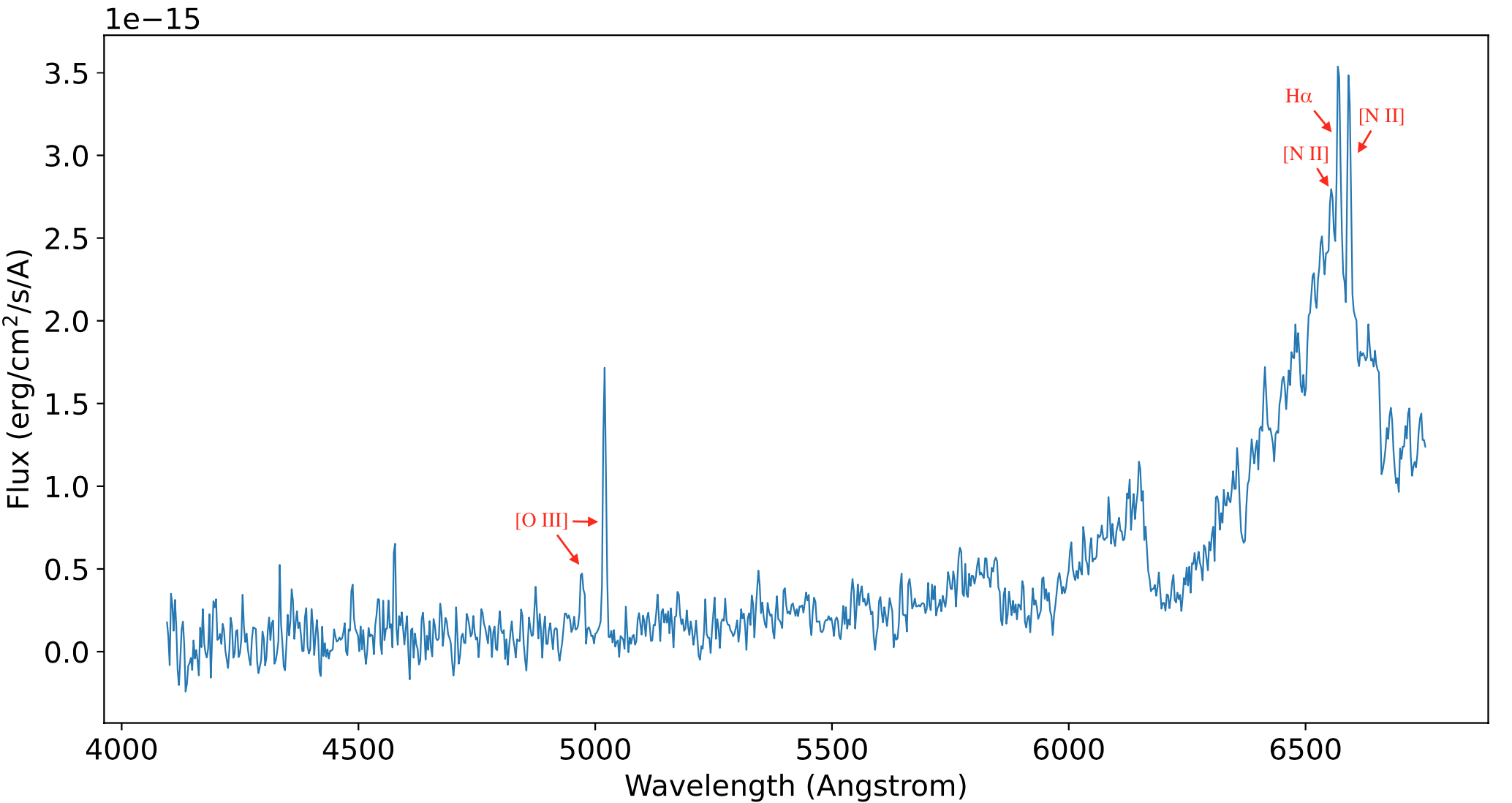}}
\end{overpic}
\includegraphics[width=0.2\textwidth]{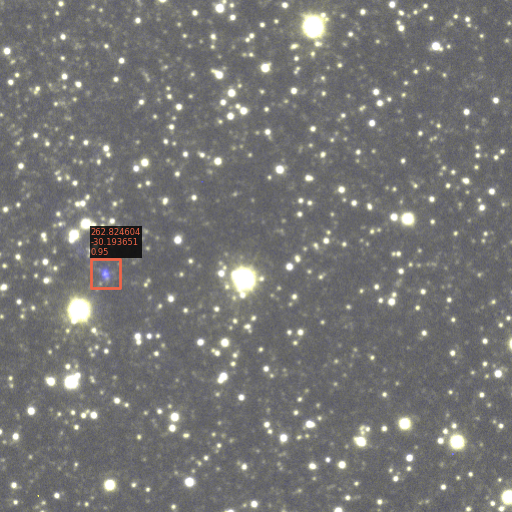}
\includegraphics[width=0.205\textwidth]{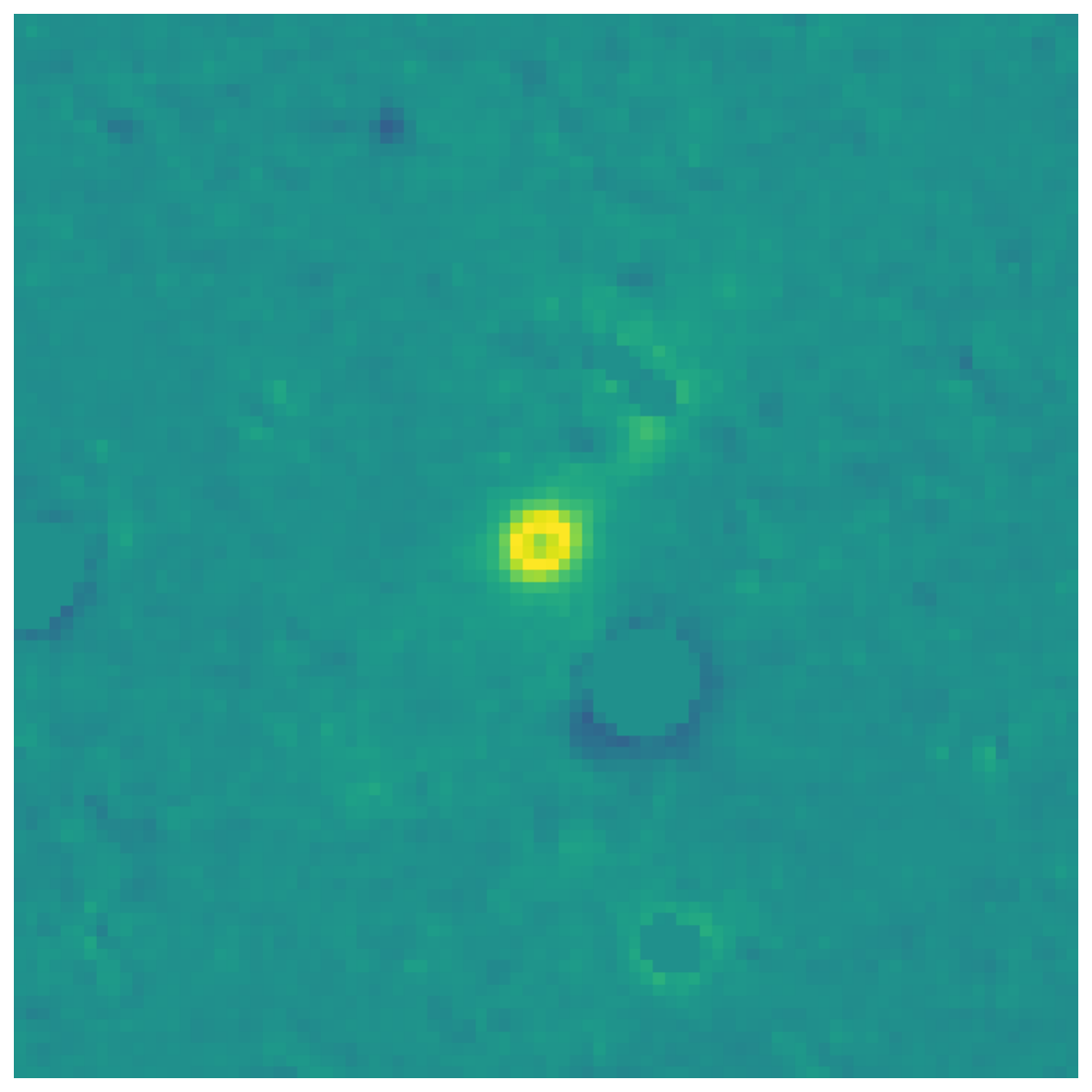}
\includegraphics[width=0.205\textwidth]{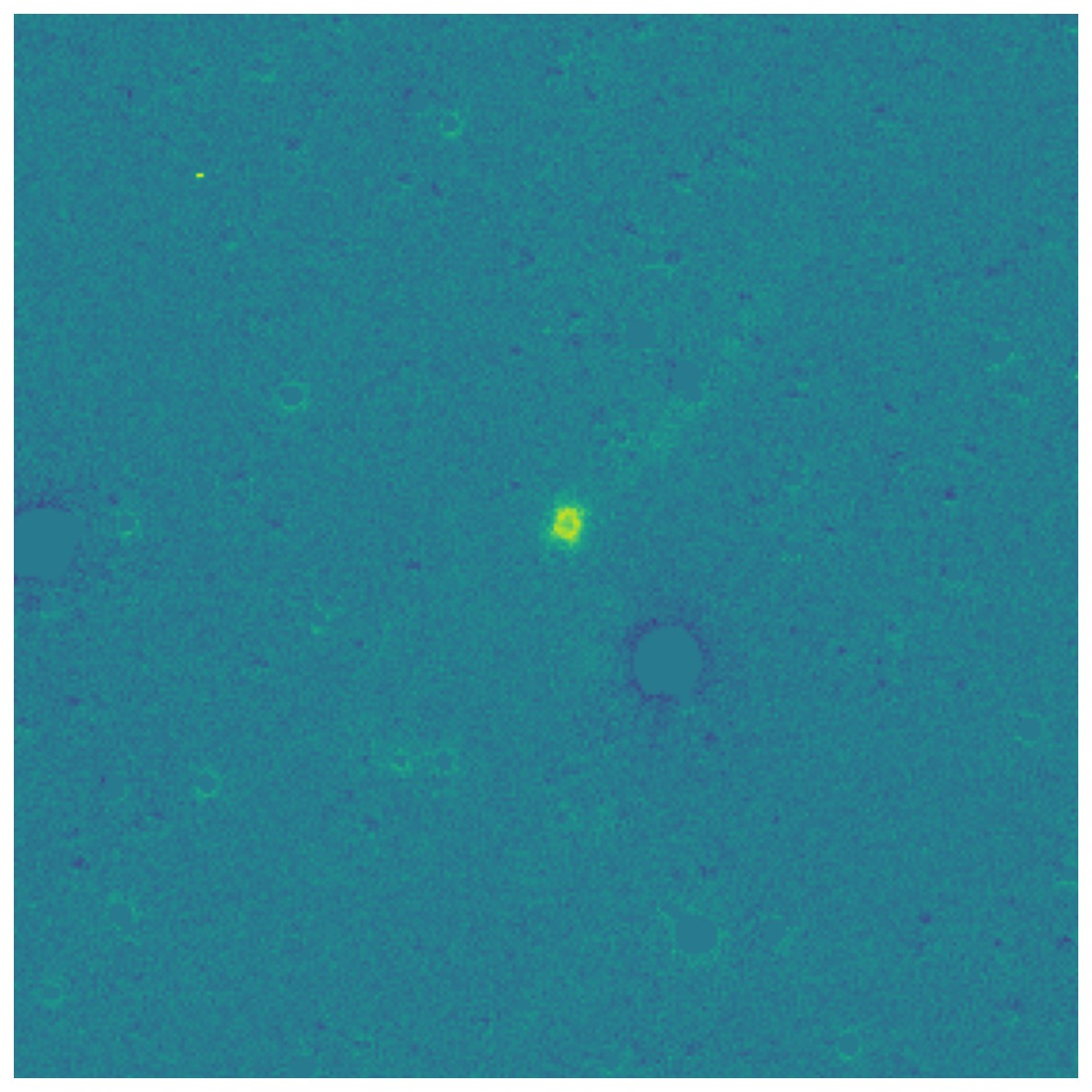}
\begin{overpic}[width=0.4\textwidth]{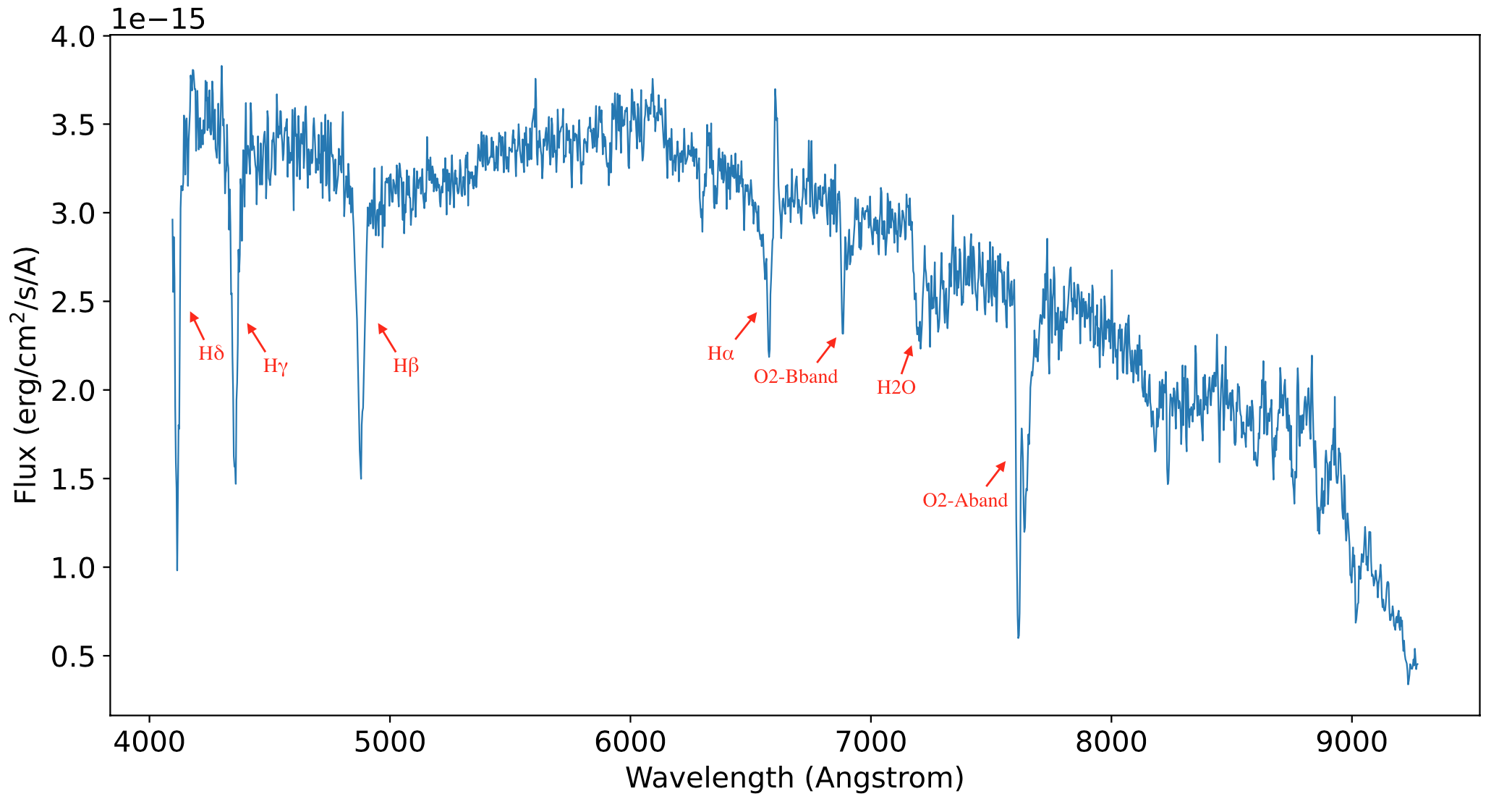}
\put(70,45){\small{YP1731-3011}}
\put(70,38){\small{White Dwarf}}
\end{overpic}
\hspace*{4.47in}
\begin{overpic}[width=0.4\textwidth]{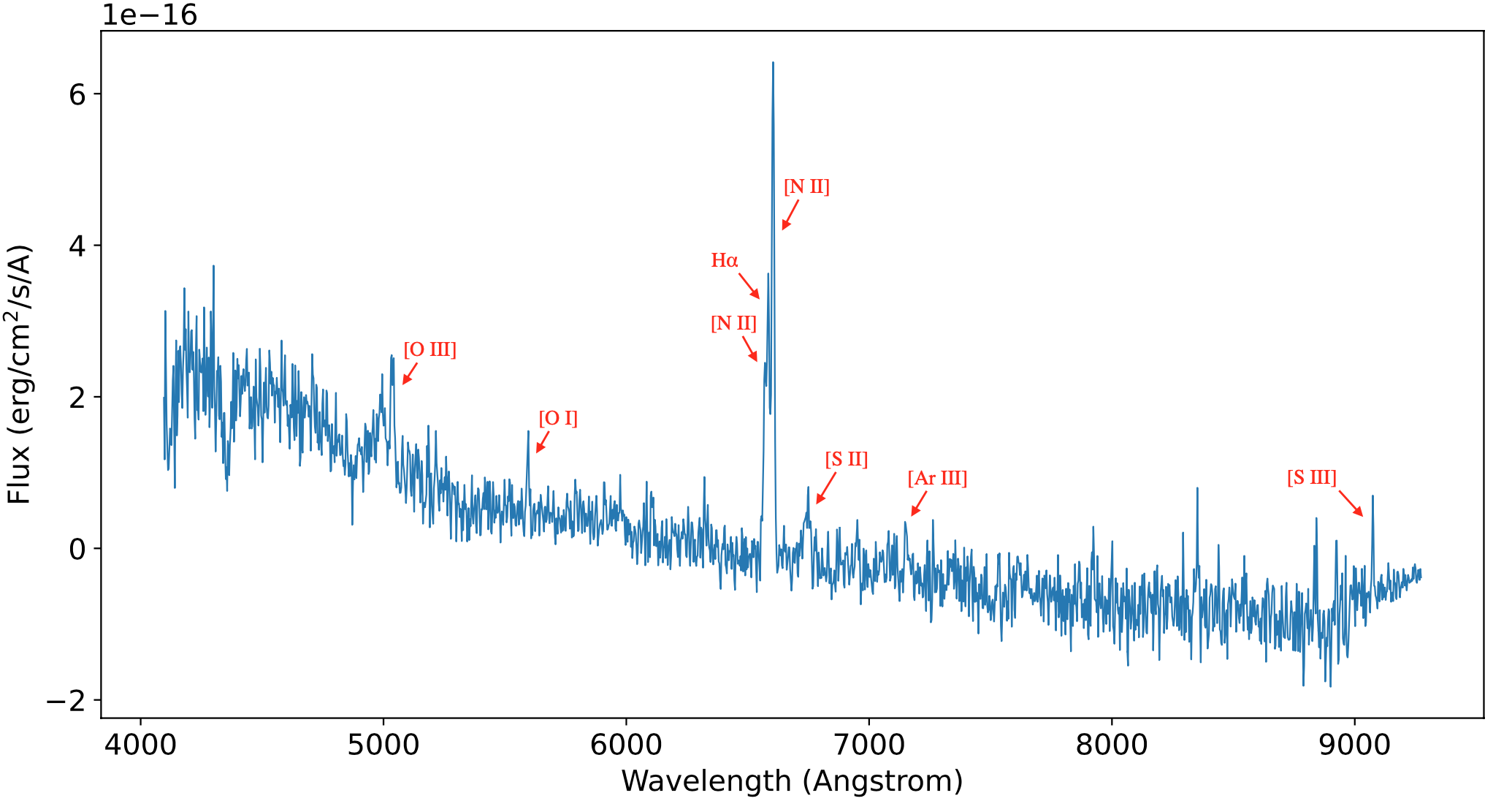}
\put(70,45){\small{YP1731-3011}}
\put(70,38){\small{True PN}}
\put(70,31){\small{Weak with a Jet}}
\end{overpic}
\caption{Selected True PNe Target Examples: VPHAS+ detected PNG images labelled by red boxes, SHS H$\alpha$-Rband quotient images, VPHAS+ H$\alpha$-Rband quotient images, SAAO spectra with spectral lines labelled.}
\label{fig:select}
\end{figure*}

Most of these PNe are either round or elliptical though some appear bipolar, with a few others 
exhibiting more irregular shapes. YP0821-4253 (the first two lines of Figure \ref{fig:select}) is the northwest part of a highly evolved bipolar 
nebula that appears to be fragmented, which we have further observation on every parts and will present in a separated paper. It is the largest target found in this work. YP1317-6513 (the third line of Figure \ref{fig:select}), 
YP1612-5209 and YP1759-2036 are bright targets with strong clear emission lines. YP1429-6214 (the forth line of Figure \ref{fig:select}) is a true bipolar PN right behind a foreground star. Typical PN emission lines of H$\alpha$, [N~II] and [O~III] are detected despite unavoidable contamination from the late-type star continuum of the assumed foreground star. YP1731-3011 (the last two lines of Figure \ref{fig:select}) is faint in the VPHAS+ detection image, but it showed an associated northwest jet and a possible southward emission bubble in the quotient images. Its 2-D spectrum can be extracted into a central white dwarf component with deep and wide Hydrogen absorption lines and a surrounding true PN spectrum. The H$\alpha$ line of the former has a possible P Cygni profile.

The rest of the PNe are small and faint, one of which approaches the limits of human visual 
resolution. They are often hidden within dense stellar fields, with some even concealed behind 
multiple stars. The discovery and confirmation of these nebulae demonstrate the powerful detection and robust interference rejection capabilities of our model that is able to find True PNe in such environments.

\begin{figure*}
\includegraphics[width=0.2\textwidth]{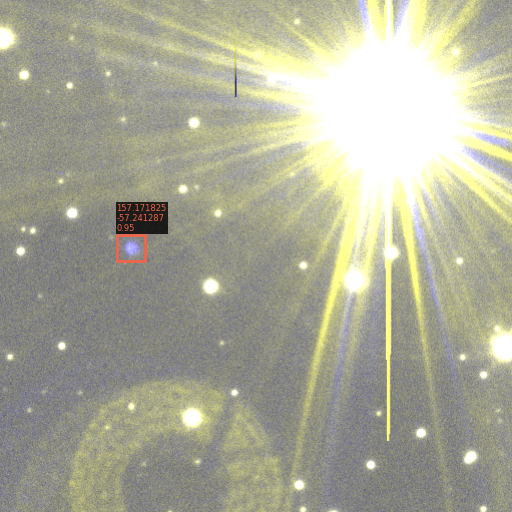}
\includegraphics[width=0.205\textwidth]{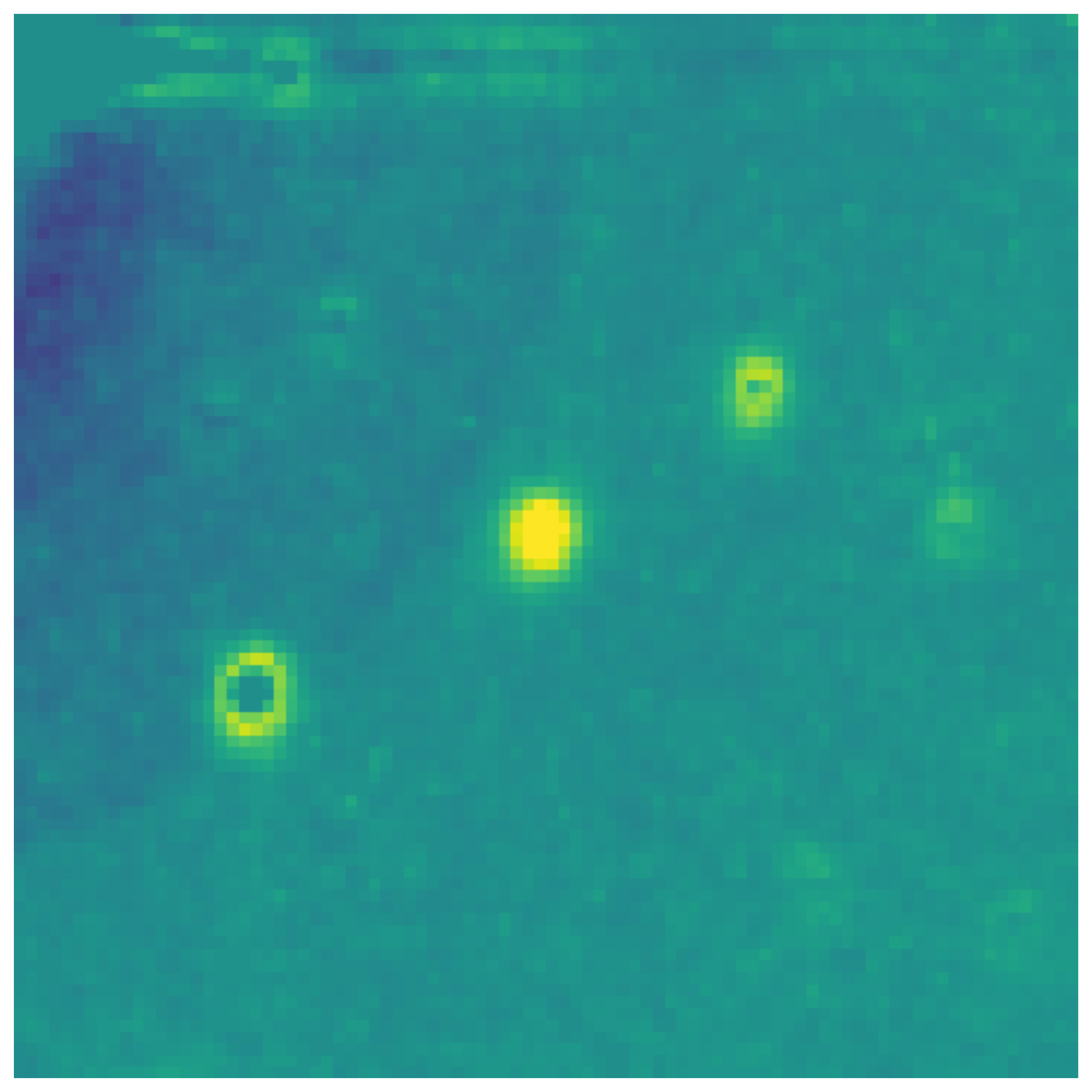}
\includegraphics[width=0.205\textwidth]{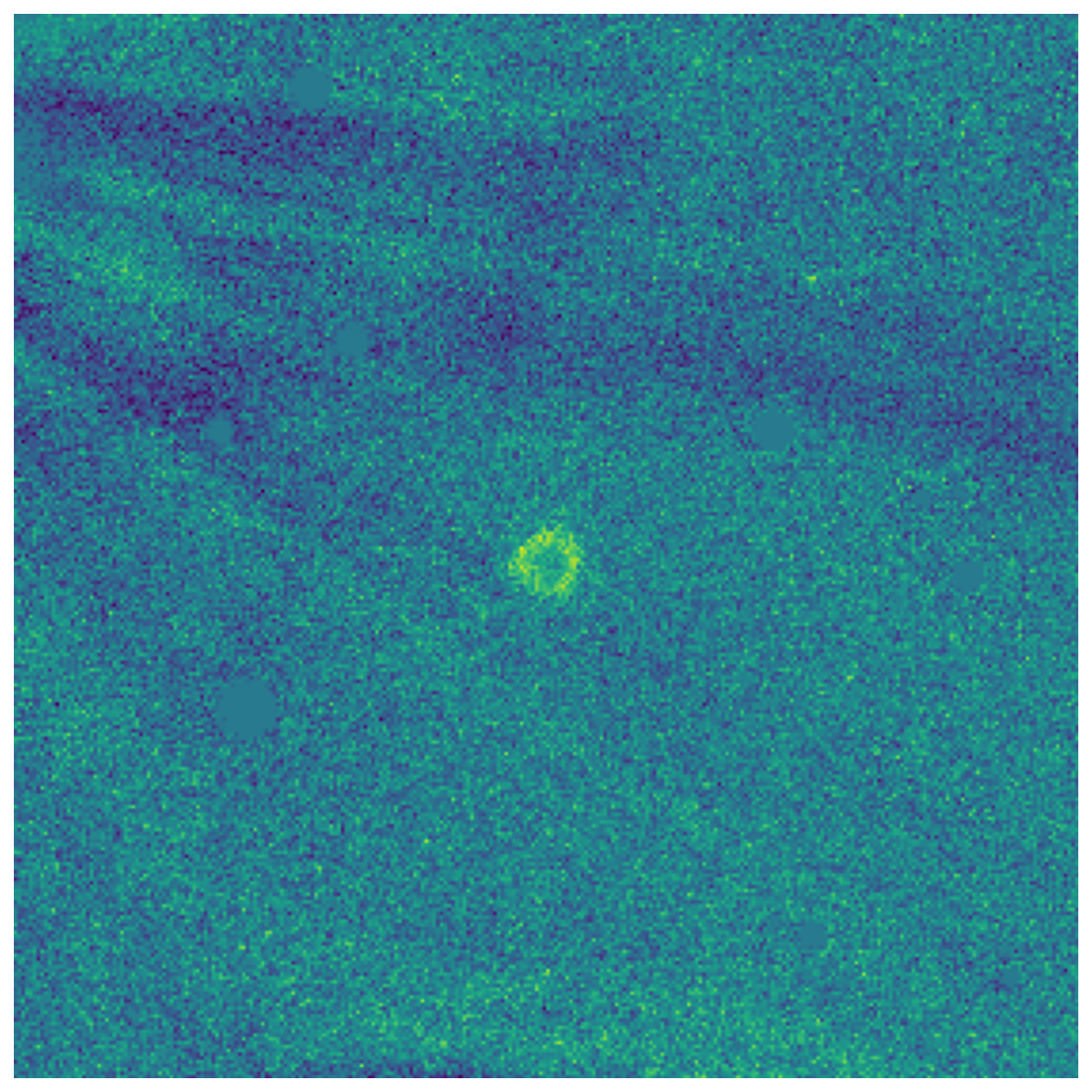}
\begin{overpic}[width=0.4\textwidth]{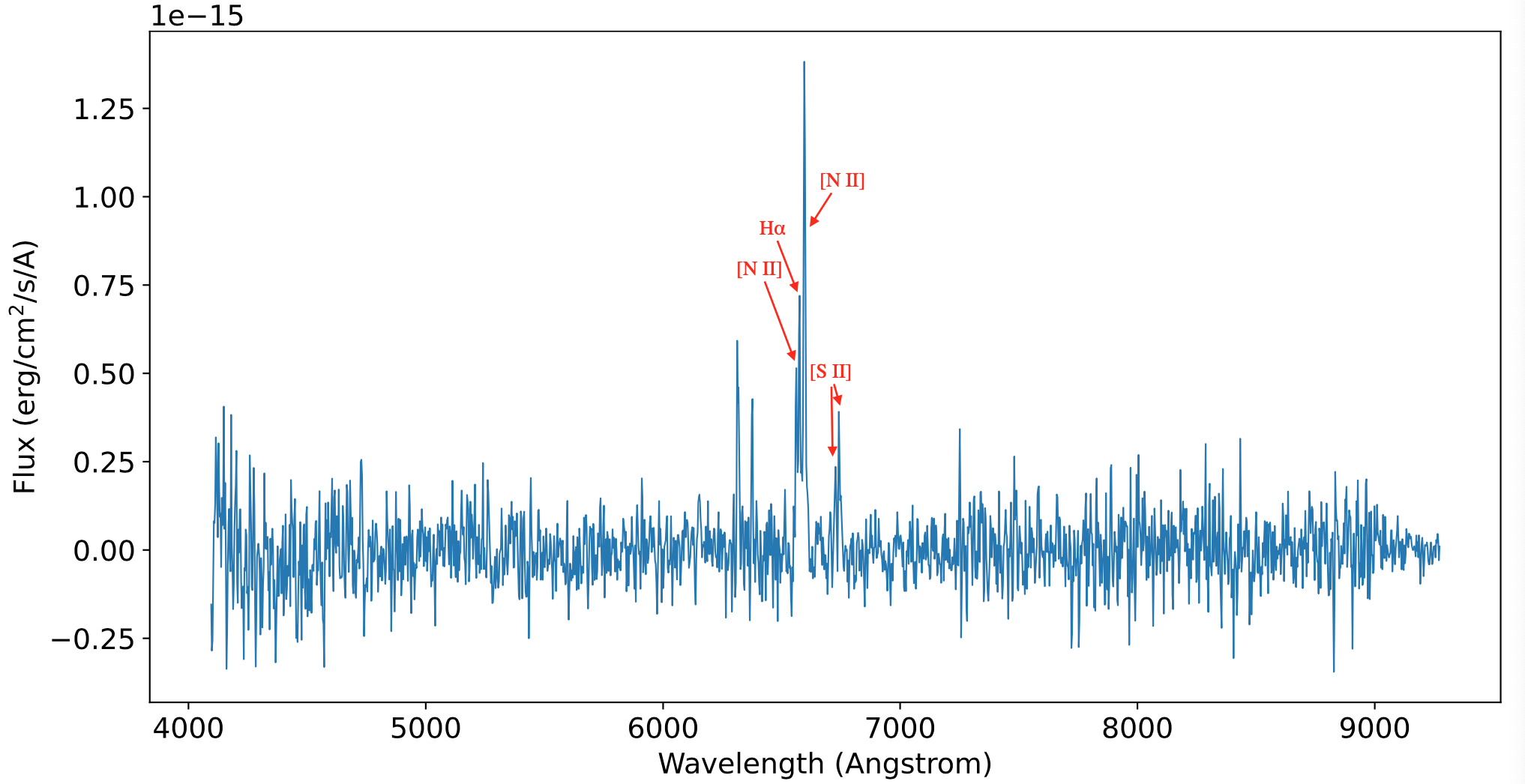}
\put(70,45){\small{YP1028-5714}}
\put(70,38){\small{Likely PN}}
\put(70,31){\small{Near a Bright Star}}
\end{overpic}
\includegraphics[width=0.2\textwidth]{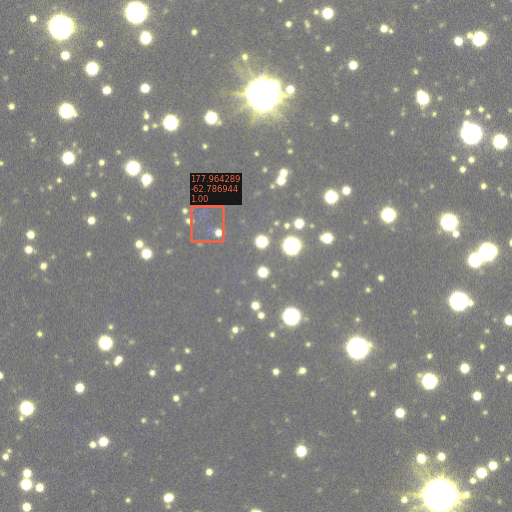}
\includegraphics[width=0.205\textwidth]{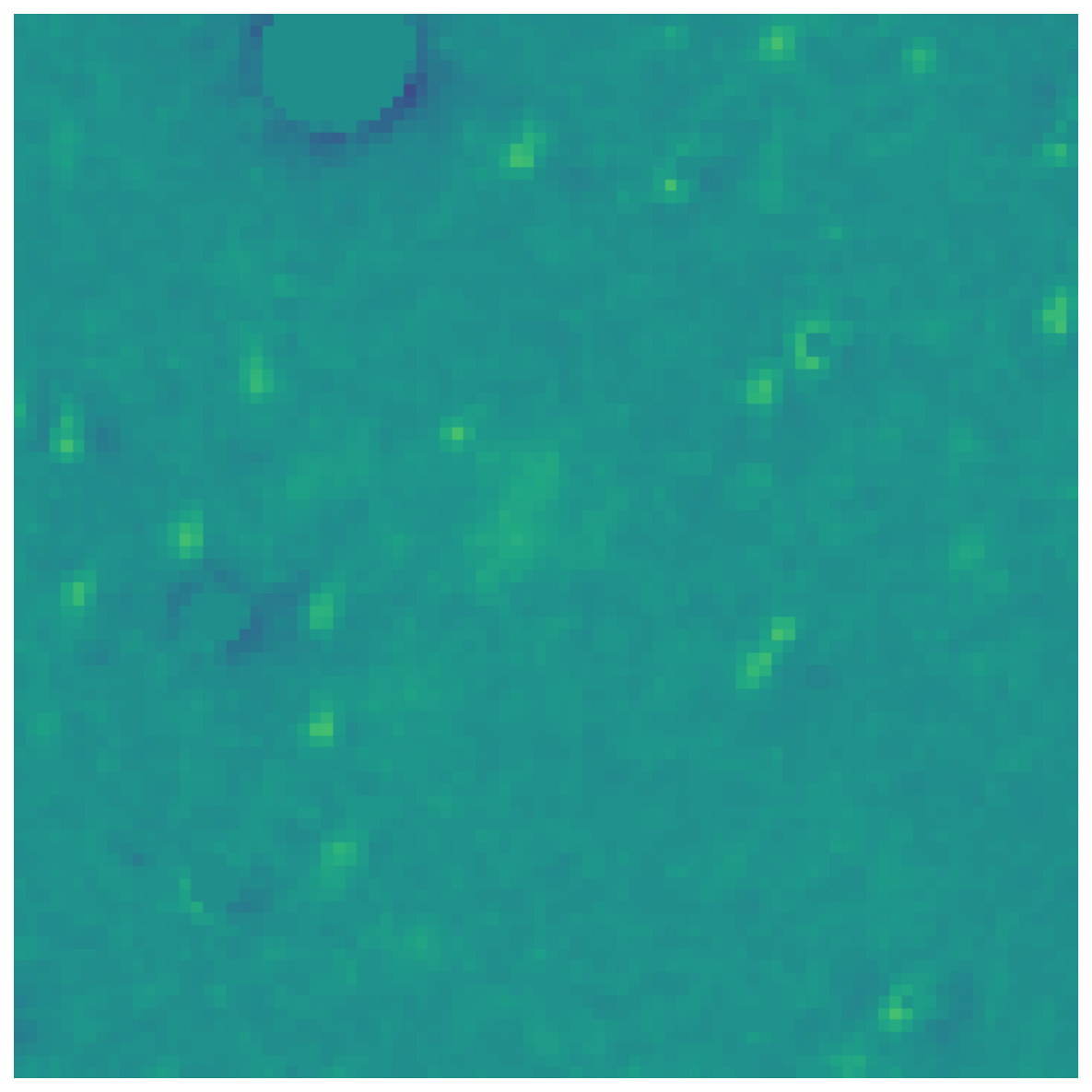}
\includegraphics[width=0.205\textwidth]{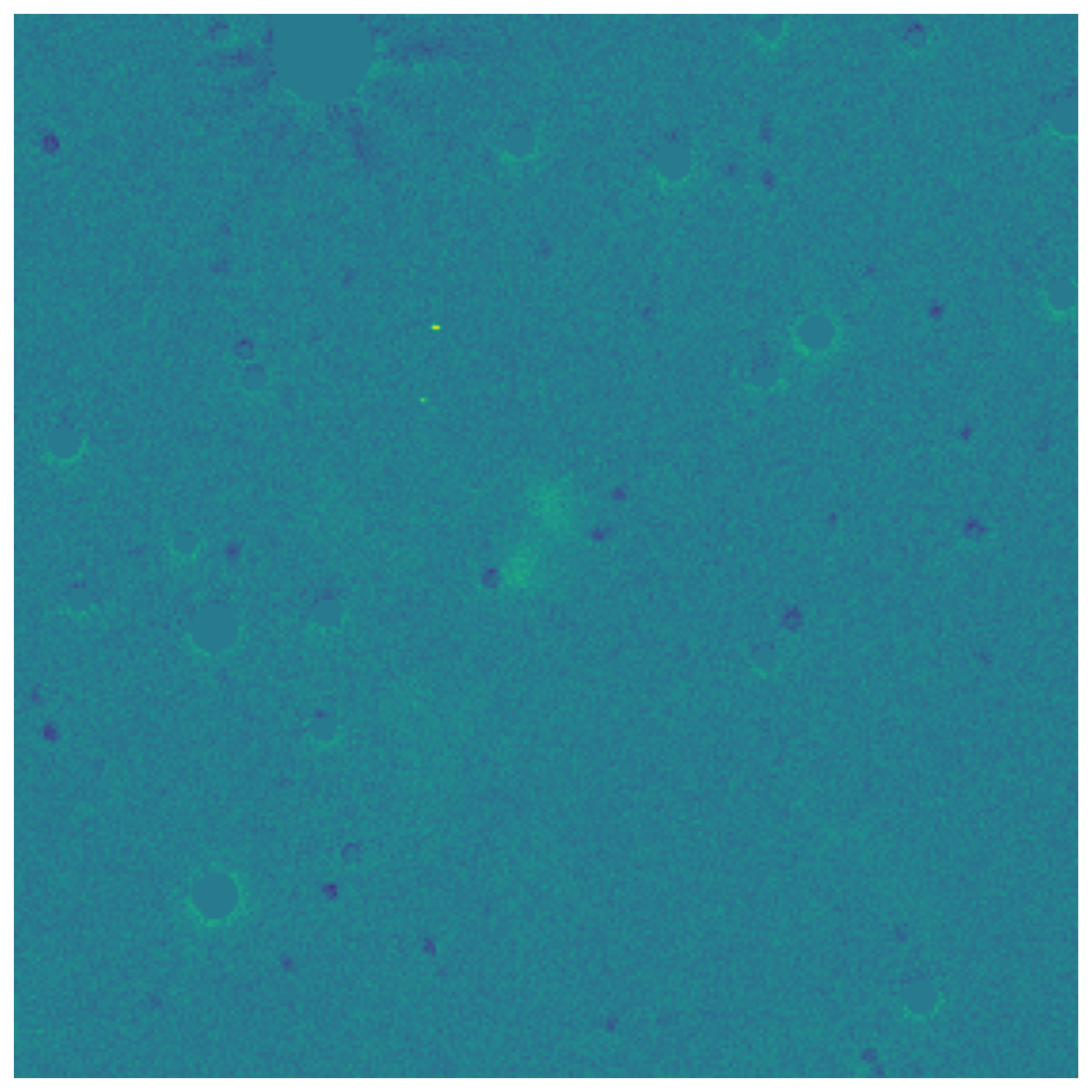}
\begin{overpic}[width=0.4\textwidth]{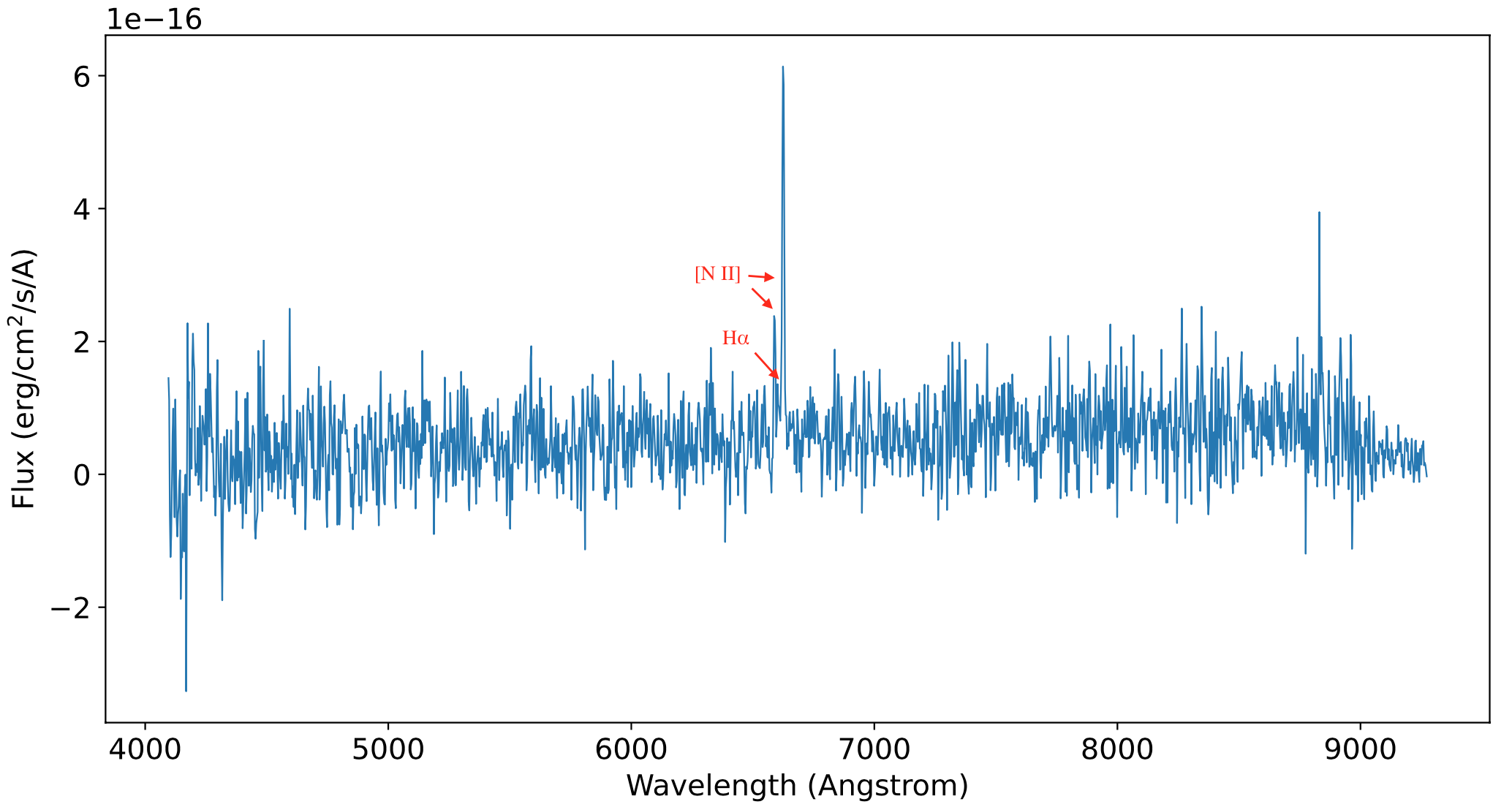}
\put(70,45){\small{YP1151-6247}}
\put(70,38){\small{EM Galaxy}}
\put(70,31){\small{With Two Blobs}}
\end{overpic}
\includegraphics[width=0.2\textwidth]{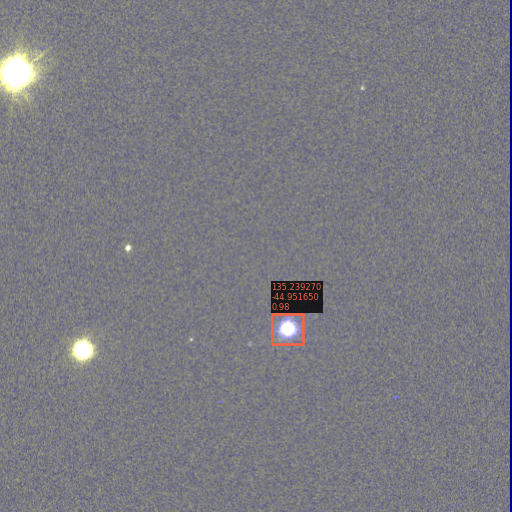}
\includegraphics[width=0.205\textwidth]{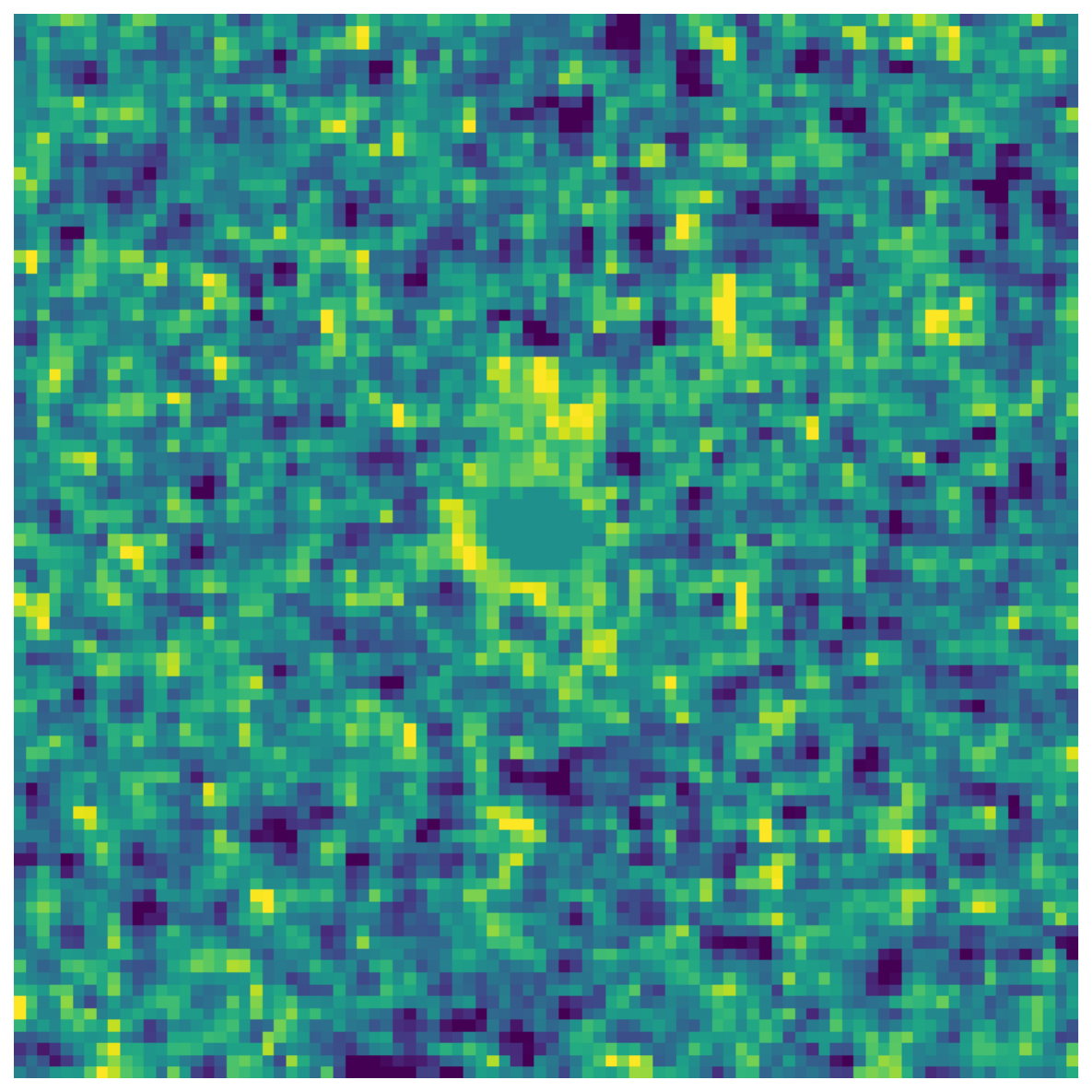}
\includegraphics[width=0.205\textwidth]{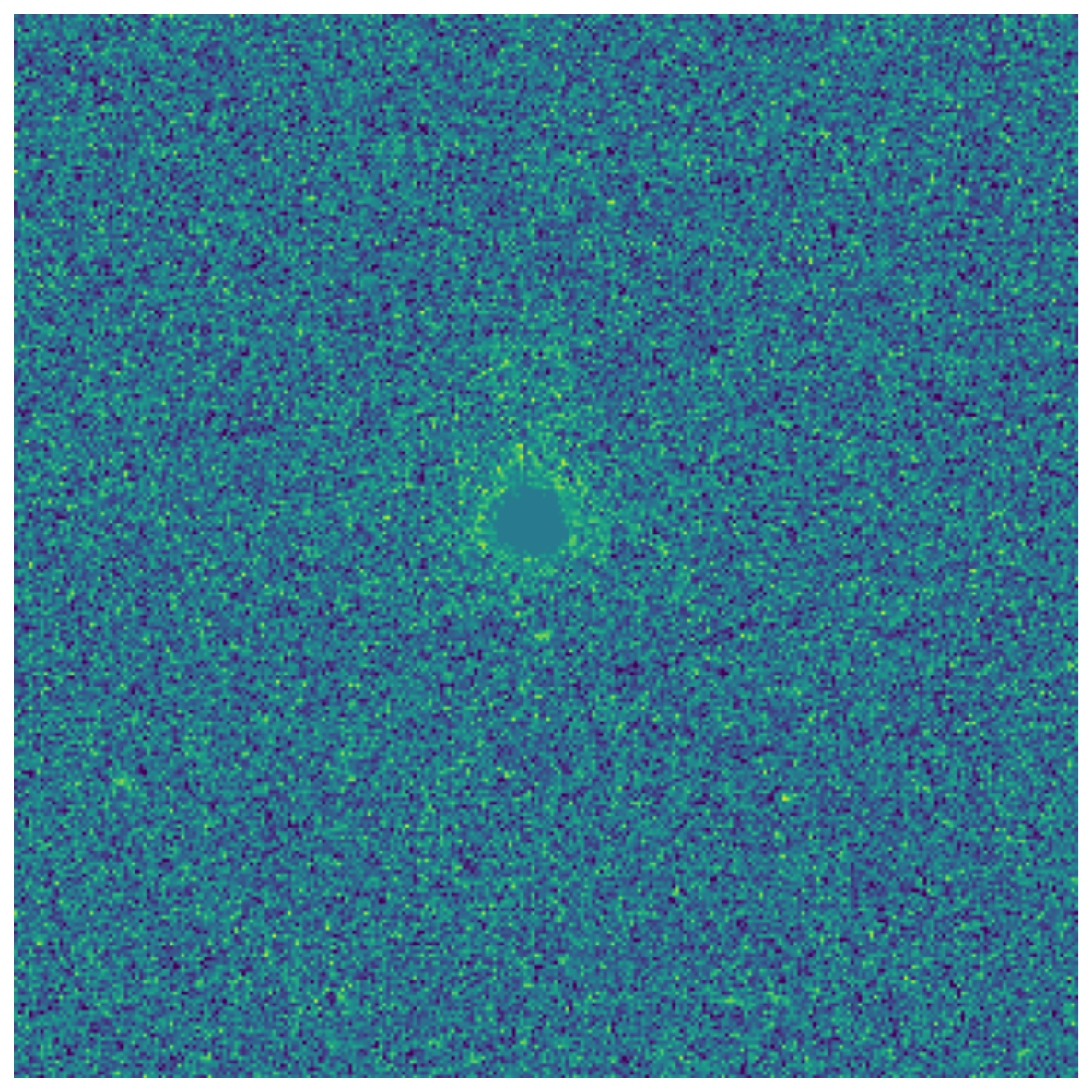}
\begin{overpic}[width=0.4\textwidth]{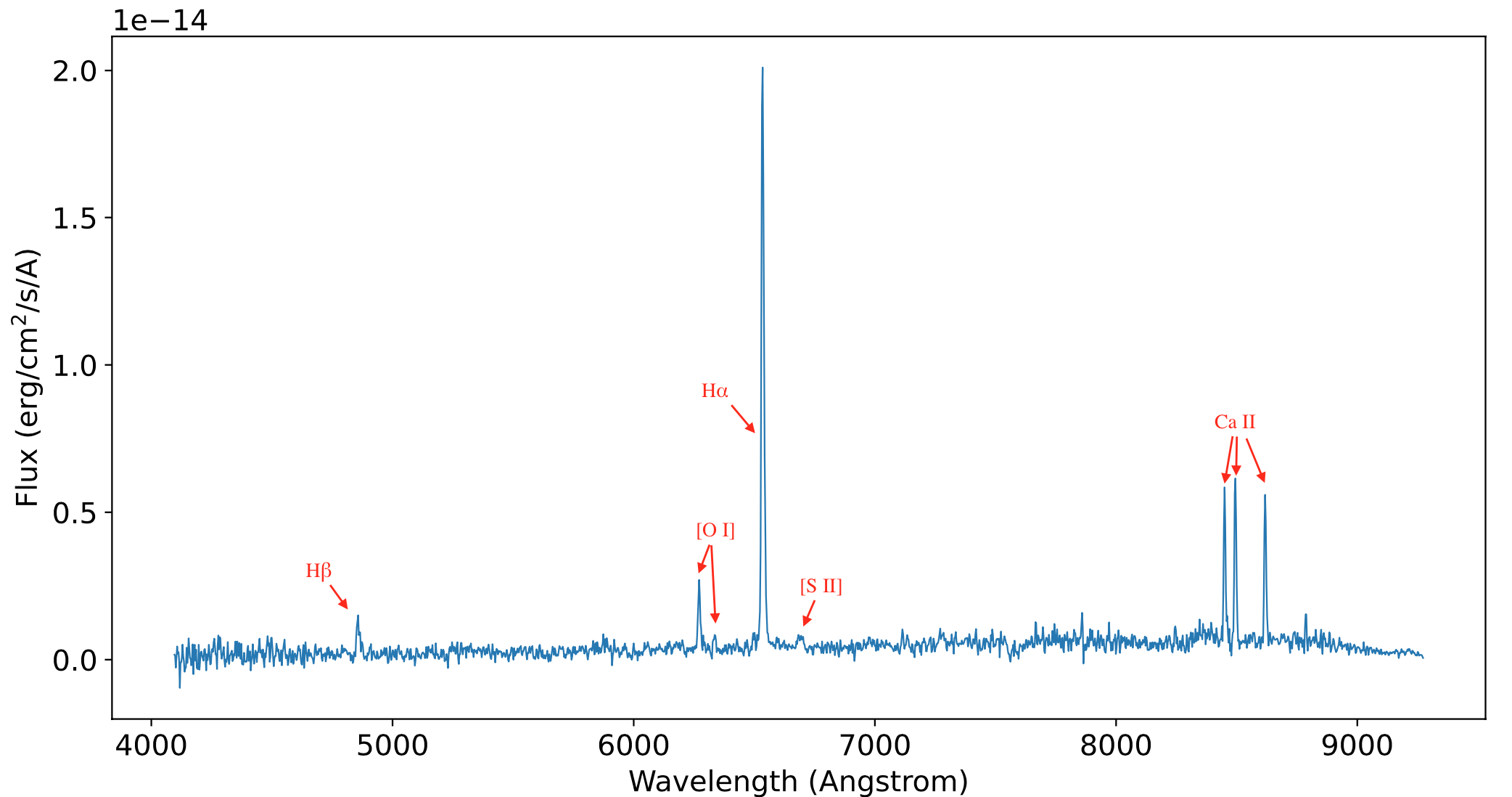}
\put(70,45){\small{YP0900-4457}}
\put(70,38){\small{EM Star}}
\put(70,31){\small{Sparse Star Field}}
\end{overpic}
\hspace*{4.47in}
\begin{overpic}[width=0.4\textwidth]{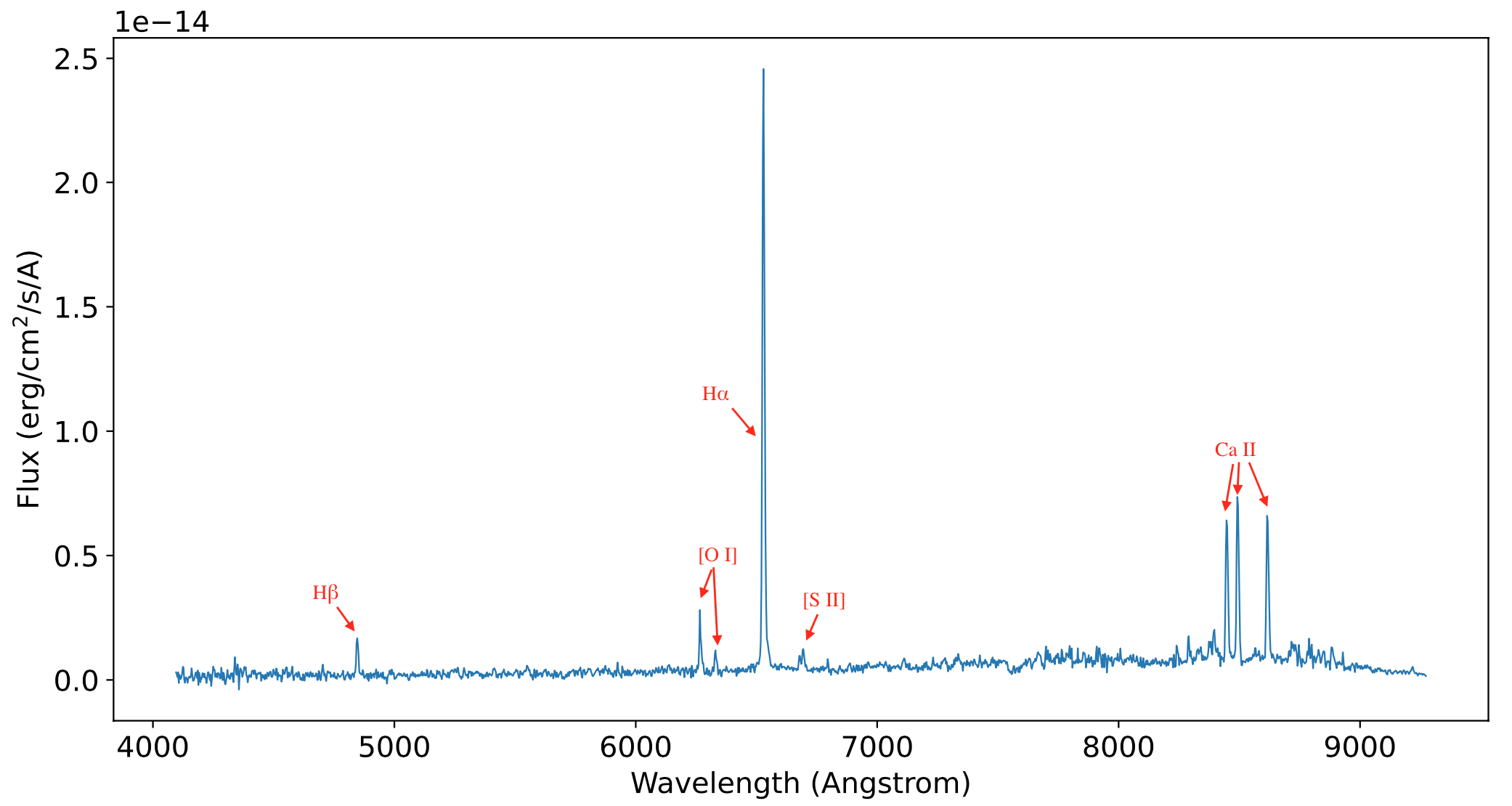}
\put(70,45){\small{YP0900-4457}}
\put(70,38){\small{Outflow}}
\end{overpic}
\includegraphics[width=0.2\textwidth]{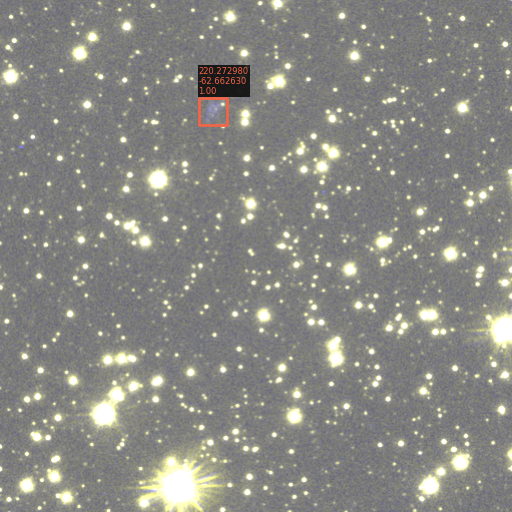}
\includegraphics[width=0.205\textwidth]{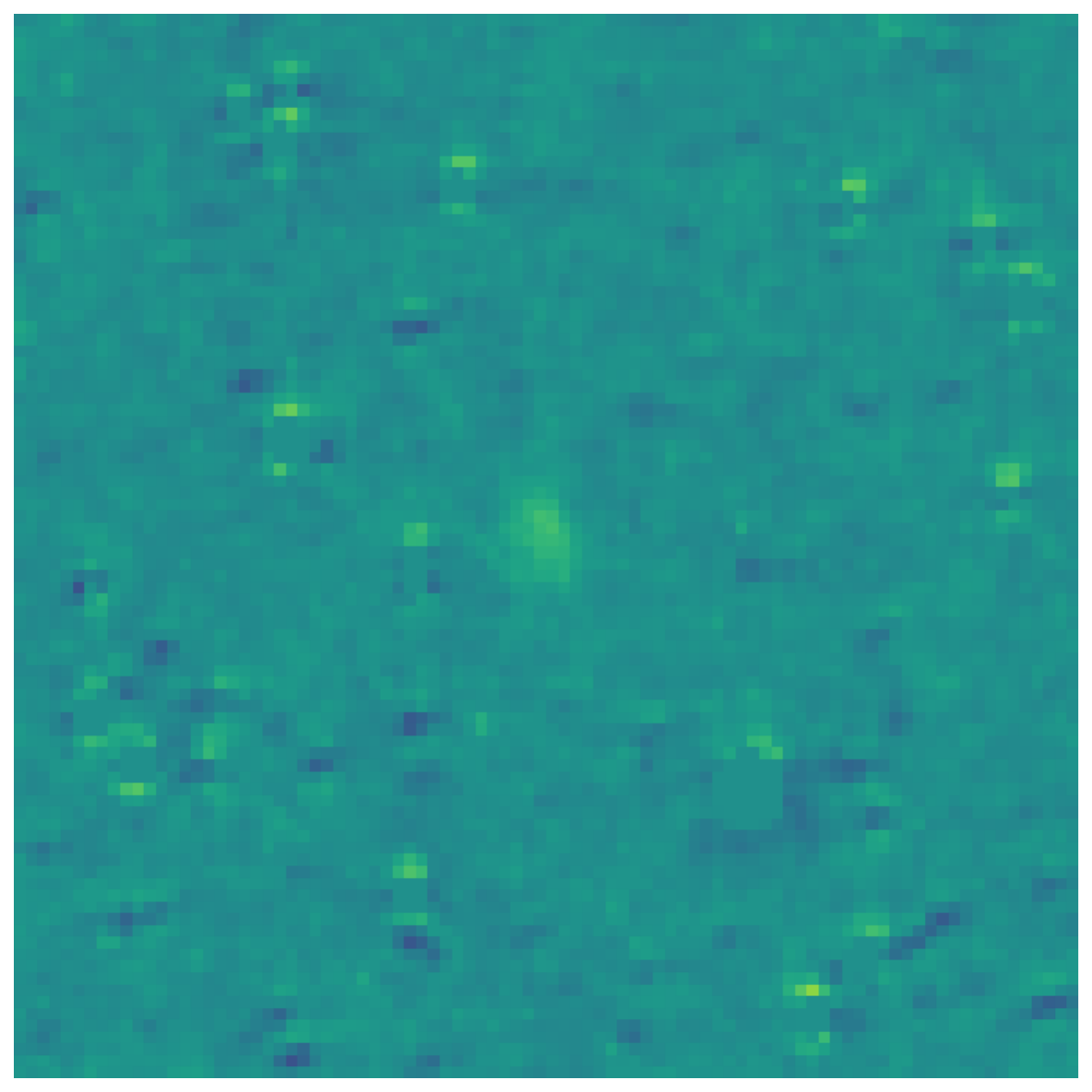}
\includegraphics[width=0.205\textwidth]{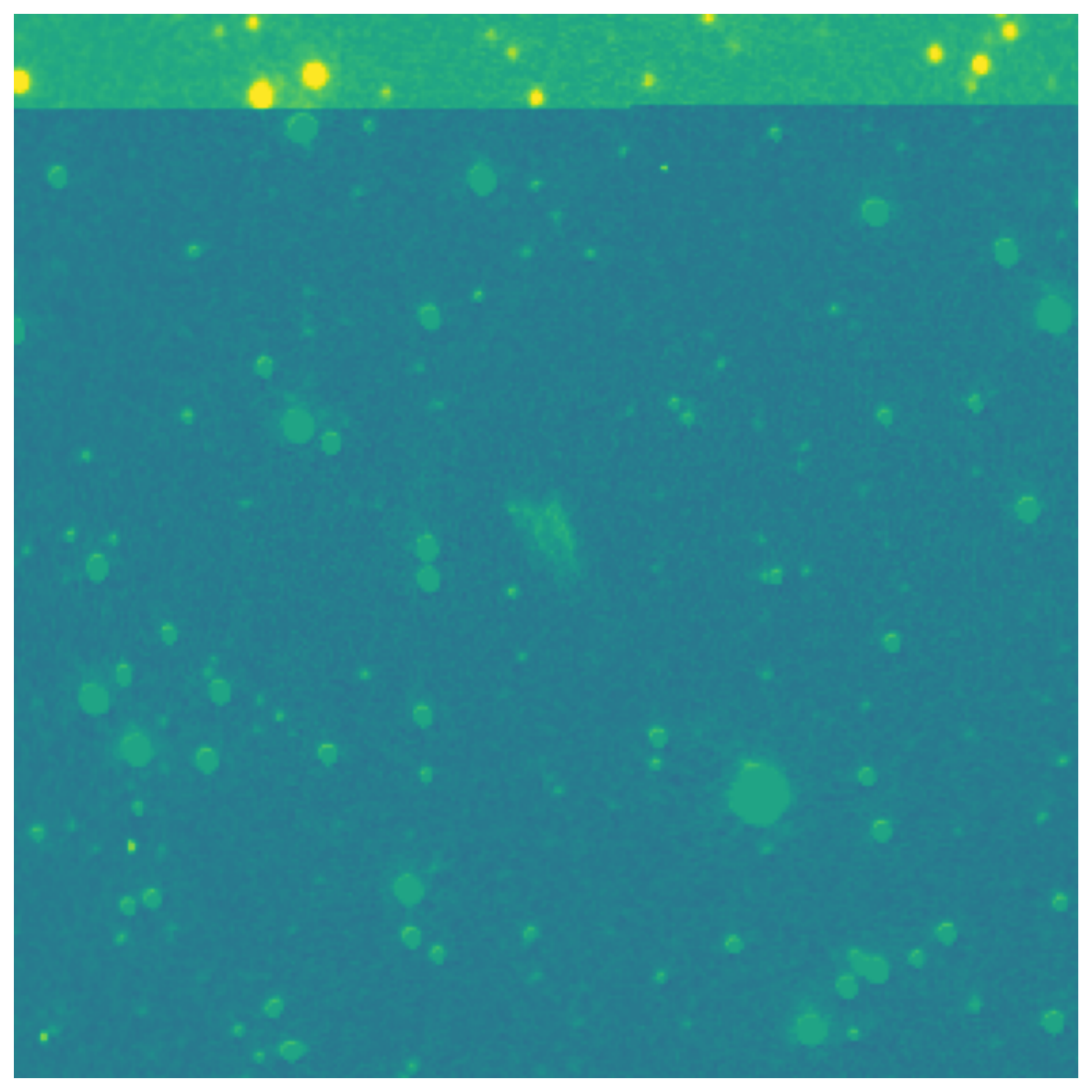}
\begin{overpic}[width=0.4\textwidth]{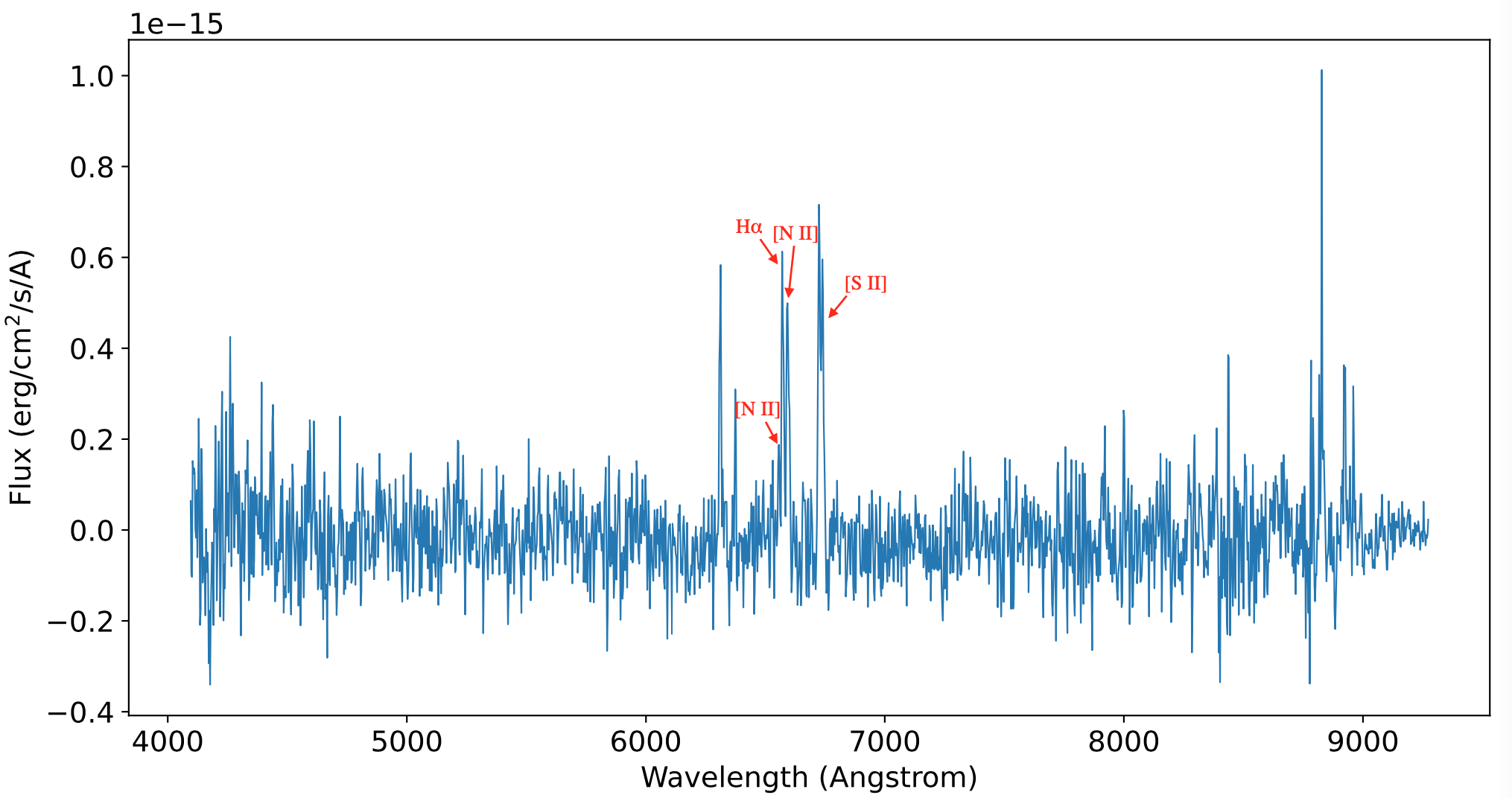}
\put(70,45){\small{YP1441-6239}}
\put(70,38){\small{SNR Fragment}}
\end{overpic}
\caption{Selected Likely \& Possible PNe, Emission-line Galaxies and Other Target Examples: VPHAS+ detected PNG images labelled by red boxes, SHS H$\alpha$-Rband quotient images, VPHAS+ H$\alpha$-Rband quotient images, SAAO spectra with spectral lines labelled.}
\label{fig:select1}
\end{figure*}

\subsection{Likely \& Possible PNe}
\label{sec:res:subsec:lp}

In this spectroscopic observation, we have further confirmed 2 likely 'L' PNe and 4 possible 'P' PNe 
again as assigned following our standard HASH precepts. This combined L, P PNe sample accounts for 
19.35\% of the total. These L and P PNe are shown in Figure \ref{fig:likely_pn} and Figure 
\ref{fig:possible_pn} respectively. These L and P classifications refer to objects that exhibit some 
standard PNe characteristics of nebular morphology and spectral features but may also exhibit other 
characteristics or uncertainties that introduce some doubt. Such objects may also possess some rare 
emission lines or line ratios atypical of PNe. Some are too faint to obtain spectra with sufficient 
signal-to-noise-ratios to accurately establish their nature, thus leading to a L or P categorization.

Most of these candidates are located within dense star fields and appear round or quasi-stellar.  
There is also one target (YP1028-5714, the first line of Figure \ref{fig:select1}) affected by the interference of an extremely bright star. 
These objects are generally very faint and difficult to discern. Their detection in complex stellar-interference environments showcases the strong performance of our model.

\subsection{Other types of emission-line sources or mimics}

Despite being trained on a core population of known PNe  our AI/ML algorithm is still able to identify all kinds of H$\alpha$ emitters in narrow-band images, regardless of whether they are PNe or compact HII regions, Wolf-Rayet shells, detached components of supernova remnants, stromgren zones,  emission-line stars, or even low-redshift emission-line galaxies whose H$\alpha$ lines fall within the filter bandpass. Hence we may need to further refine the precise classification of targets through subsequent mult-wavelength image inspections and spectroscopic observations. Therefore, our sample also includes a small number of other types of celestial objects.

\subsubsection{Emission-line galaxies}
\label{sec:res:subsec:gal}
We uncovered 3 narrow-line emission line galaxies in this work (see Figure \ref{fig:em_gal} and Table 
\ref{tab:redshift}) via the observed $30-50$ \AA\ redshifts exhibited by the emission lines (also typically seen in 
PNe) in their spectra. One of the sources, YP0827-3033, with Z=0.0062 is very compact, YP1151-6247 (the second line of Figure \ref{fig:select1}) with Z=0.00414 
appears as two slightly separated oval emission blobs initially considered as a bi-polar PN but now likely two 
interacting galaxies. Interestingly the [N~II]/H$\alpha$ ratio is 8.6. The third example is YP1701-2519, a small 
but clearly resolved emission region with a Z=0.0059. The ability of the VPHAS+ survey to be sensitive to emission 
line galaxies depends on the filter bandpass and the redshift of the emission line galaxy. The filter has central 
wavelength and FWHM of 6588\AA\ and 107\AA, respectively \citep{2005MNRAS.362..753D}.

\begin{table}[!htbp]
\centering
\caption{Redshift and estimated distance of the 3 emission-line galaxies from spectral measurement. The Hubble 
constant value assumed here is 70 (km/s)/Mpc.}
\label{tab:redshift}
\renewcommand\arraystretch{1.18}
\begin{tabular}{lcc} 
 \hline
Common Name & Redshift Z & Distance (Mpc) \\
 \hline
YP0827-3033 & 0.0062 & 26.57 \\
YP1151-6247 & 0.00414 & 17.74 \\
YP1701-2519 & 0.0059 & 25.29 \\
\hline
\end{tabular}
\end{table}

\subsection{Emission-line stars and late-type stars}
\label{sec:res:subsec:star}
As part of our sample verification strategy we selected a series of compact, apparently unresolved sources for 
observation that appeared to have obvious H$\alpha$ emission given the disparity in intensity between the H$\alpha$ 
image and the broad-band red equivalent. These were suspected of being either emission-line stars or perhaps late-
type star contaminants. This selection was done blindly from the AI/ML provided VPHAS+ target lists without 
reference to other imagery (which could have been indicative of likely source type). Spectral analysis of these 
selected cases revealed that two (6.45\% of the total) were indeed late-type stars dominated by strong molecular 
bands (see Figure \ref{fig:lt_star}). As described in \cite{2006MNRAS.373...79P}, these arise because of how the 
molecular band distribution falls within the H$\alpha$ filter bandpass while simultaneously increasing in strength 
of peak intensity as one moves further to the red. This gets selected by the H$\alpha$ filter whereas the band 
influence is diluted by their inclusion in the broad-band red-filter. The apparent image shape of YP1705-2524 (one 
of these late type star contaminants) is slightly elongated 
due to the presence of another relatively faint star nearby. Two emission line stars (EM stars) were also found, 
again accounting for 6.45\% of the total (see Figure \ref{fig:em_star}). One of the EM stars (YP0900-4457, the third and forth lines of Figure \ref{fig:select1}) is located in a 
sparsely populated stellar field region indicating a strong absorption screen behind the stars that are visible. 
The spectrum exhibits multiple very strong emission lines with an nebular outflow to the north. A spectrum of this 
outflow was also taken which has a very similar spectrum to that of the assumed host star. However, there is 
another possibility that our spectroscopic observation missed the outflow and instead captured the edge of the 
star, which requires further examination. Both emission line stars discovered show Ca~II triplet lines in the 
infrared in emission. These results show that selection of these types of apparent emitter could provide a rich 
vein of faint emission lines stars for further study - more examples are needed to statistically establish the true 
emission versus late-type star fractions which here are nominally 50\% each. 

\subsection{Others}
\label{sec:res:subsec:other}

In this spectroscopic observation, we also identified 2 other targets, accounting for 6.45\% of the total (see Figure \ref{fig:others}). 

One of them turns out to be a small detached fragment of the famous supernova remnant RCW 86 (YP1441-6239, the last line of Figure \ref{fig:select1}). Our AI network 
detected multiple similar targets and the exclusion of them from our PNe searching requires much larger-scale 
surrounding images for judgment.

The other remaining observed target independently identified turns out to be a known H$\alpha$ emission object 
located within the super star cluster Westerlund 1. It has been previously studied using VPHAS+, HST, and VLT data, 
see \cite{2014MNRAS.437L...1W}, and identified as an ionized hydrogen gas cloud surrounding a red super-giant.

\section{Conclusions}
\label{sec:con}
We have presented the first discovery and confirmation of PNe candidates found from deep learning techniques 
applied to high-resolution VPHAS+ survey data. We have proven our techniques are able to independently and 
automatically uncover faint, resolved PNe in very dense star fields near the Galactic centre using the high-resolution H$\alpha$ survey VPHAS+. These targets are often beyond 
the reach of traditional detection software and often visual scrutiny. Our automated techniques, once trained as 
described in Paper~I save the significant manpower required of eye-inspection applied to
the large imaging survey datasets. To achieve these very encouraging results we developed a novel Swin-Transformer 
algorithm (see \cite{2024MNRAS.528.4733S}) with an IPHAS training sample based on the existing inventory of PNe 
curated within the HASH PNe database \citep{2016JPhCS.728c2008P}. We found more than 800 high-quality candidates 
and selected 31 of them for initial spectral observation at SAAO. We spectroscopically confirmed that 22 are true 
(T), likely (L) and possible (P) PNe giving a formal confirmation level of 70.97\%. These PNe
are mostly very faint and located in dense fields, with several directly partially obscured behind stars or under 
the glow/interference of a very bright nearby stars. We also found a CSPN white dwarf with an interesting H$\alpha$ 
jet. Of the remaining 9 sources 7 were emission line sources comprising 3 emission line galaxies, 2 emission lines 
stars, a component of a supernova and an emission blob in a star cluster. Only 2/31 objects were non-emitters being classed as late-type stars with string red molecular bands. This give an overall success rate for detection of emission sources of all kinds at 29/31 or 93.5\%.

For the future we will continue working on the other half of VPHAS+ fields, stacking them for fainter nebulae, and improved searching techniques based on the fusion of all available the wavebands. And a pixel-to-pixel H$\alpha$ nebulae map is also planned for a more complete search for faint PNe and other kinds of nebulae.

\begin{acknowledgements}
QAP thanks the Hong Kong Research Grants Council for GRF research support under grants 17326116 and 17300417. YL thanks HKU and QAP for provision of a PhD scholarship from RMGS funds awarded to the LSR. This work is also supported by National Natural Science Foundation of China (NSFC) with funding number of 12303105, 12173027 and 12173062 and Civil Aerospace Technology Research Project (D050105). We acknowledge the science research grants from the China Manned Space Project with NO. CMS-CSST-2021-A01 and science research grants from the Square Kilometre Array (SKA) Project with NO. 2020SKA0110102. We thank the SAAO for observing time. 
\end{acknowledgements}


\begin{appendix}
\begin{figure*}
\includegraphics[width=0.2\textwidth]{fig/YP0821-4253.png}
\includegraphics[width=0.205\textwidth]{fig/YP0821-4253_shs_q.pdf}
\includegraphics[width=0.205\textwidth]{fig/YP0821-4253_vphas_q.pdf}
\begin{overpic}[width=0.4\textwidth]{fig/YP0821-4253_spec.png}
\put(70,45){\small{YP0821-4253}}
\end{overpic}
\hspace*{1.44in}
\includegraphics[width=0.205\textwidth]{fig/YP0821-4253_shs_q_l1.pdf}
\includegraphics[width=0.205\textwidth]{fig/YP0821-4253_vphas_q_l.pdf}\\
\includegraphics[width=0.2\textwidth]{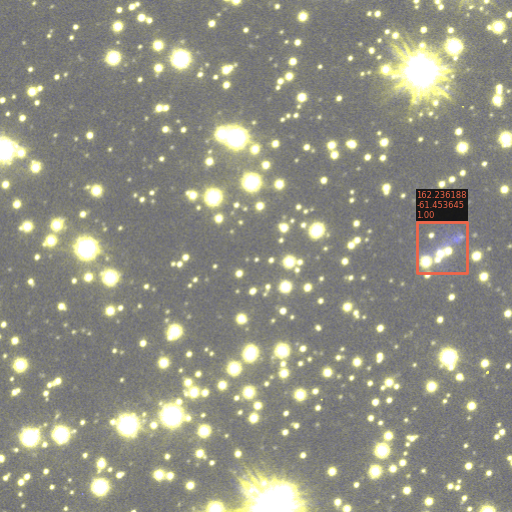}
\includegraphics[width=0.205\textwidth]{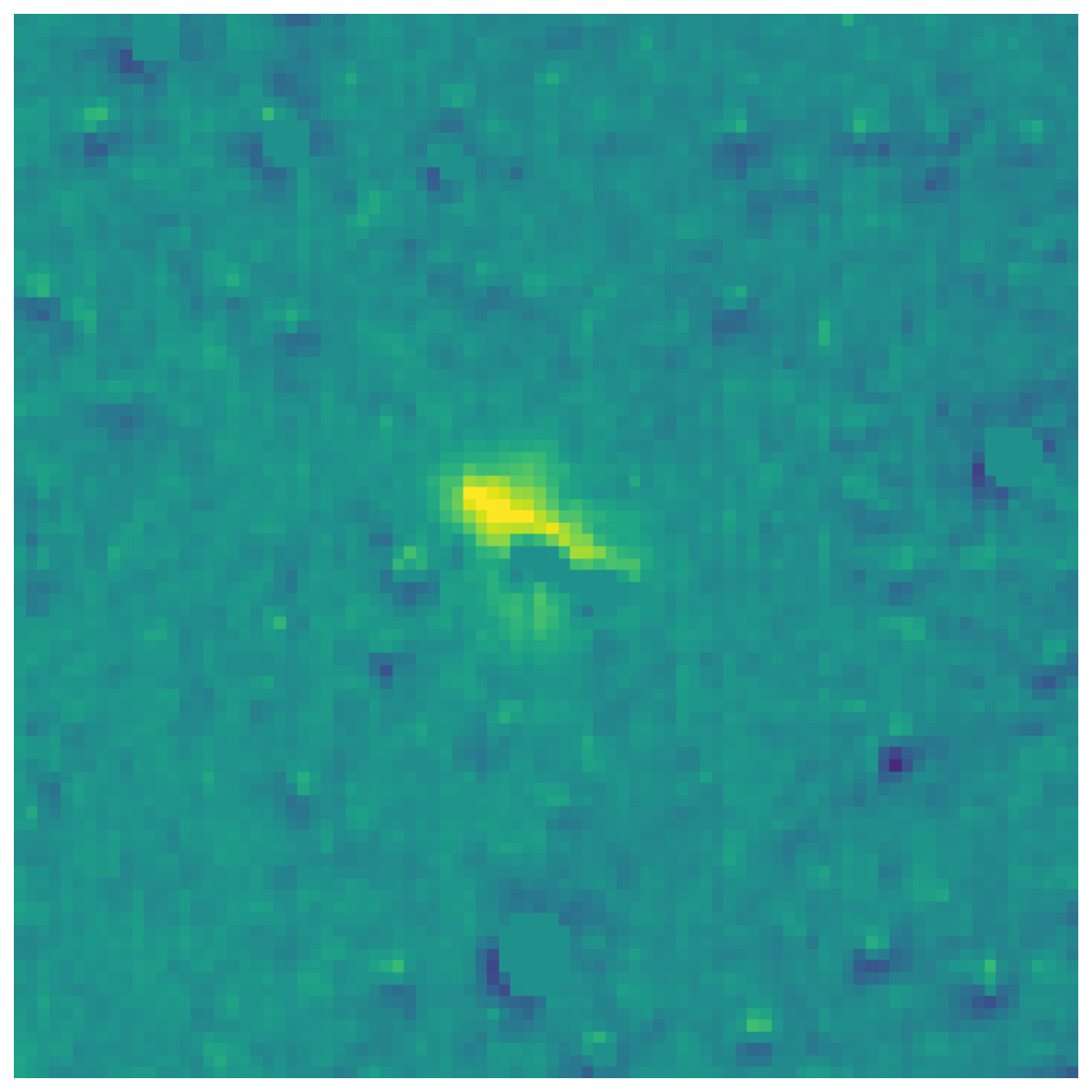}
\includegraphics[width=0.205\textwidth]{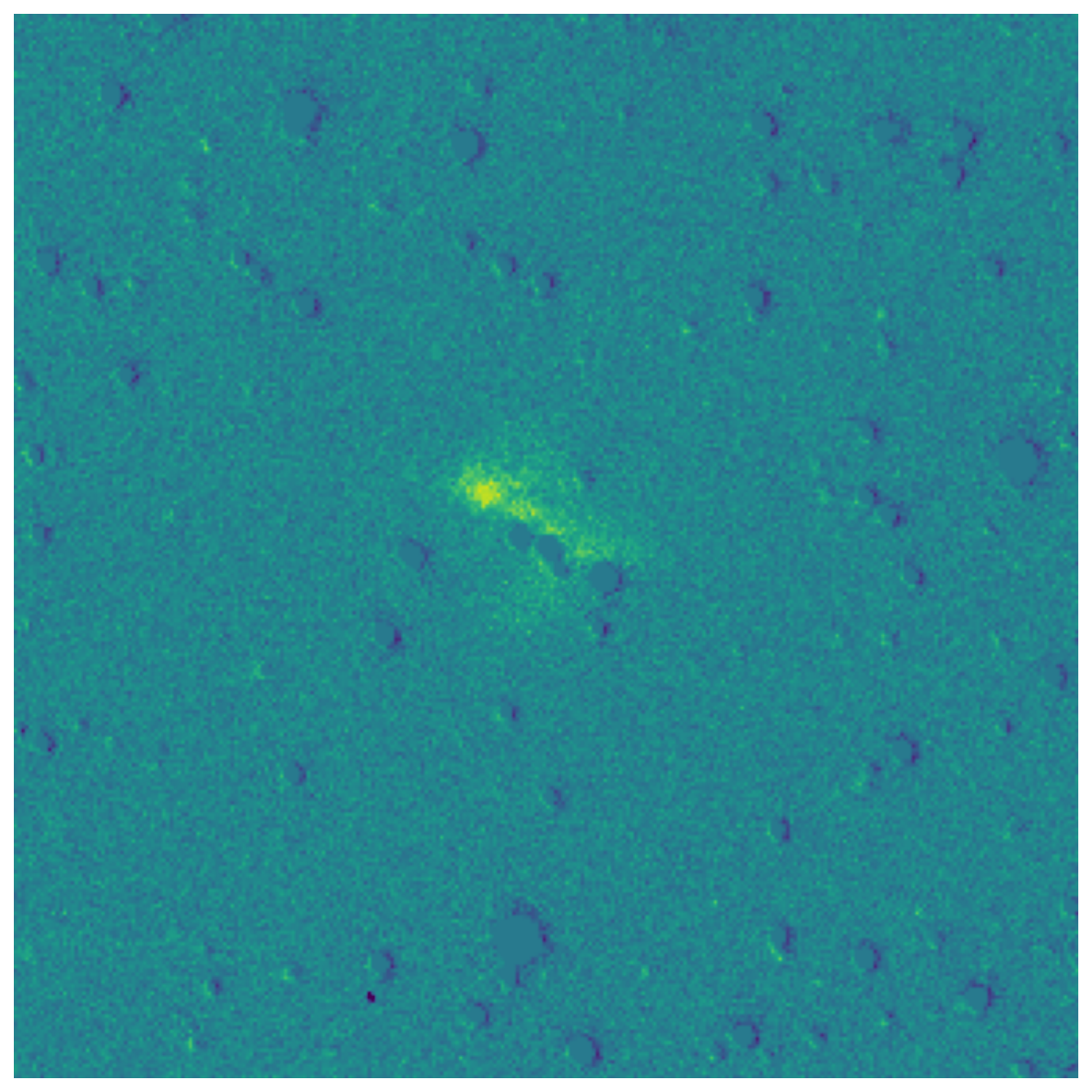}
\begin{overpic}[width=0.4\textwidth]{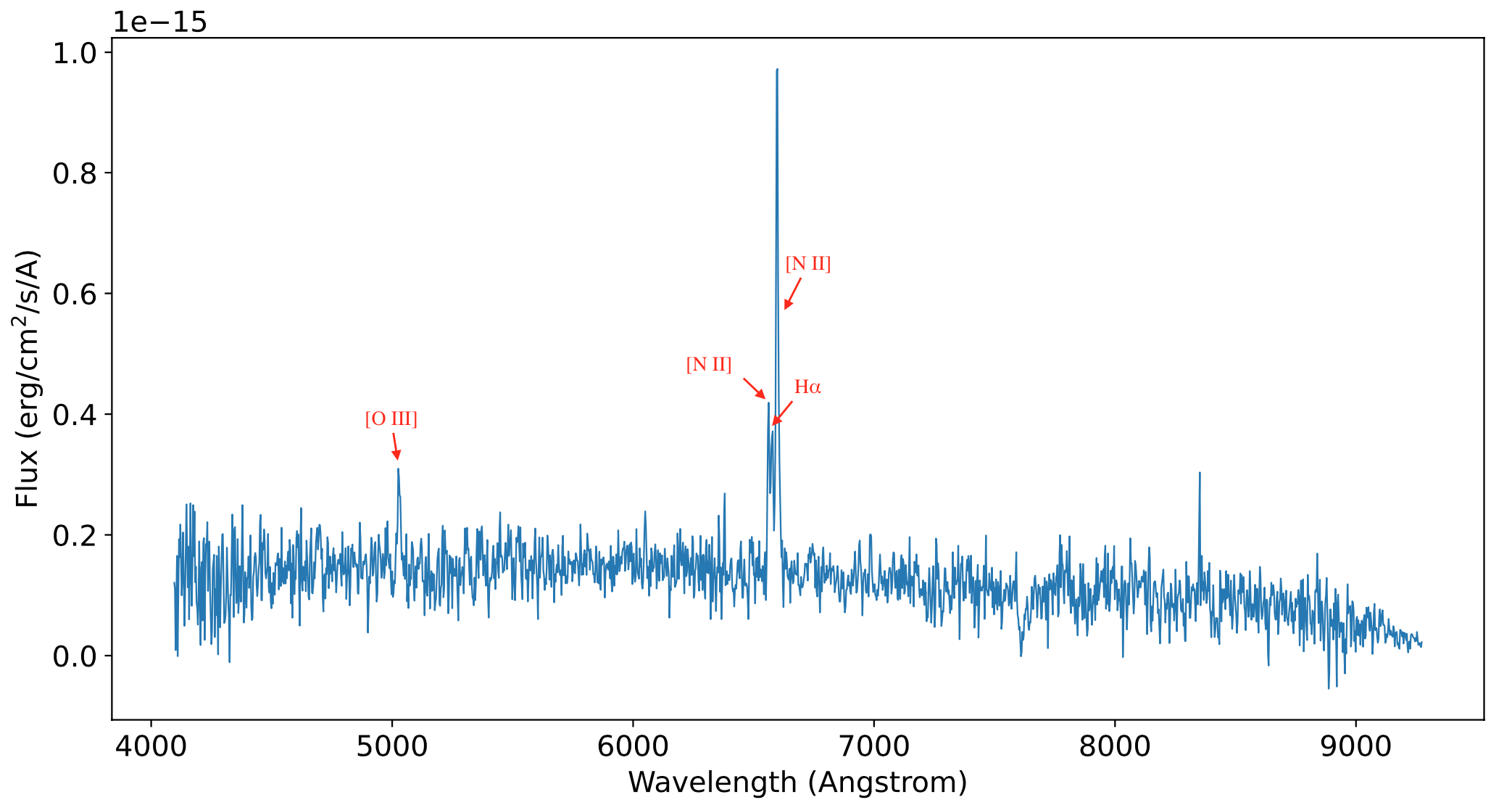}
\put(70,45){\small{YP1048-6127}}
\end{overpic}
\includegraphics[width=0.2\textwidth]{fig/YP1317-6513.png}
\includegraphics[width=0.205\textwidth]{fig/YP1317-6513_shs_q.pdf}
\includegraphics[width=0.205\textwidth]{fig/YP1317-6513_vphas_q.pdf}
\begin{overpic}[width=0.4\textwidth]{fig/YP1317-6513_spec.png}
\put(70,45){\small{YP1317-6513}}
\end{overpic}
\includegraphics[width=0.2\textwidth]{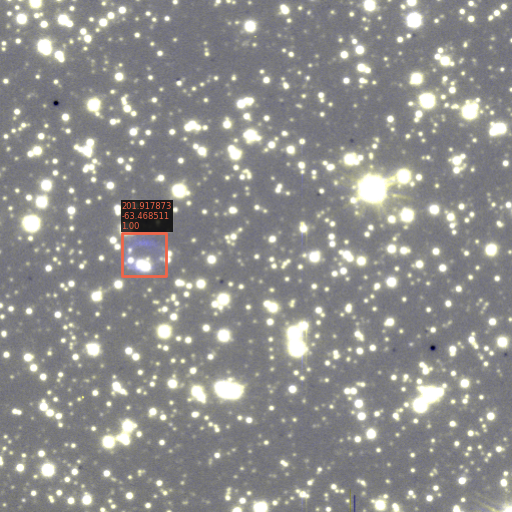}
\includegraphics[width=0.205\textwidth]{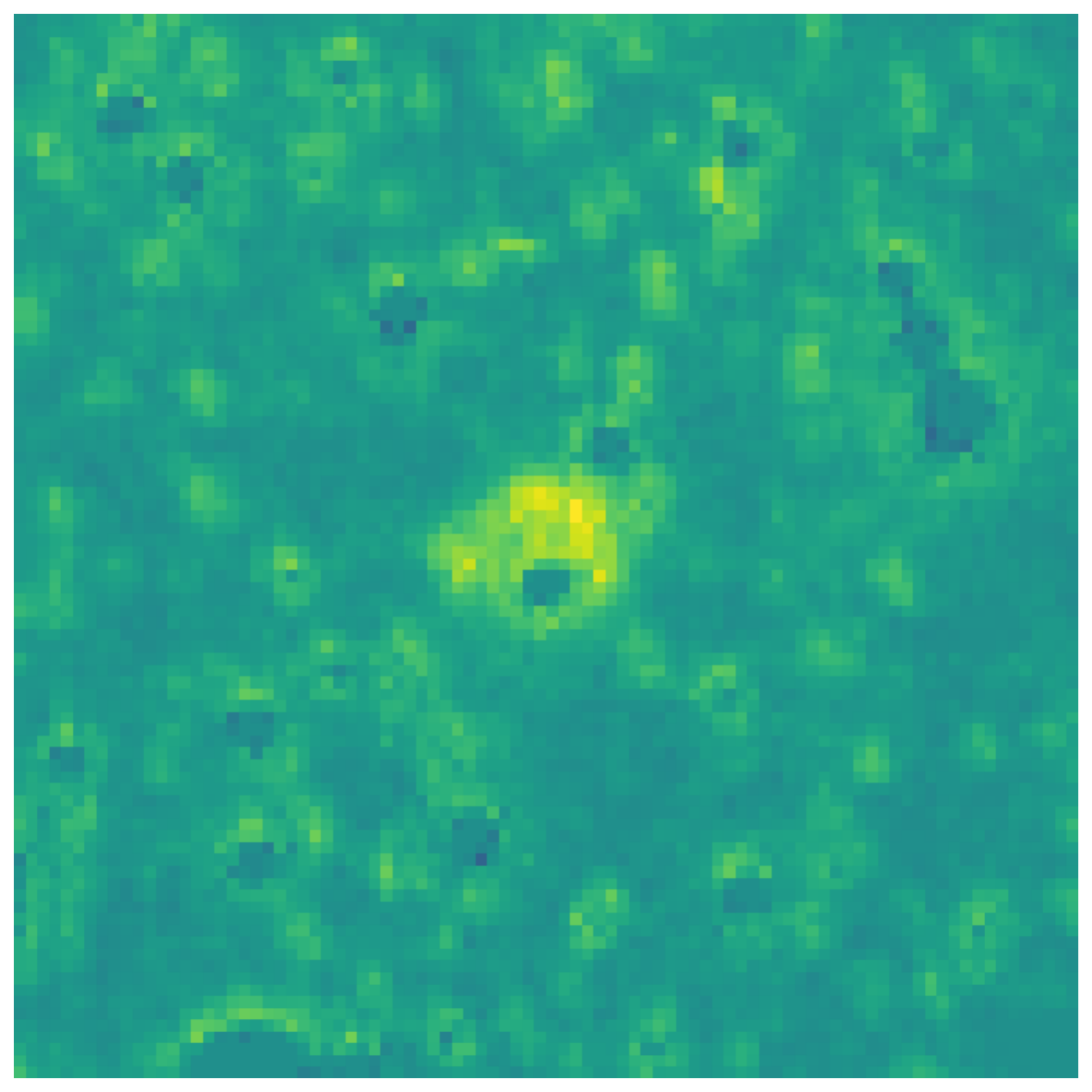}
\includegraphics[width=0.205\textwidth]{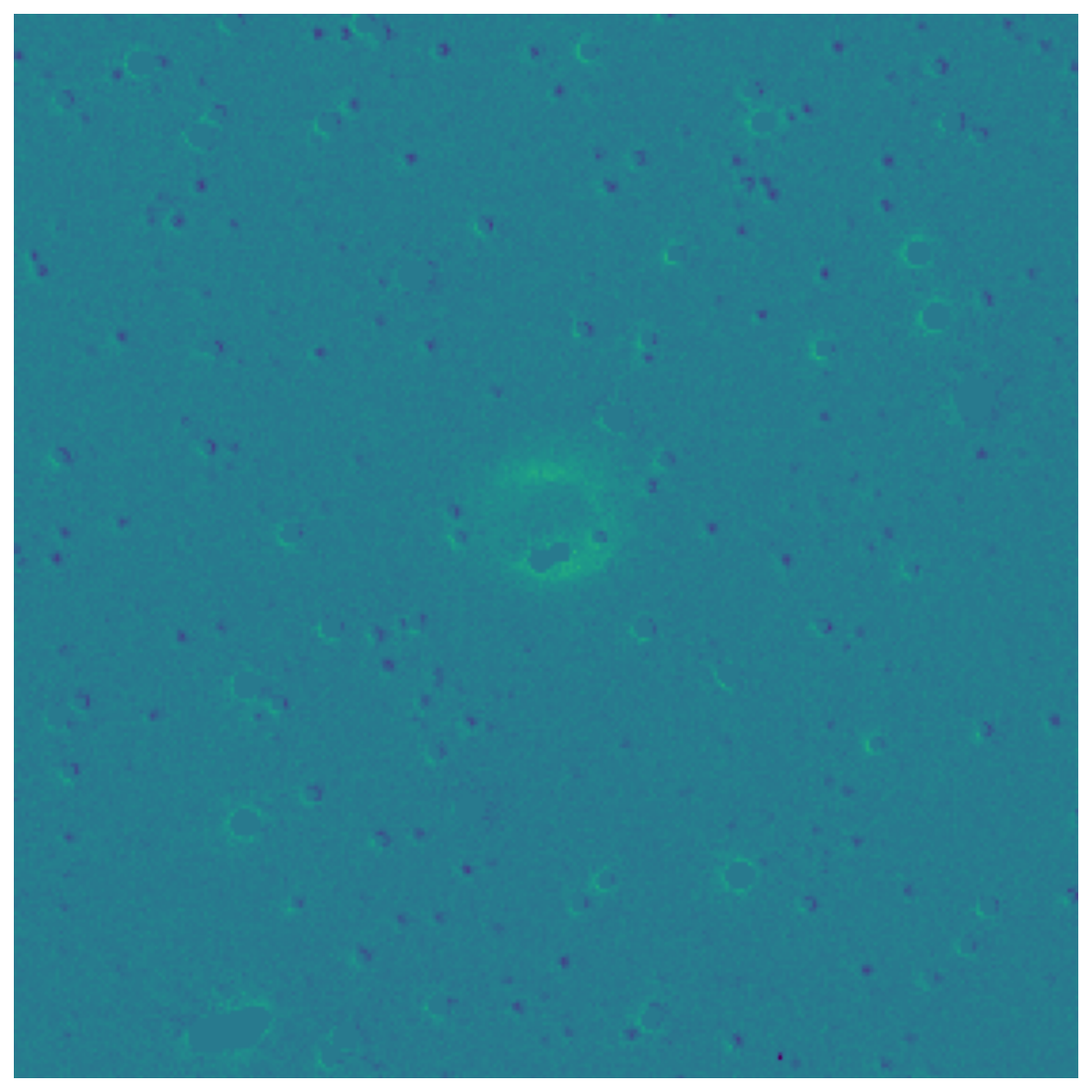}
\begin{overpic}[width=0.4\textwidth]{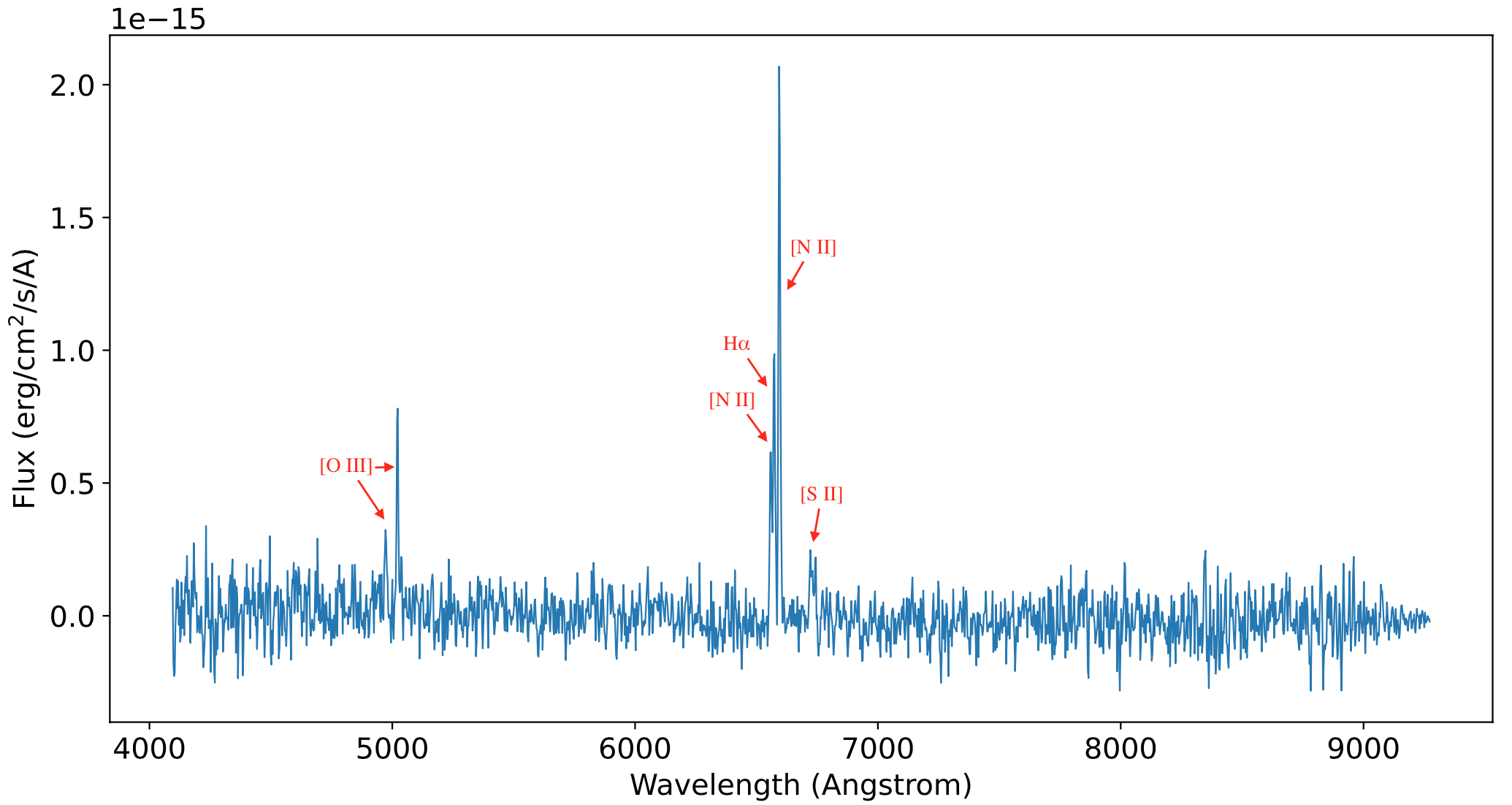}
\put(70,45){\small{YP1327-6328}}
\end{overpic}
\includegraphics[width=0.2\textwidth]{fig/YP1429-6214.png}
\includegraphics[width=0.205\textwidth]{fig/YP1429-6214_shs_q.pdf}
\includegraphics[width=0.205\textwidth]{fig/YP1429-6214_vphas_q.pdf}
\begin{overpic}[width=0.4\textwidth]{fig/YP1429-6214_spec.png}
\put(70,45){\small{YP1429-6214}}
\put(8,18){\includegraphics[scale=0.12]{fig/YP1429-6214a_spec.png}}
\end{overpic}
\caption{True PNe: VPHAS+ detected PNG images labelled by red boxes, SHS H$\alpha$-Rband quotient images, VPHAS+ H$\alpha$-Rband quotient images, SAAO spectra with spectral lines labelled.}
\label{fig:true_pn}
\end{figure*}

\begin{figure*}
\ContinuedFloat
\includegraphics[width=0.2\textwidth]{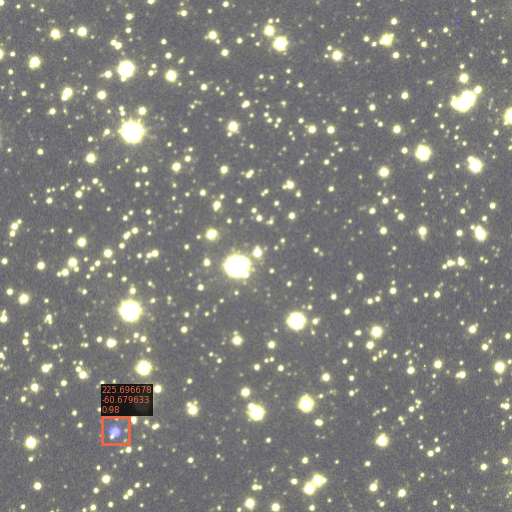}
\includegraphics[width=0.205\textwidth]{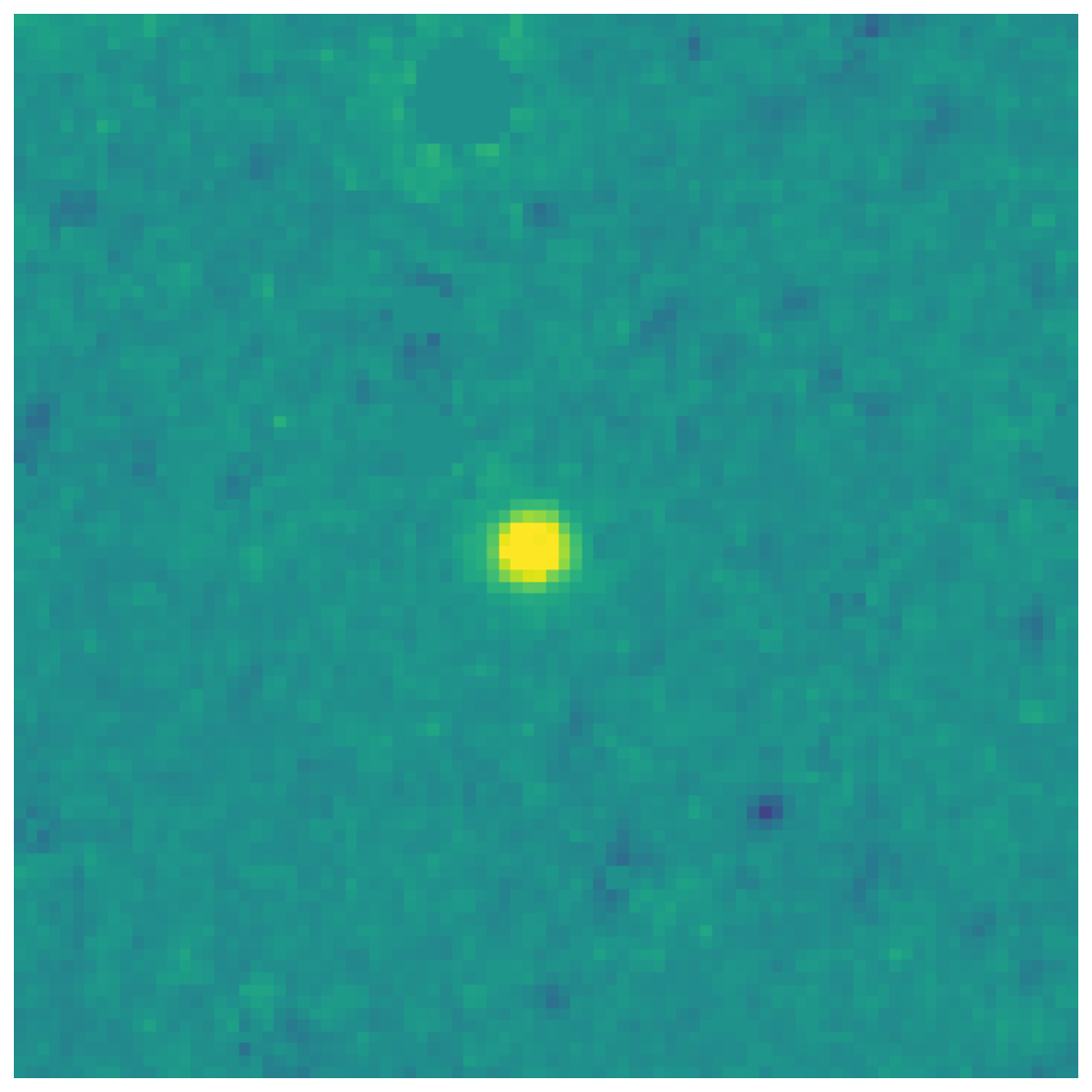}
\includegraphics[width=0.205\textwidth]{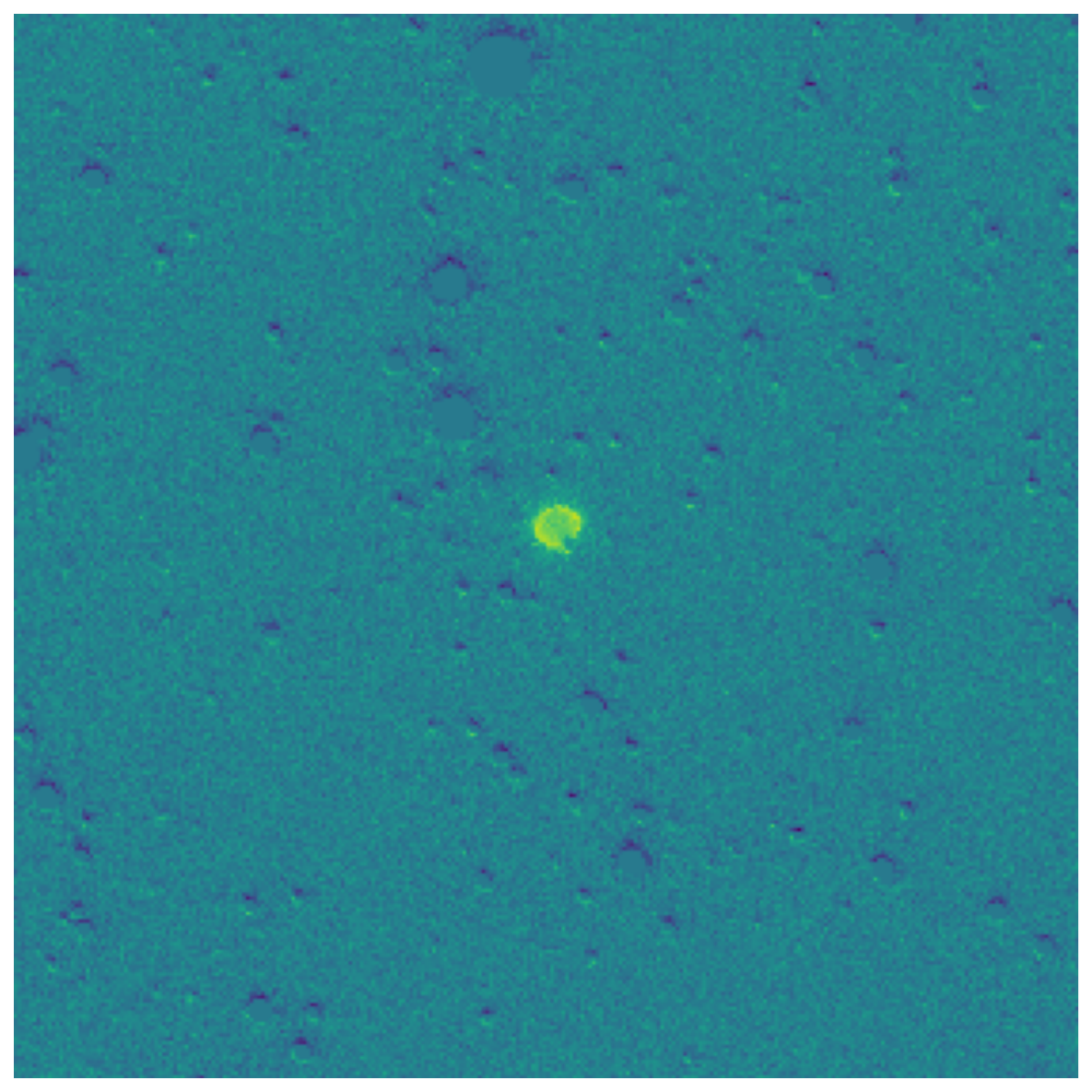}
\begin{overpic}[width=0.4\textwidth]{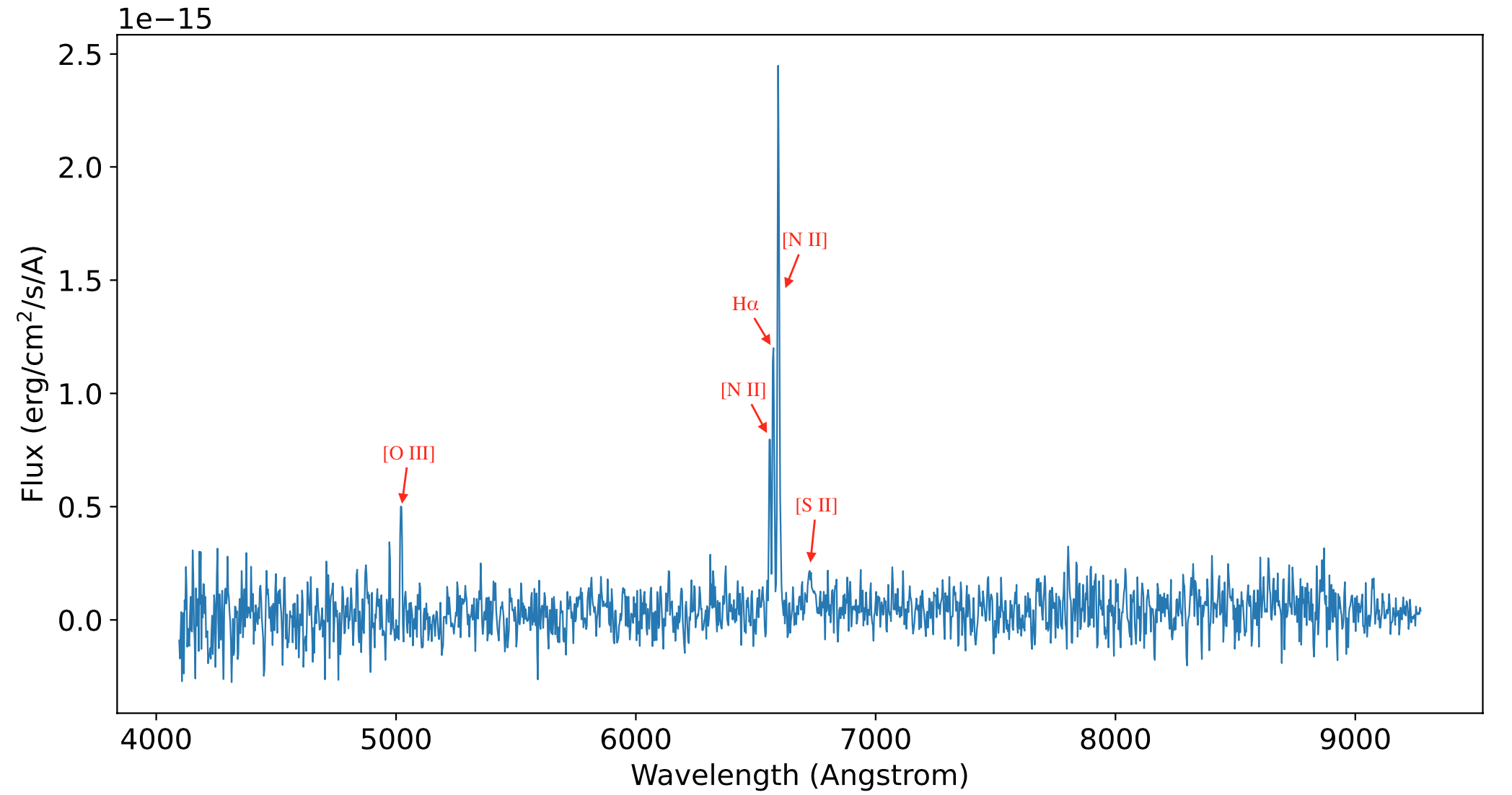}
\put(70,45){\small{YP1502-6040}}
\end{overpic}
\includegraphics[width=0.2\textwidth]{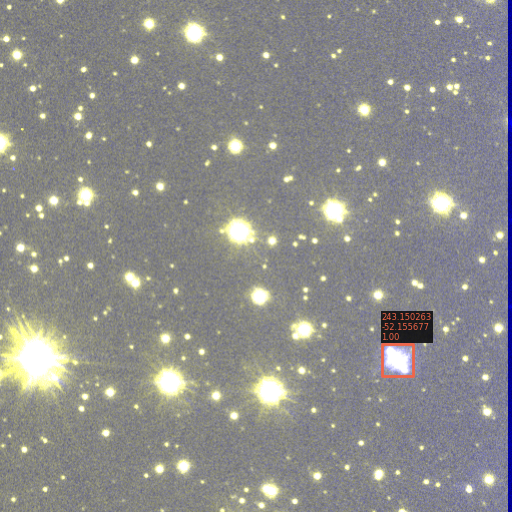}
\includegraphics[width=0.205\textwidth]{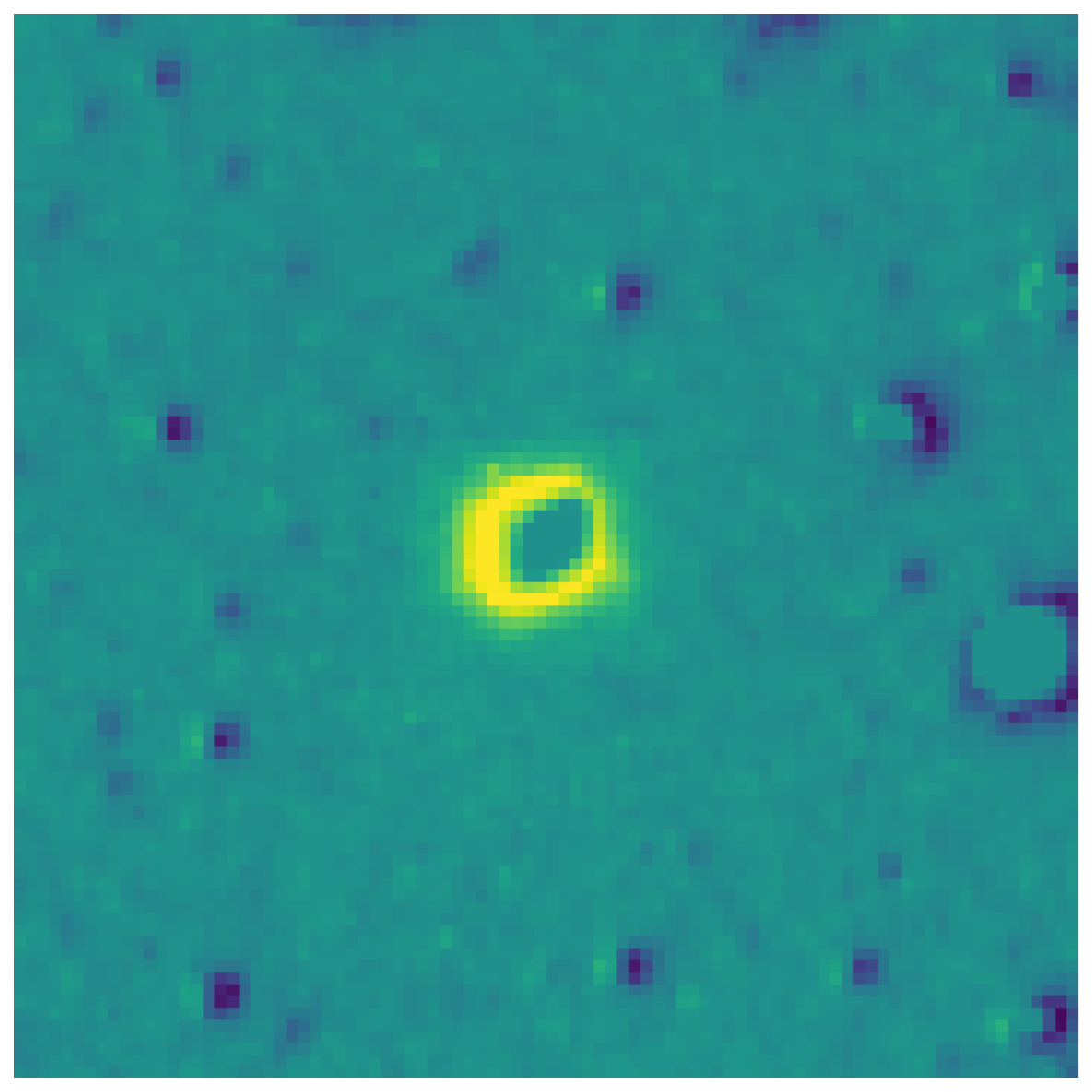}
\includegraphics[width=0.205\textwidth]{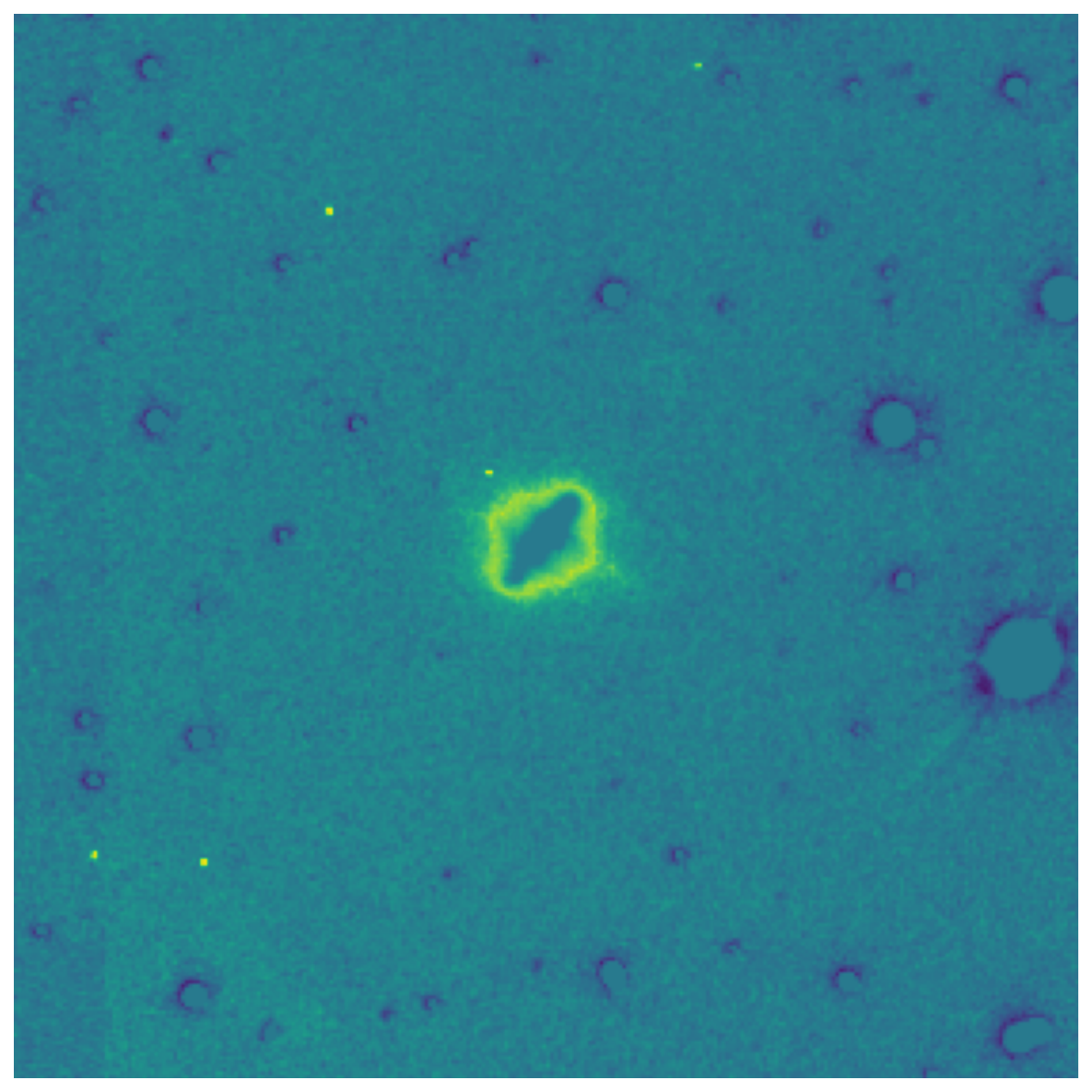}
\begin{overpic}[width=0.4\textwidth]{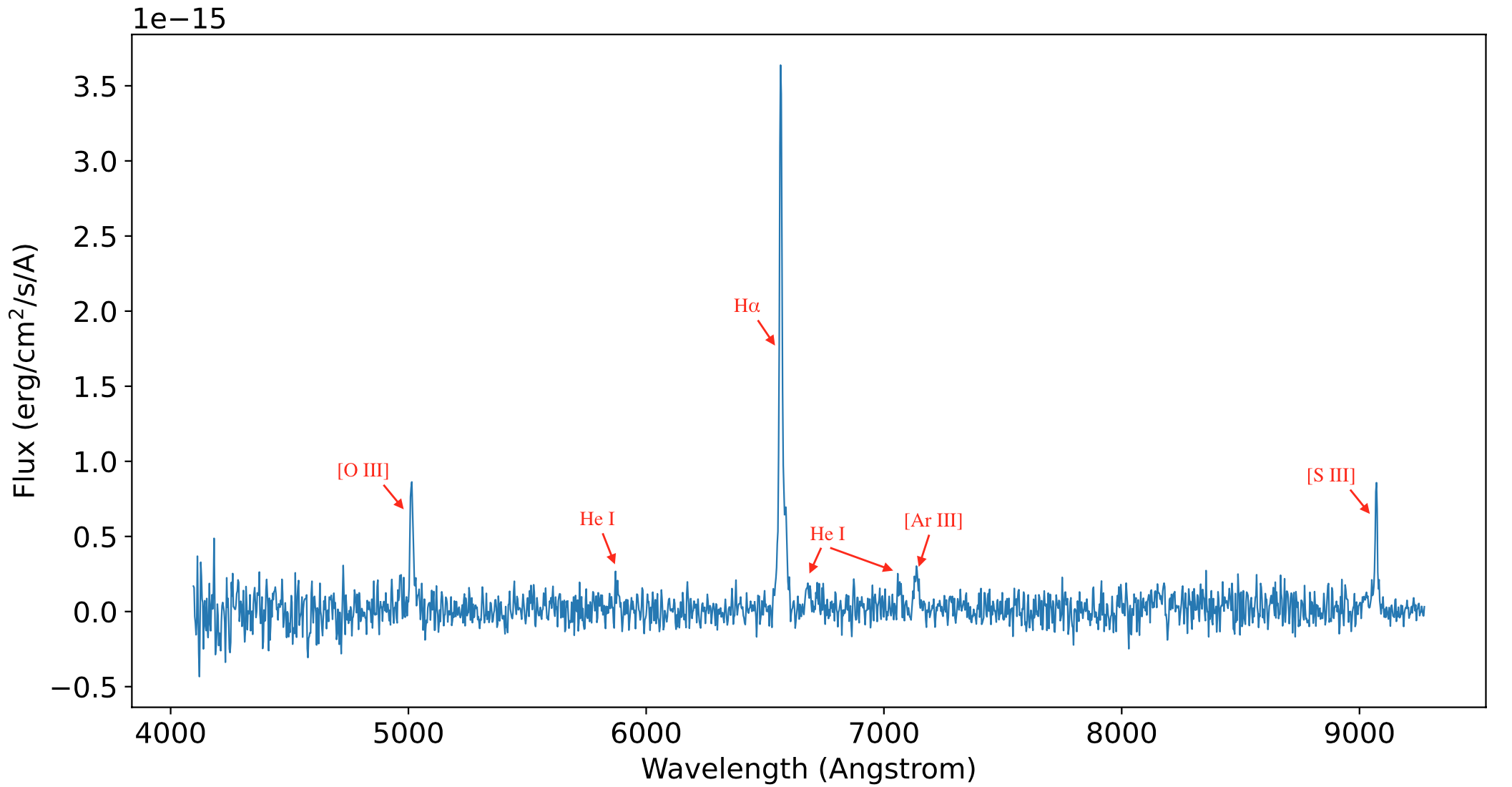}
\put(70,45){\small{YP1612-5209}}
\end{overpic}
\includegraphics[width=0.2\textwidth]{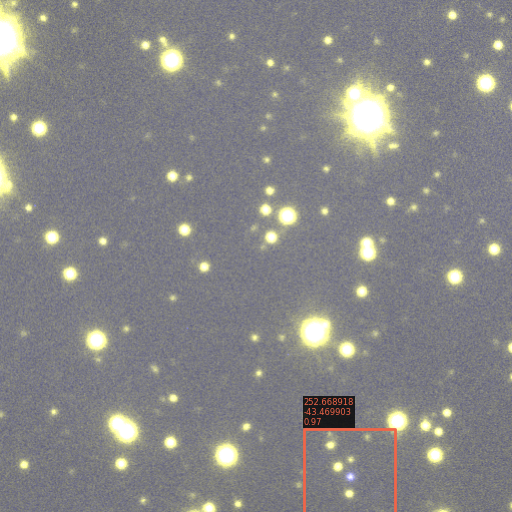}
\includegraphics[width=0.205\textwidth]{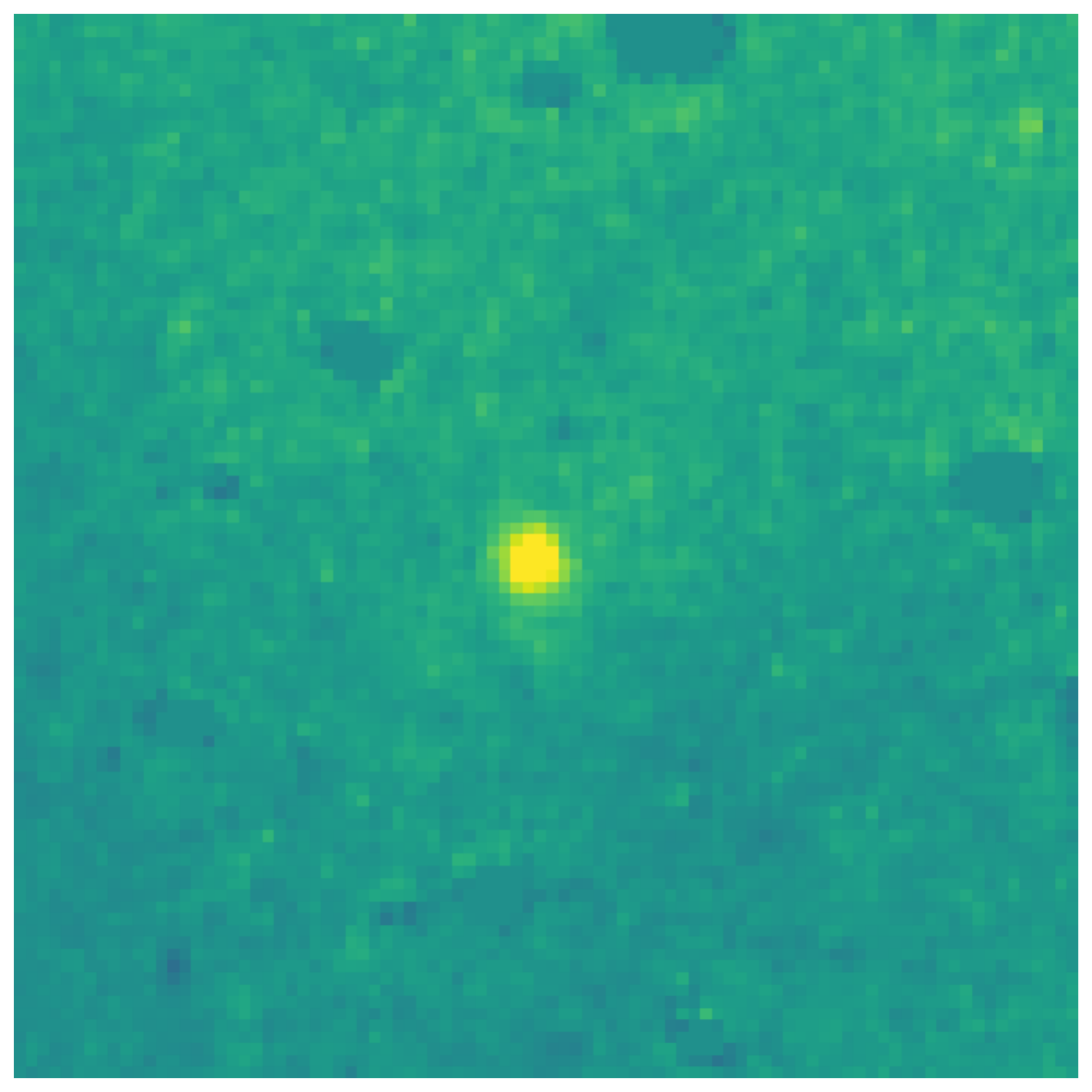}
\includegraphics[width=0.205\textwidth]{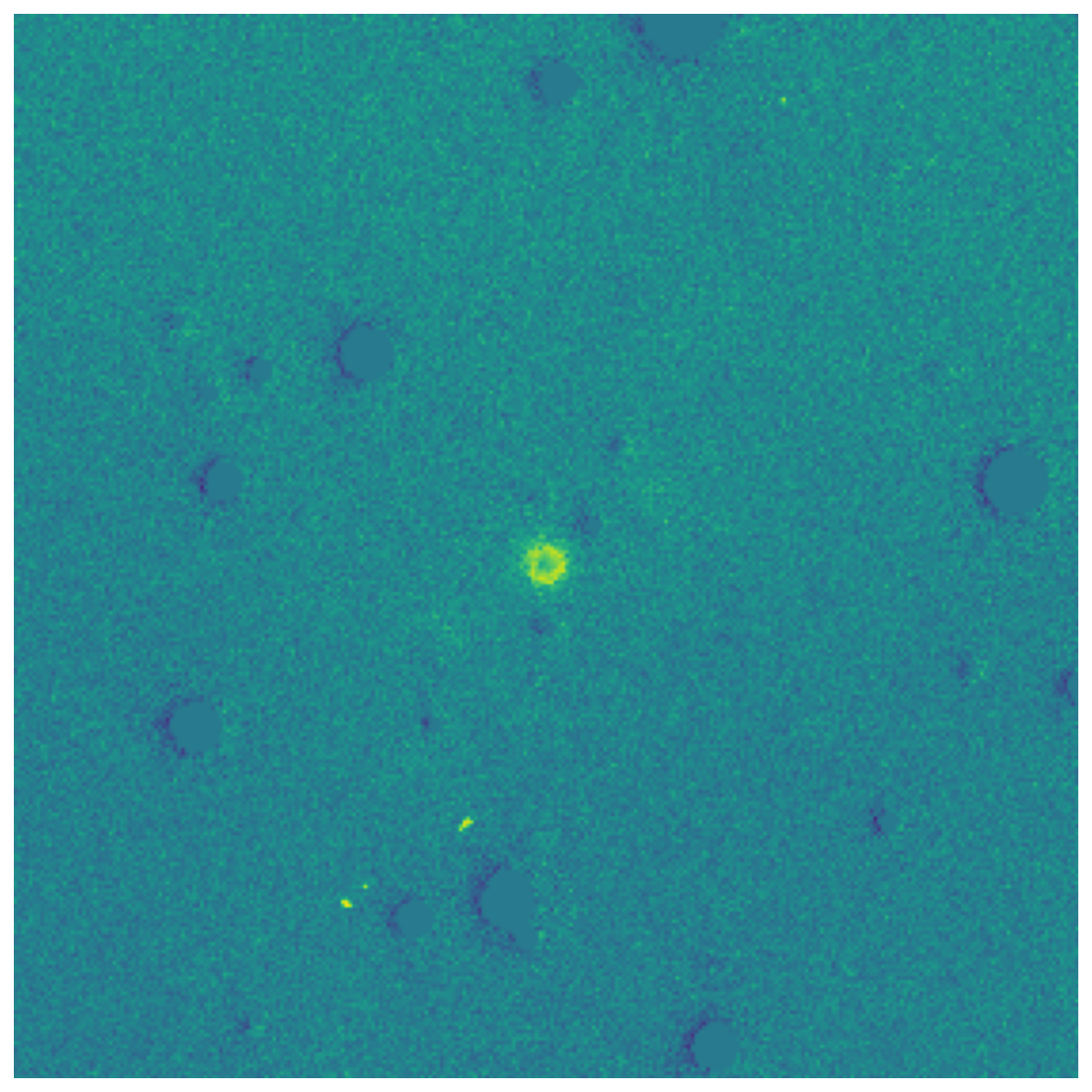}
\begin{overpic}[width=0.4\textwidth]{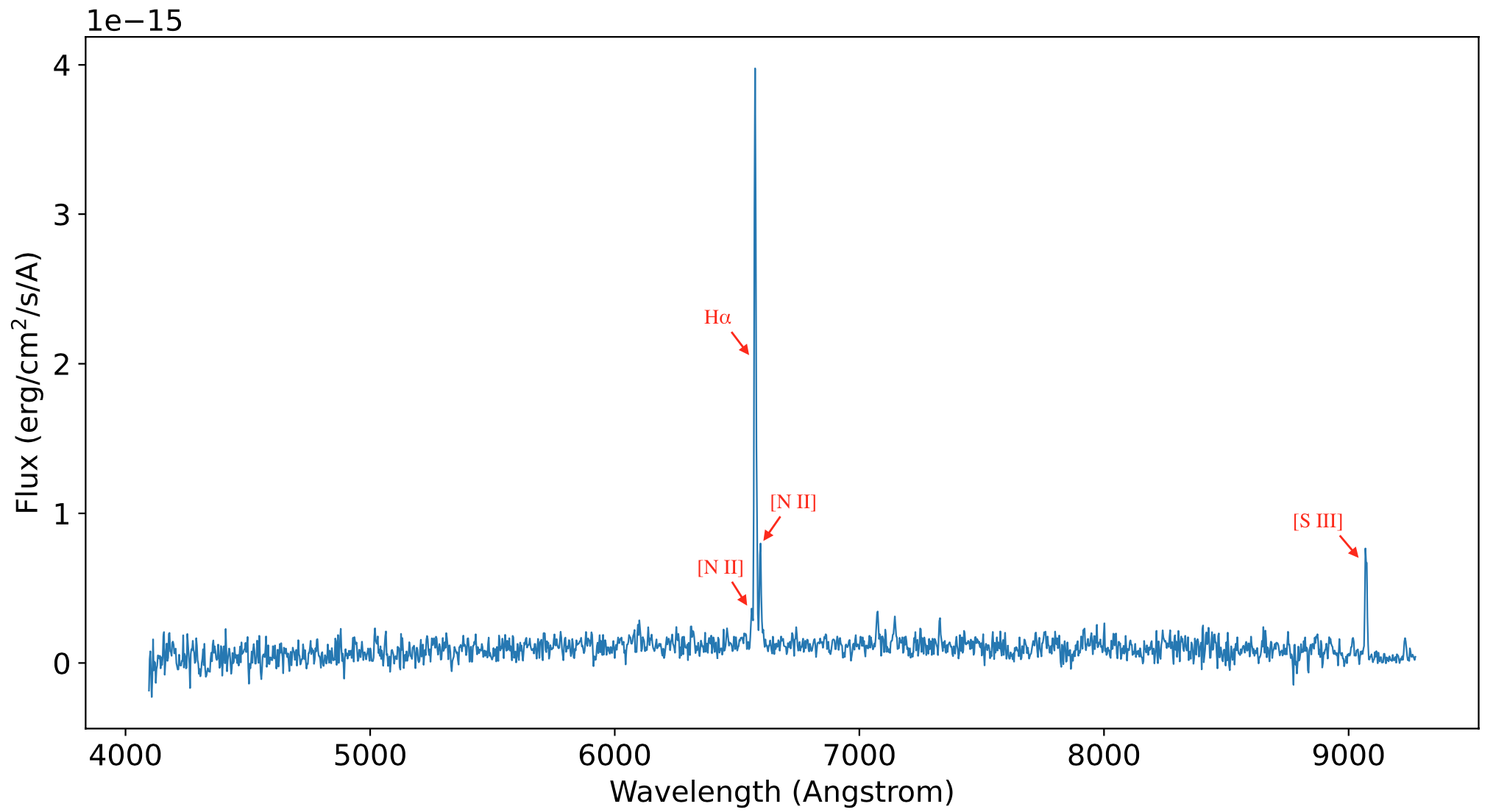}
\put(70,45){\small{YP1650-4328}}
\end{overpic}
\includegraphics[width=0.2\textwidth]{fig/YP1731-3011.png}
\includegraphics[width=0.205\textwidth]{fig/YP1731-3011_shs_q.pdf}
\includegraphics[width=0.205\textwidth]{fig/YP1731-3011_vphas_q.pdf}
\begin{overpic}[width=0.4\textwidth]{fig/YP1731-3011a_spec.png}
\put(70,45){\small{YP1731-3011}}
\put(70,40){\small{White Dwarf}}
\end{overpic}
\hspace*{4.47in}
\begin{overpic}[width=0.4\textwidth]{fig/YP1731-3011b_spec.png}
\put(70,45){\small{YP1731-3011}}
\put(70,40){\small{True PN}}
\end{overpic}
\includegraphics[width=0.2\textwidth]{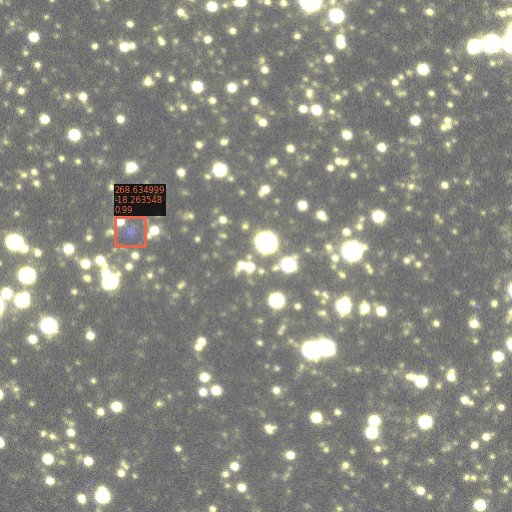}
\includegraphics[width=0.205\textwidth]{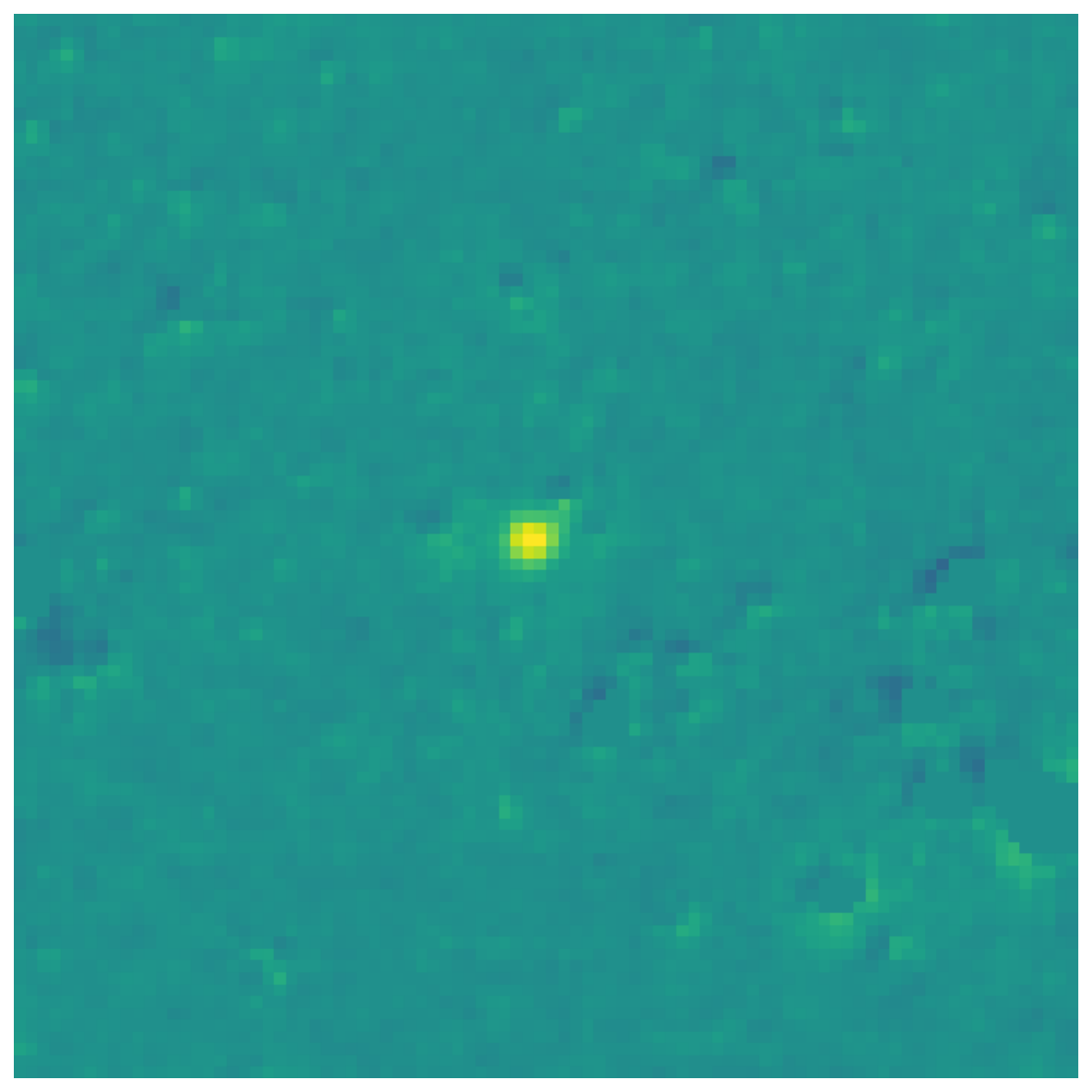}
\includegraphics[width=0.205\textwidth]{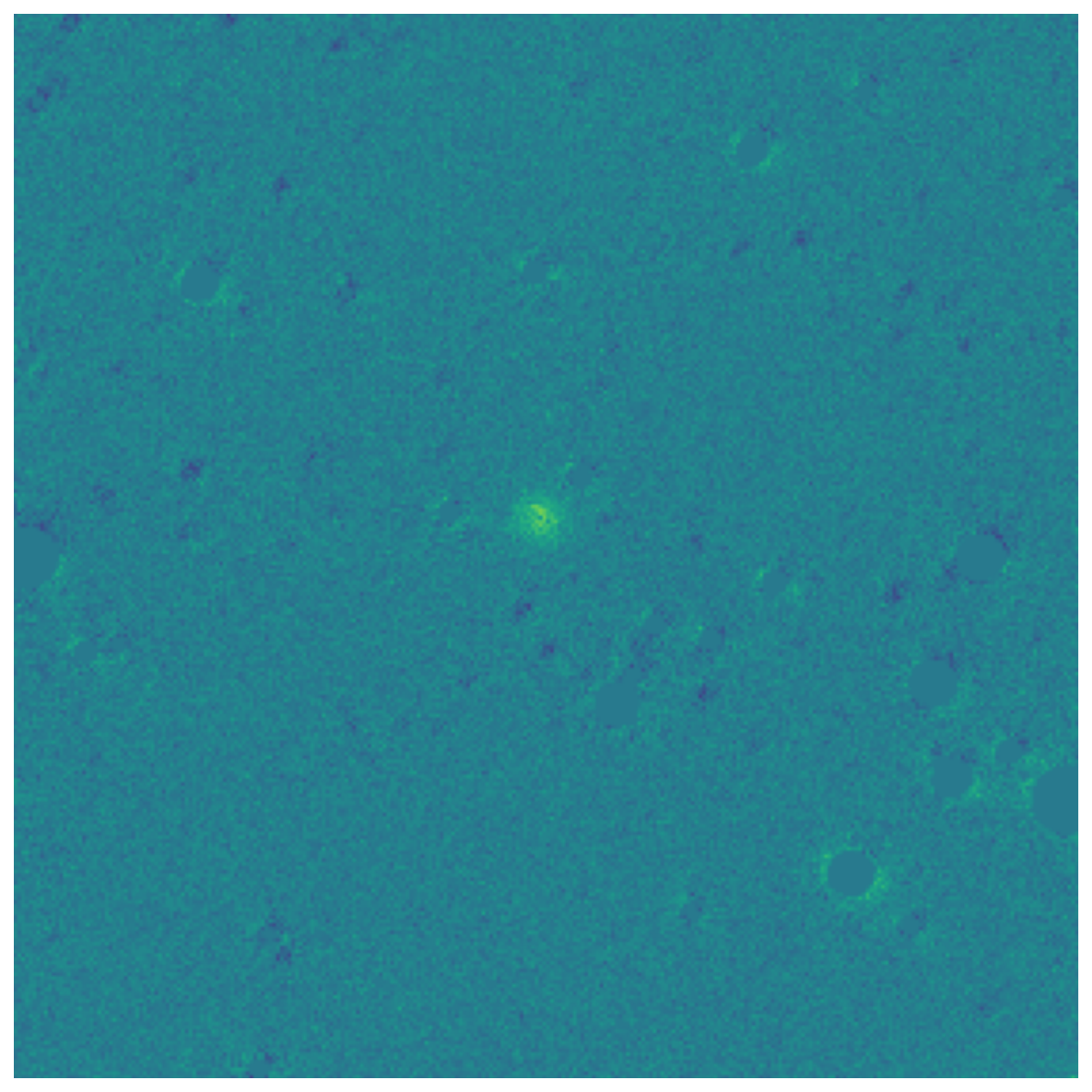}
\begin{overpic}[width=0.4\textwidth]{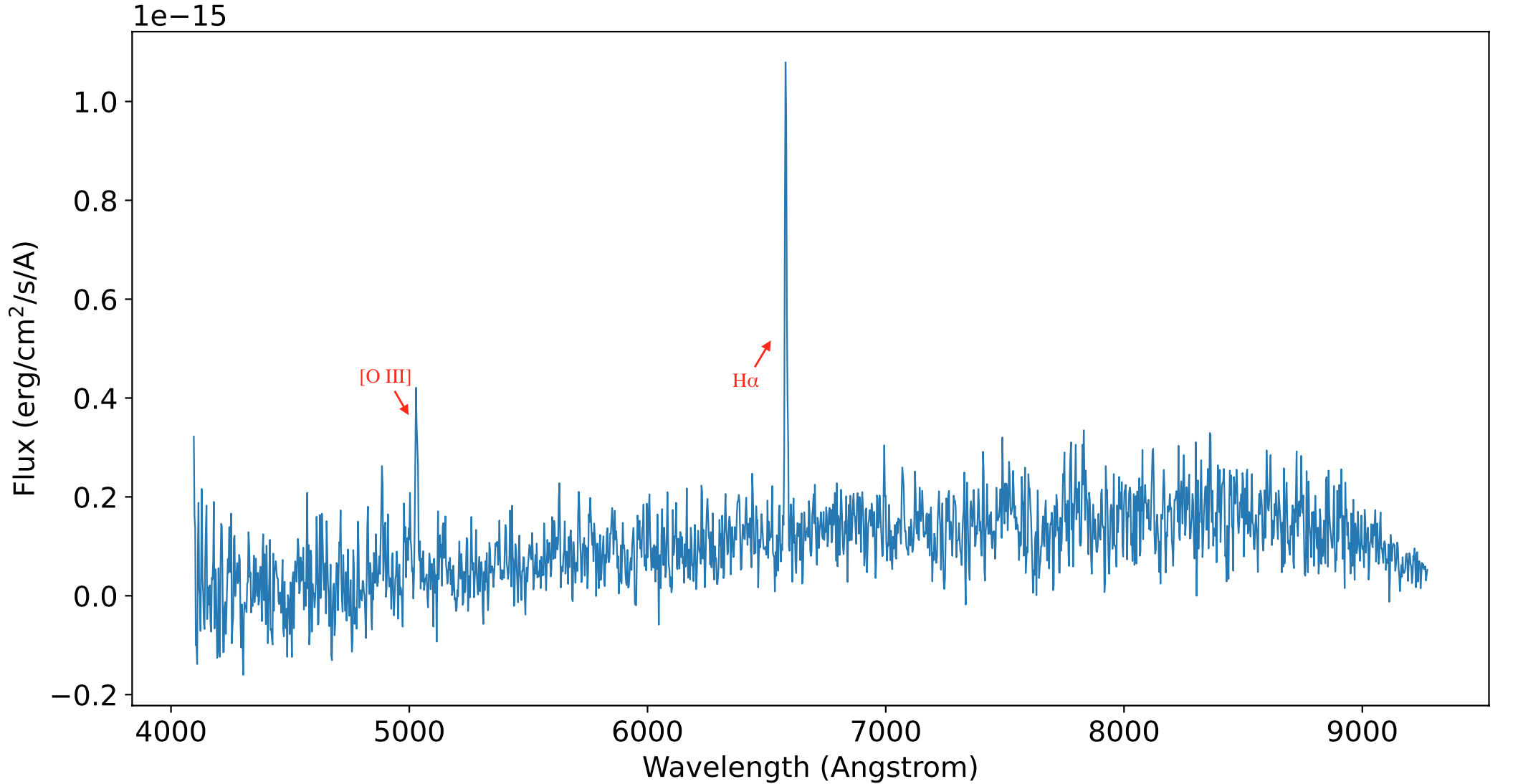}
\put(70,45){\small{YP1754-1815}}
\end{overpic}
\caption{True PNe: VPHAS+ detected PNG images labelled by red boxes, SHS H$\alpha$-Rband quotient images, VPHAS+ H$\alpha$-Rband quotient images, SAAO spectra with spectral lines labelled.}
\label{fig:true_pn}
\end{figure*}

\begin{figure*}
\ContinuedFloat
\includegraphics[width=0.2\textwidth]{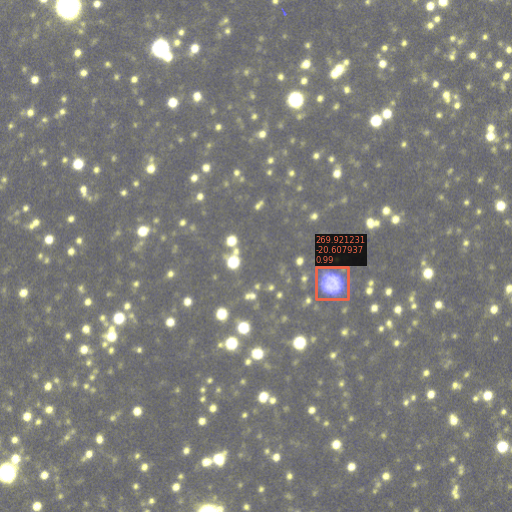}
\includegraphics[width=0.205\textwidth]{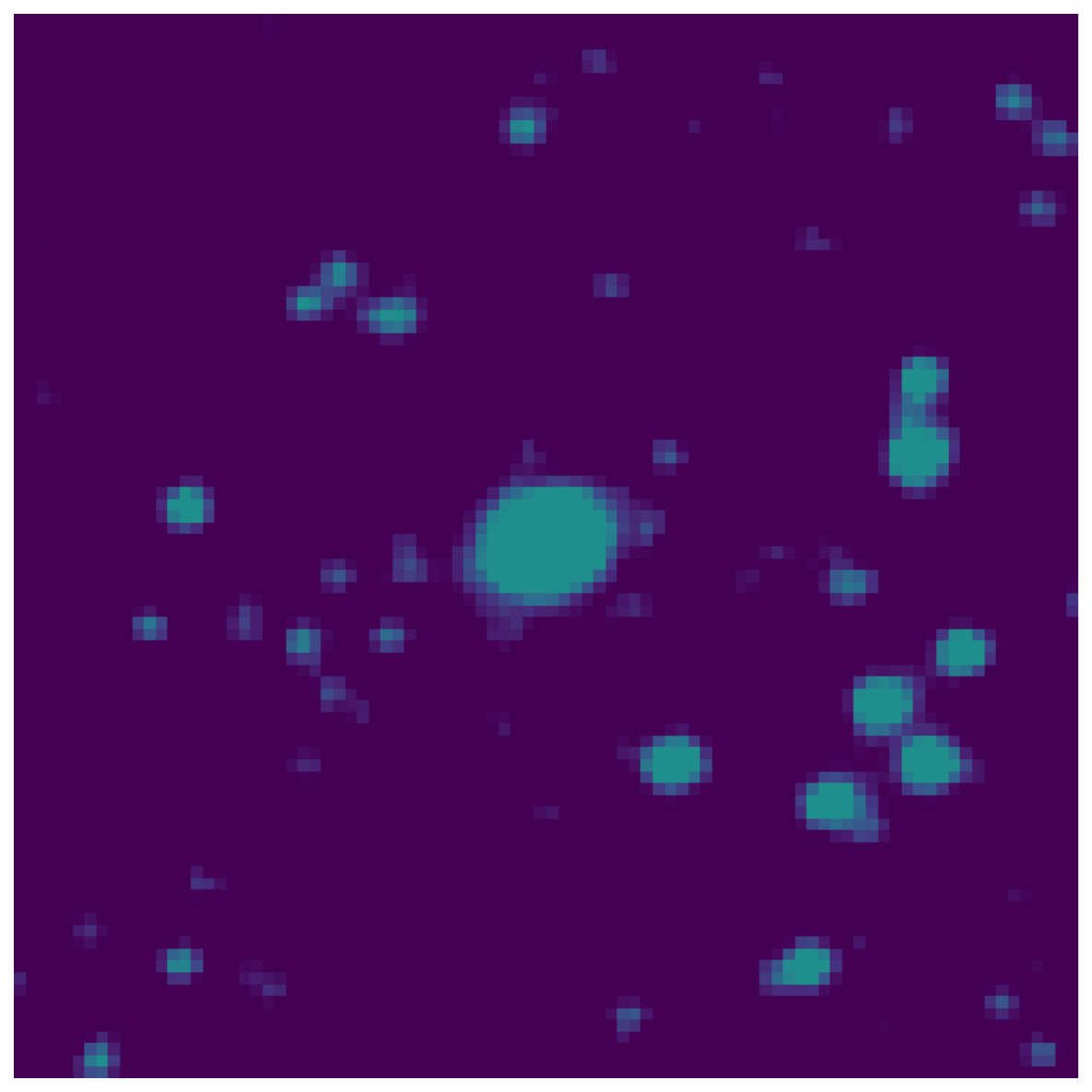}
\includegraphics[width=0.205\textwidth]{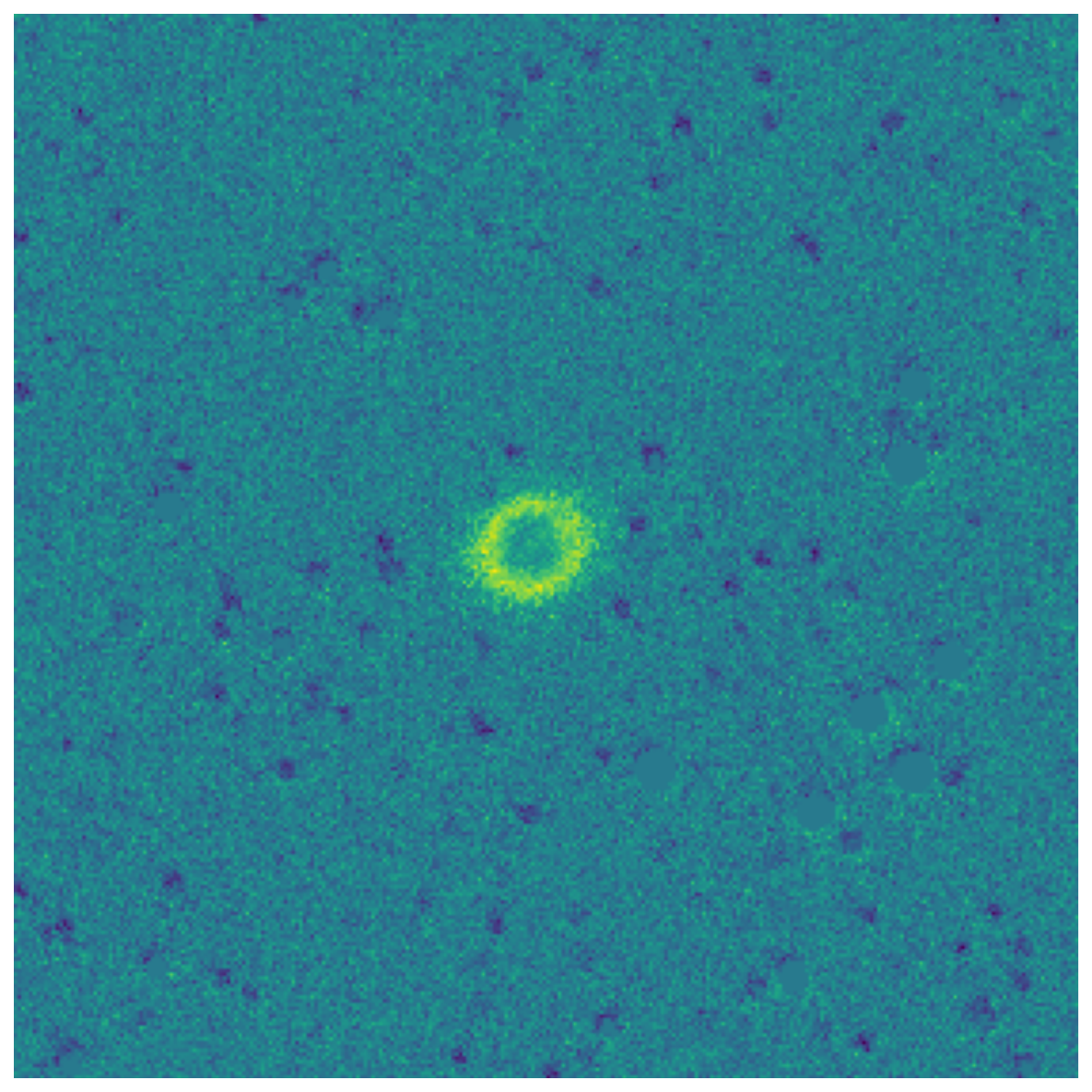}
\begin{overpic}[width=0.4\textwidth]{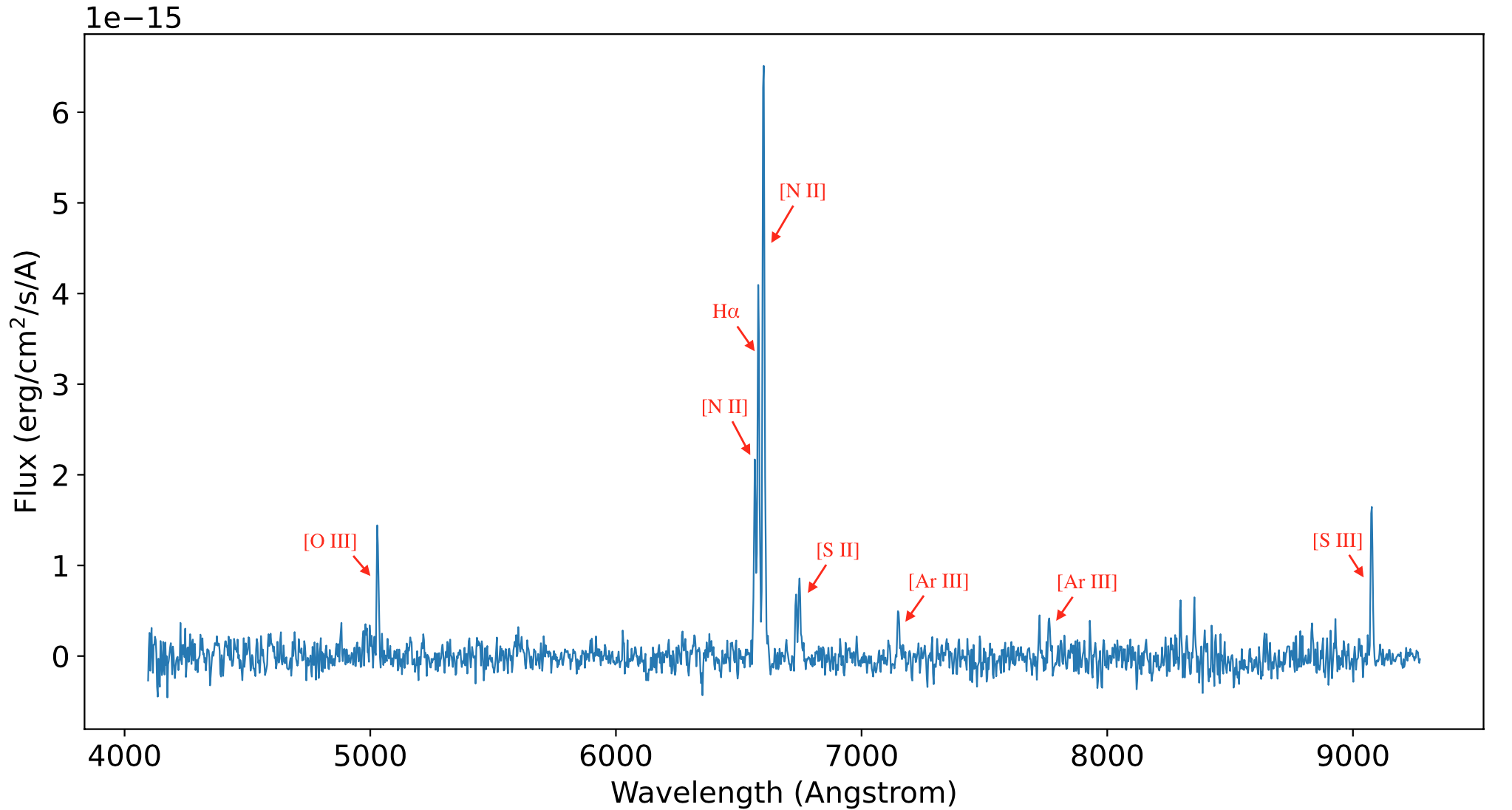}
\put(70,45){\small{YP1759-2036}}
\end{overpic}
\includegraphics[width=0.2\textwidth]{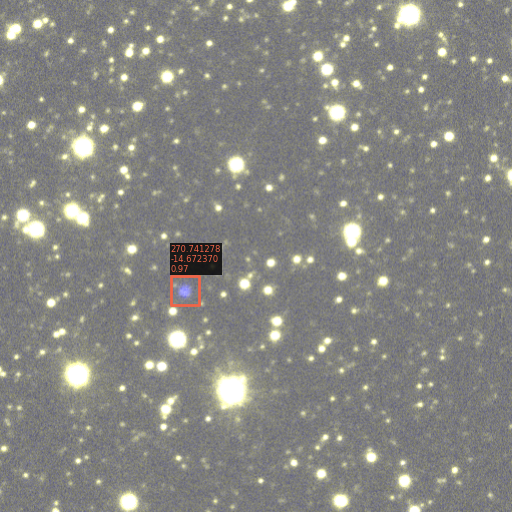}
\includegraphics[width=0.205\textwidth]{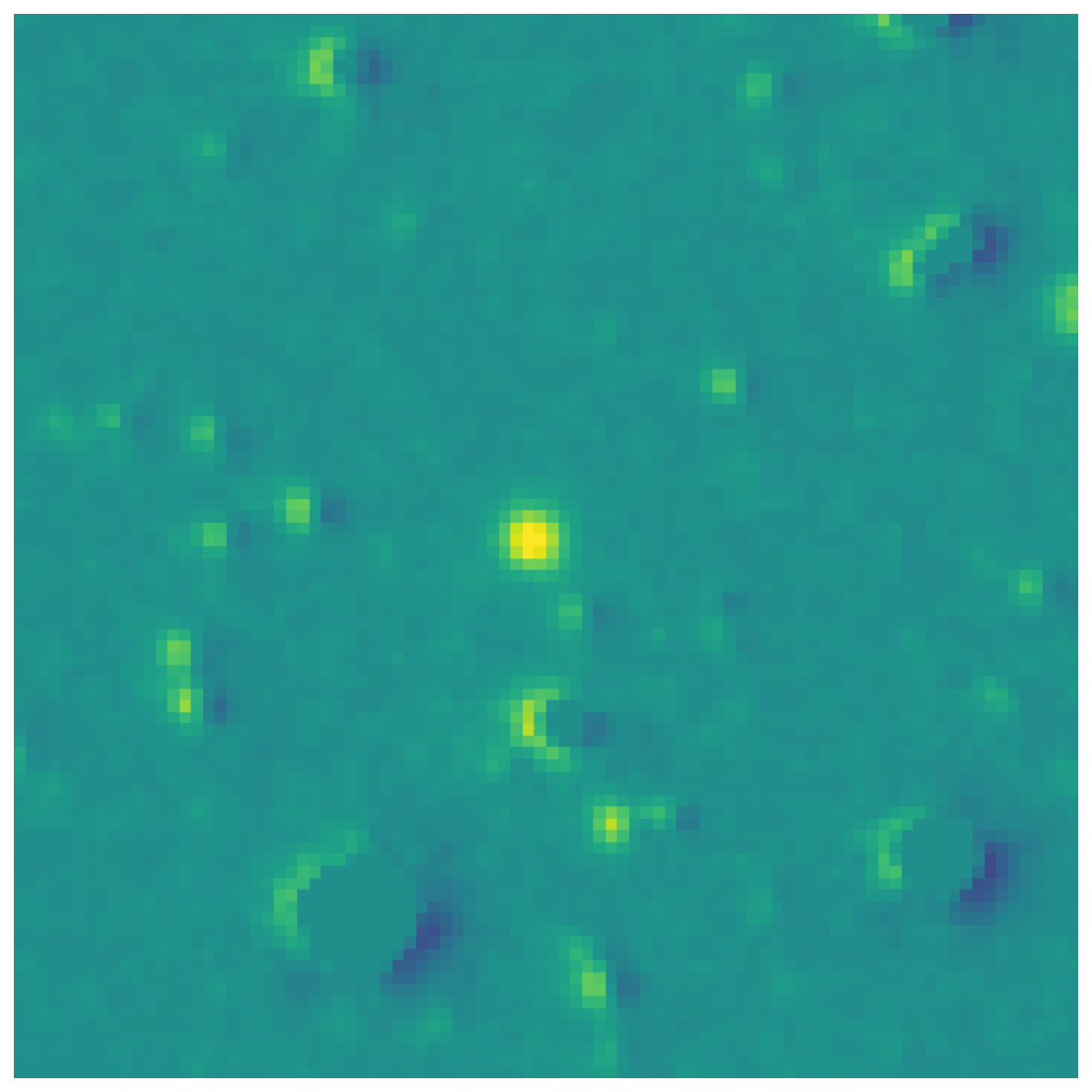}
\includegraphics[width=0.205\textwidth]{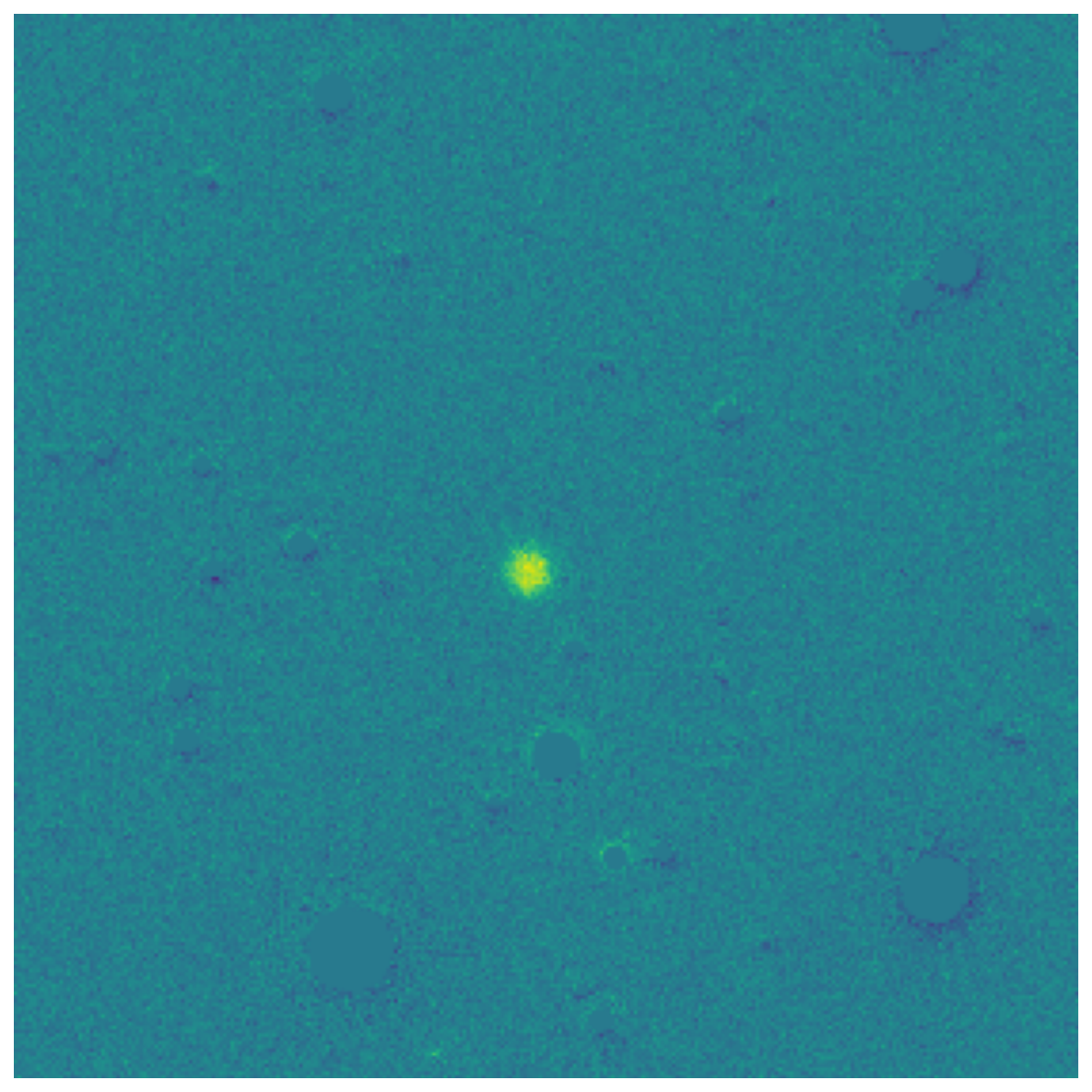}
\begin{overpic}[width=0.4\textwidth]{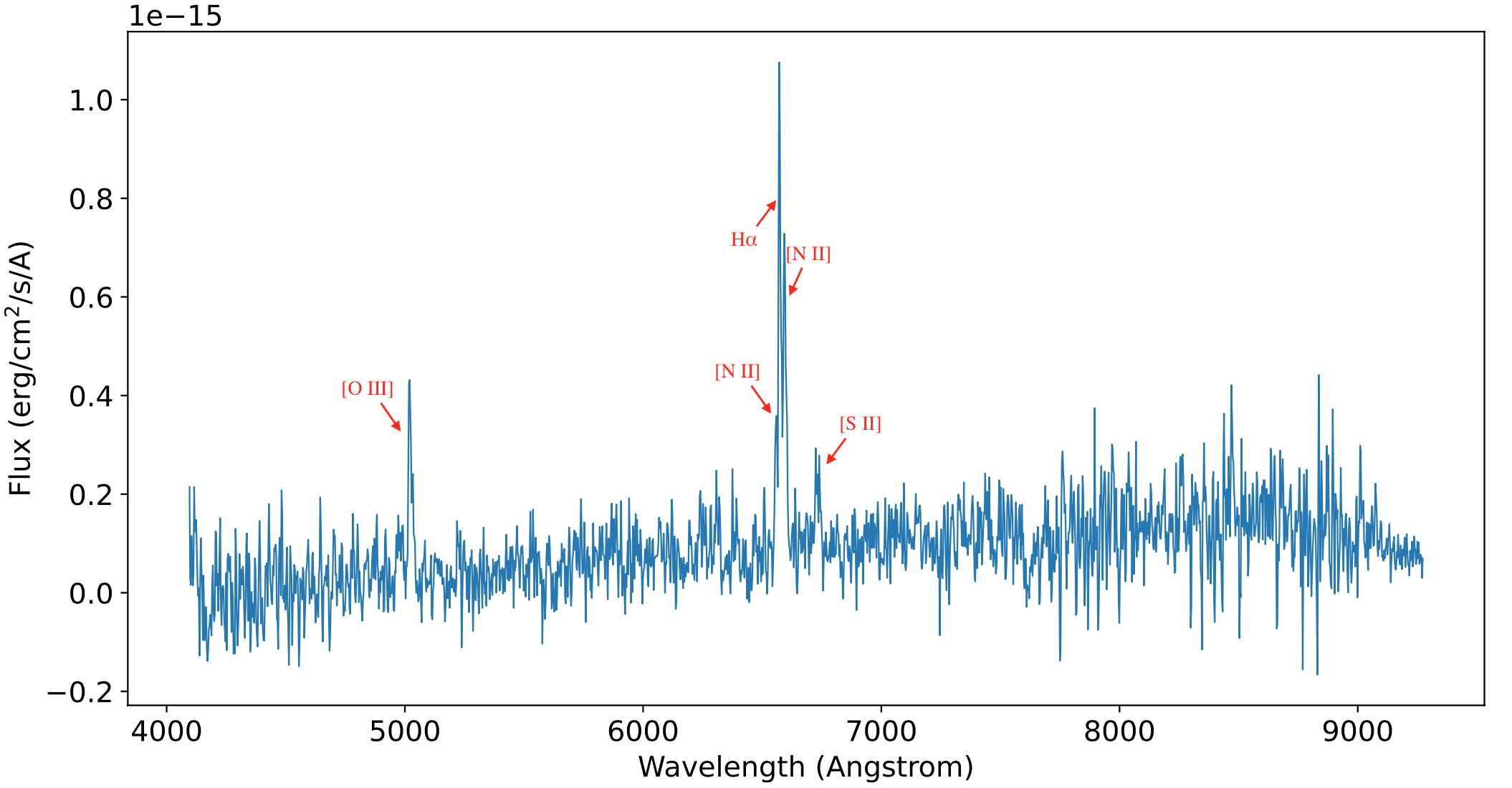}
\put(70,45){\small{YP1802-1440}}
\end{overpic}
\includegraphics[width=0.2\textwidth]{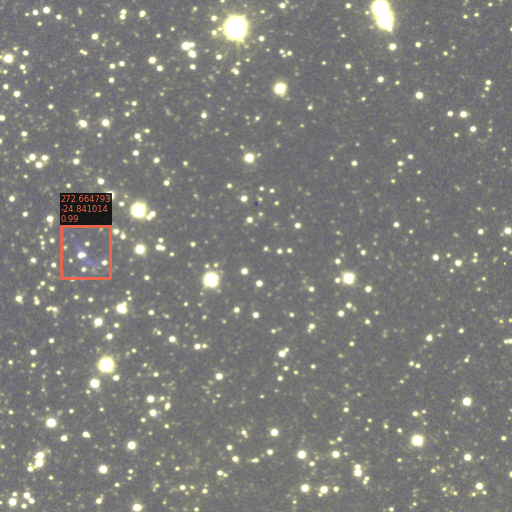}
\includegraphics[width=0.205\textwidth]{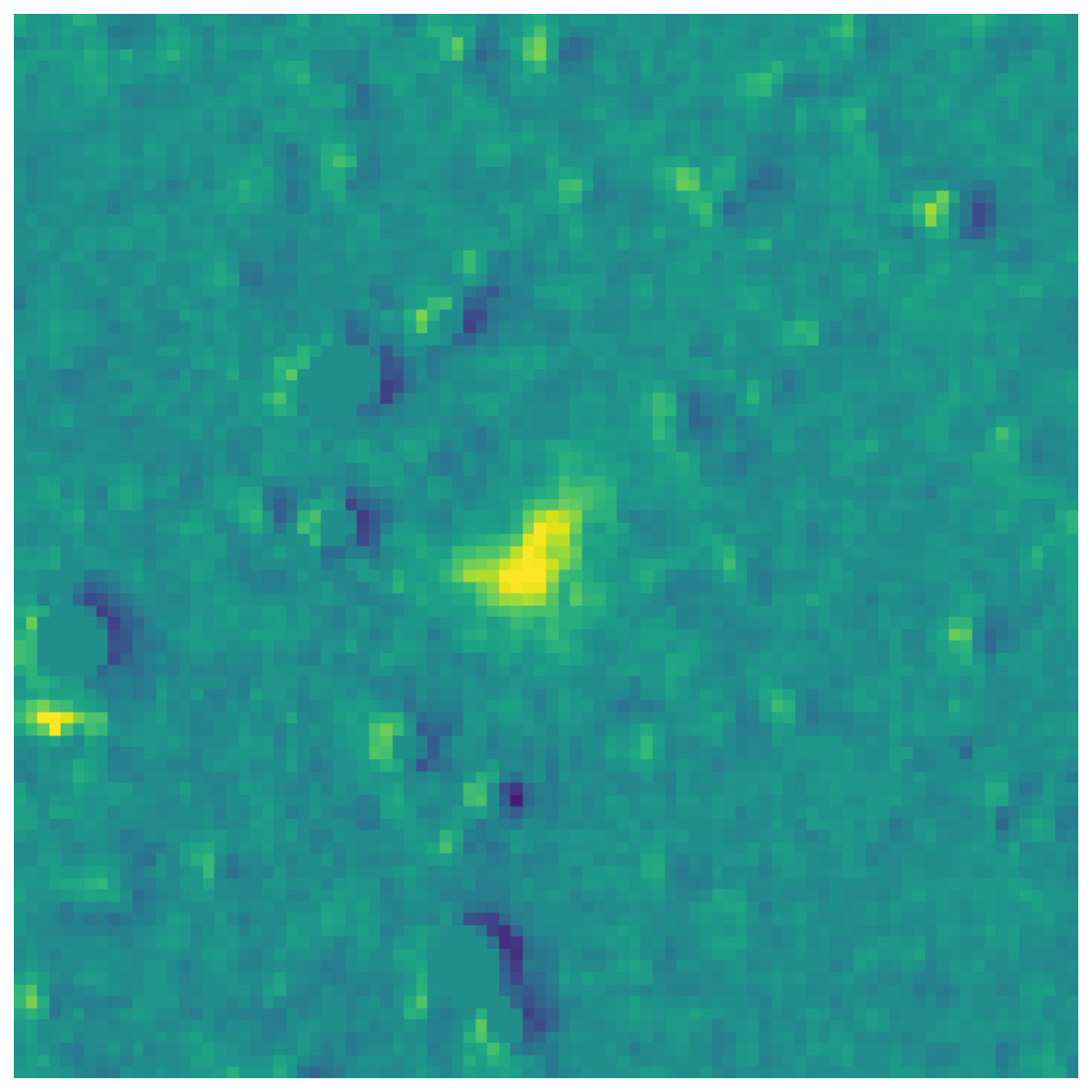}
\includegraphics[width=0.205\textwidth]{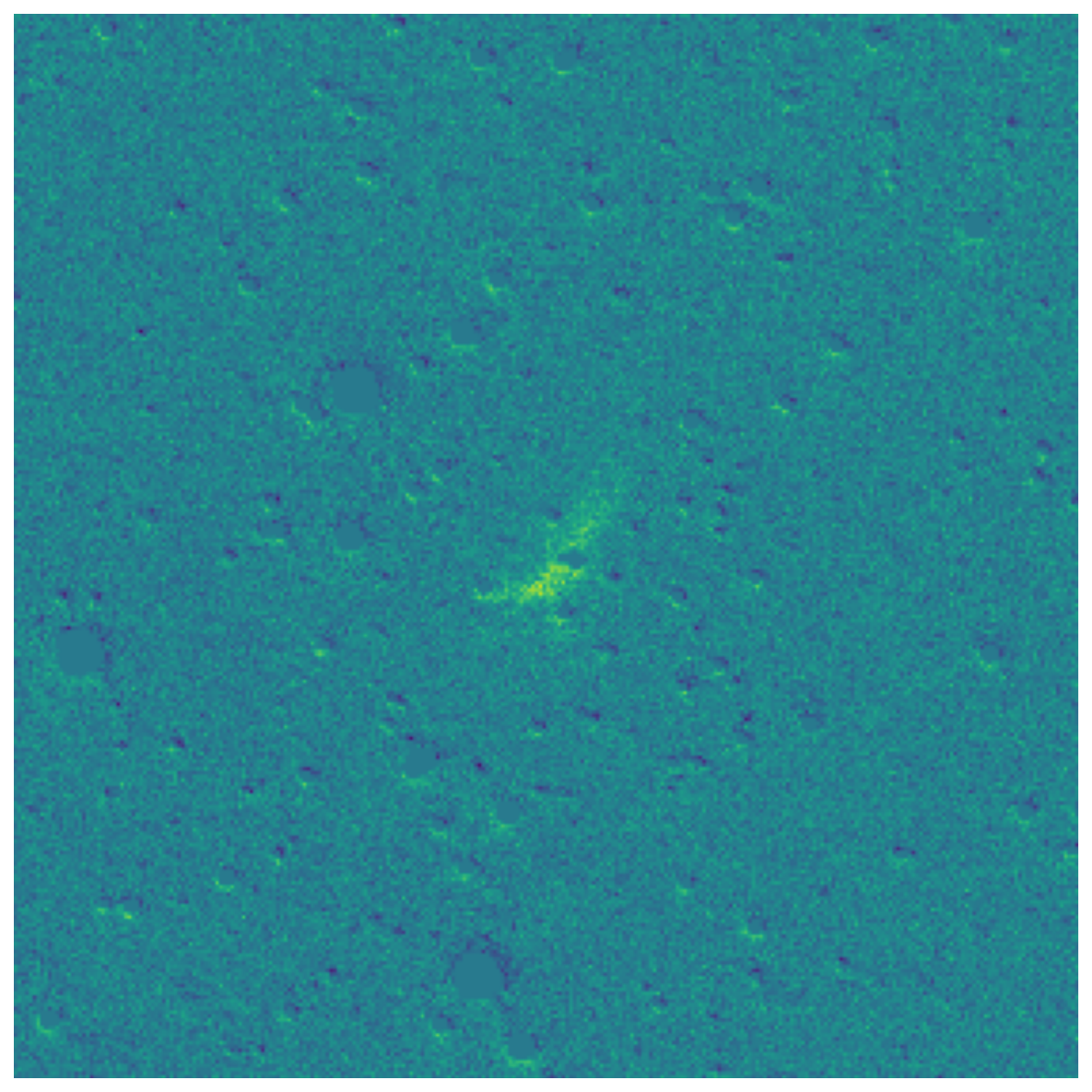}
\begin{overpic}[width=0.4\textwidth]{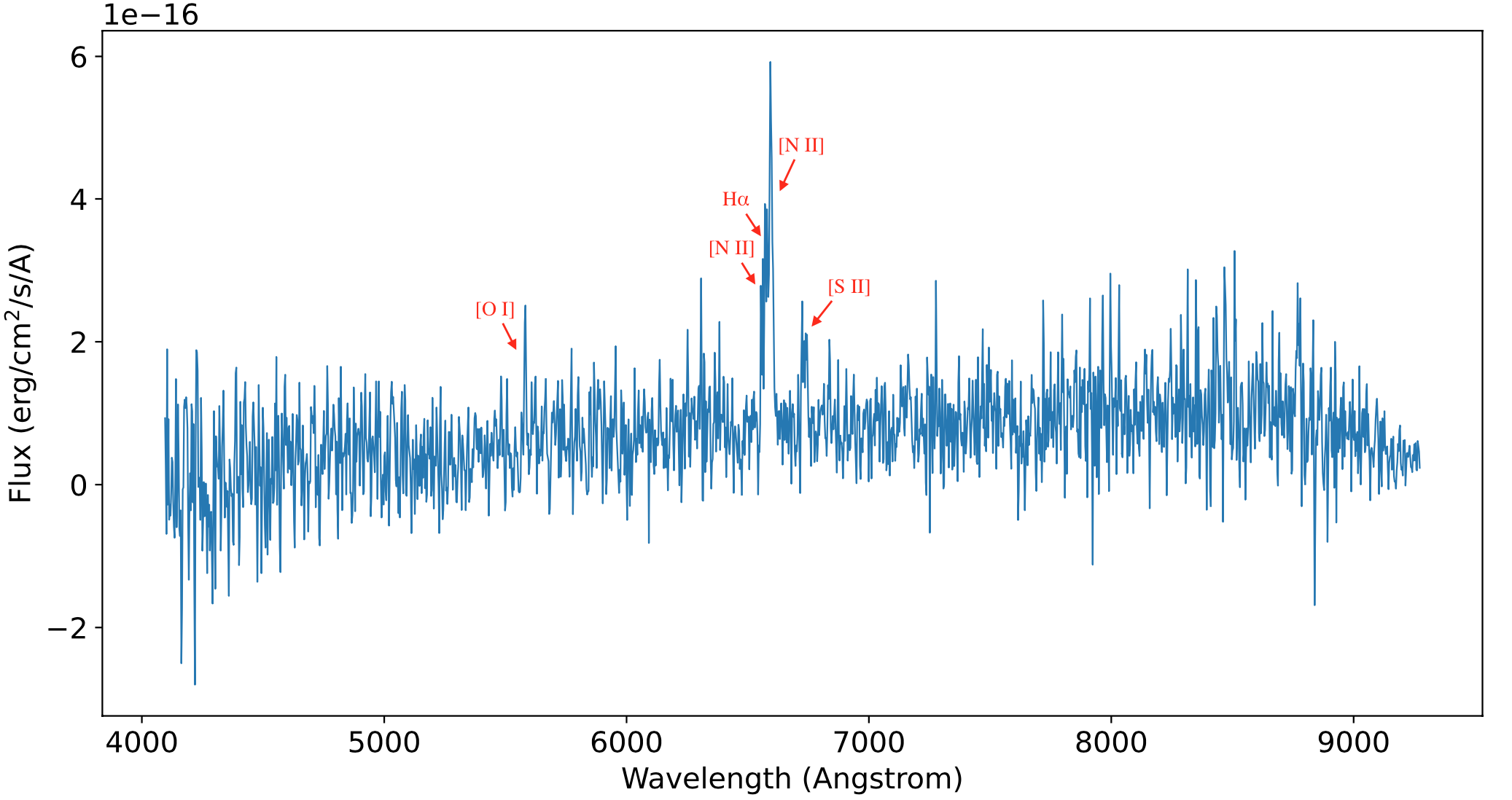}
\put(70,45){\small{YP1810-2450}}
\end{overpic}
\includegraphics[width=0.2\textwidth]{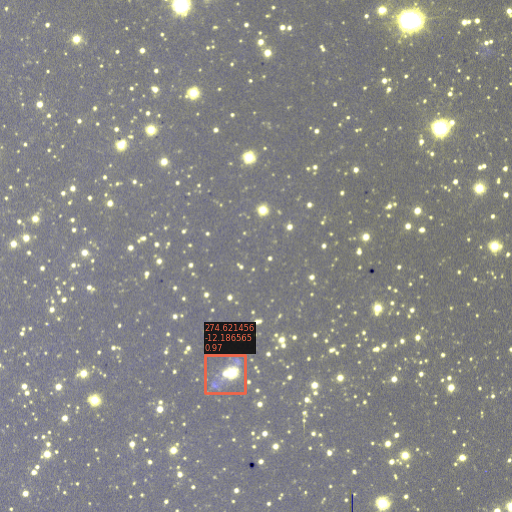}
\includegraphics[width=0.205\textwidth]{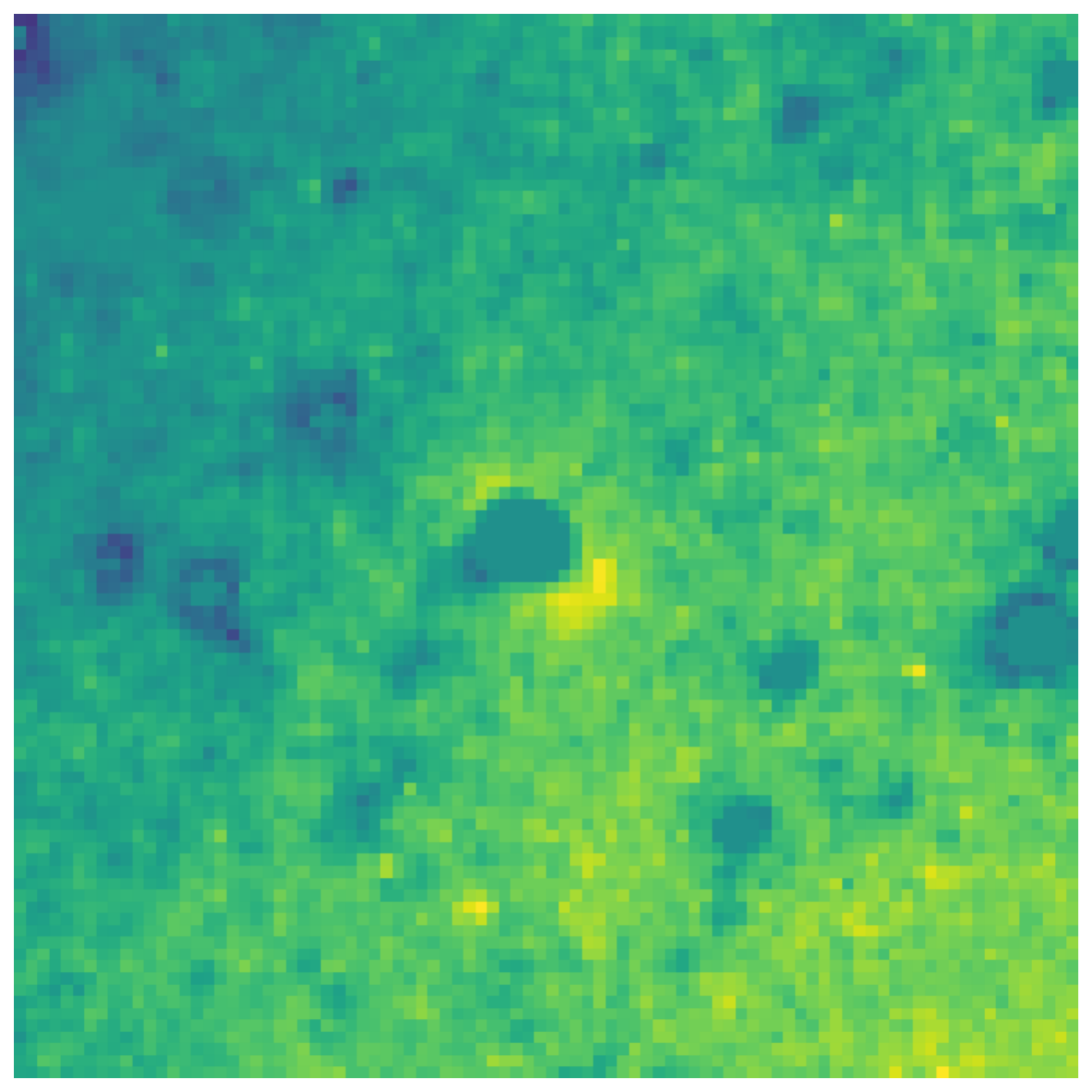}
\includegraphics[width=0.205\textwidth]{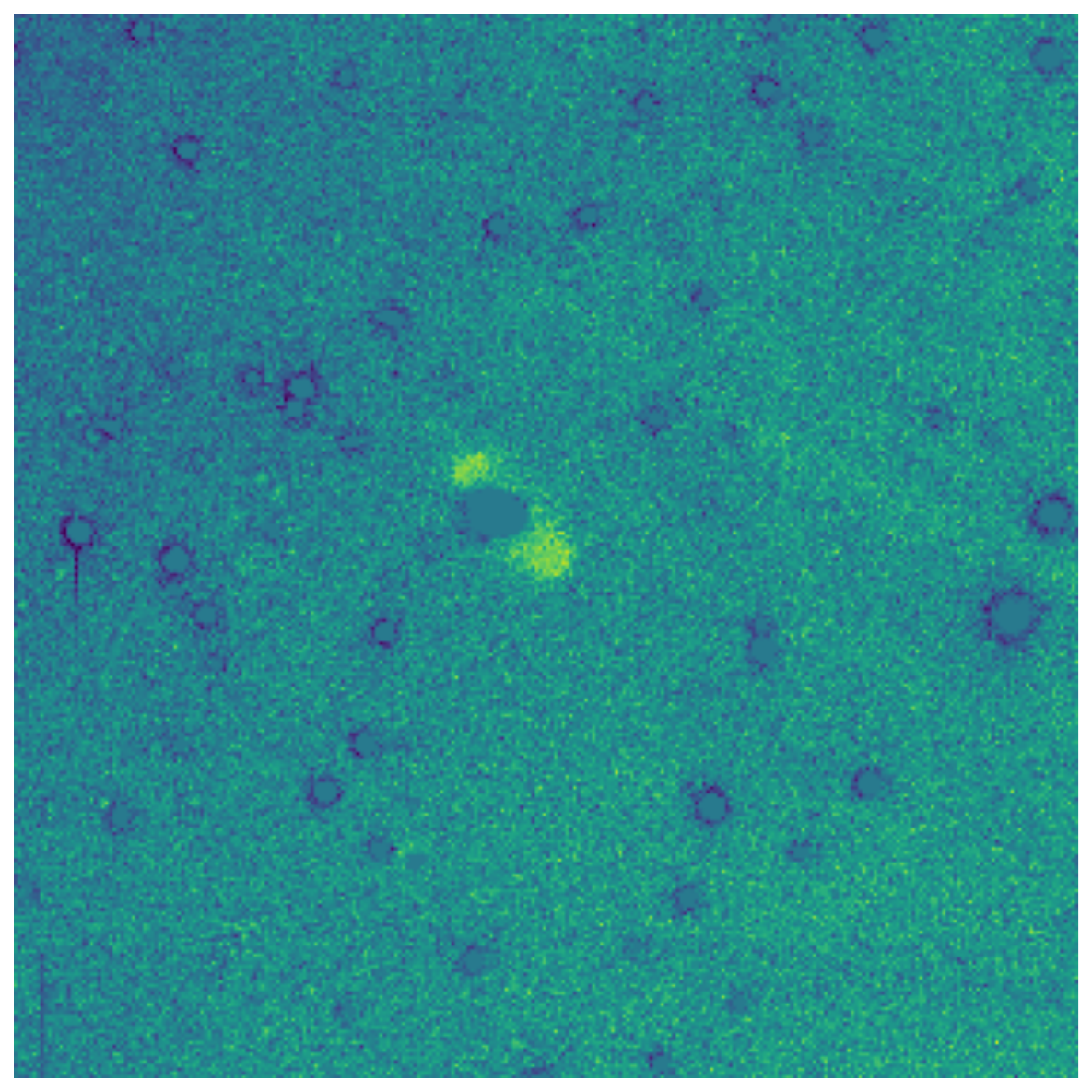}
\begin{overpic}[width=0.4\textwidth]{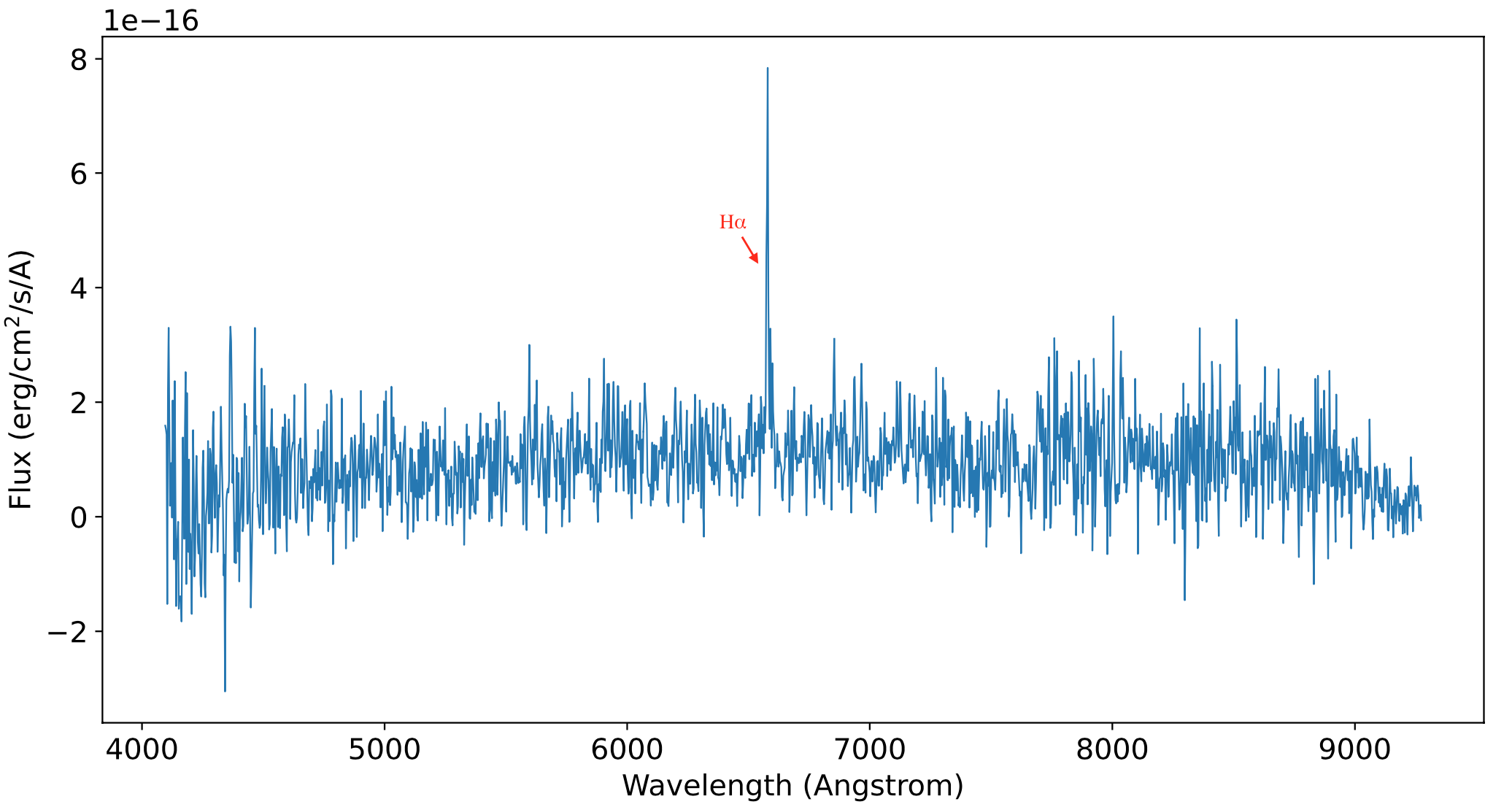}
\put(70,45){\small{YP1818-1211}}
\end{overpic}
\includegraphics[width=0.2\textwidth]{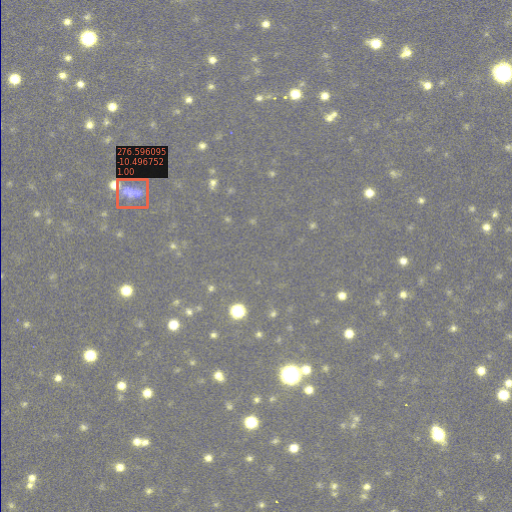}
\includegraphics[width=0.205\textwidth]{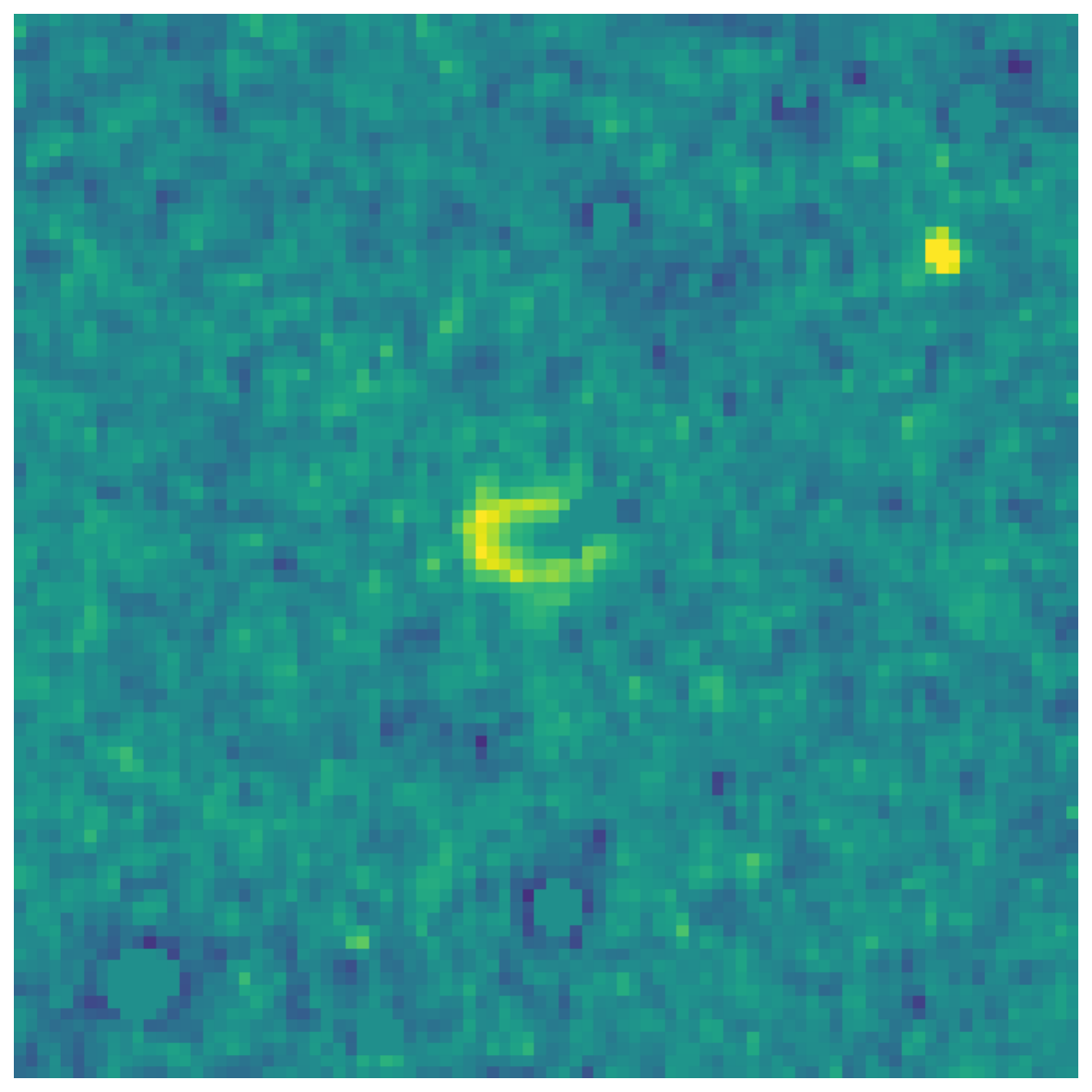}
\includegraphics[width=0.205\textwidth]{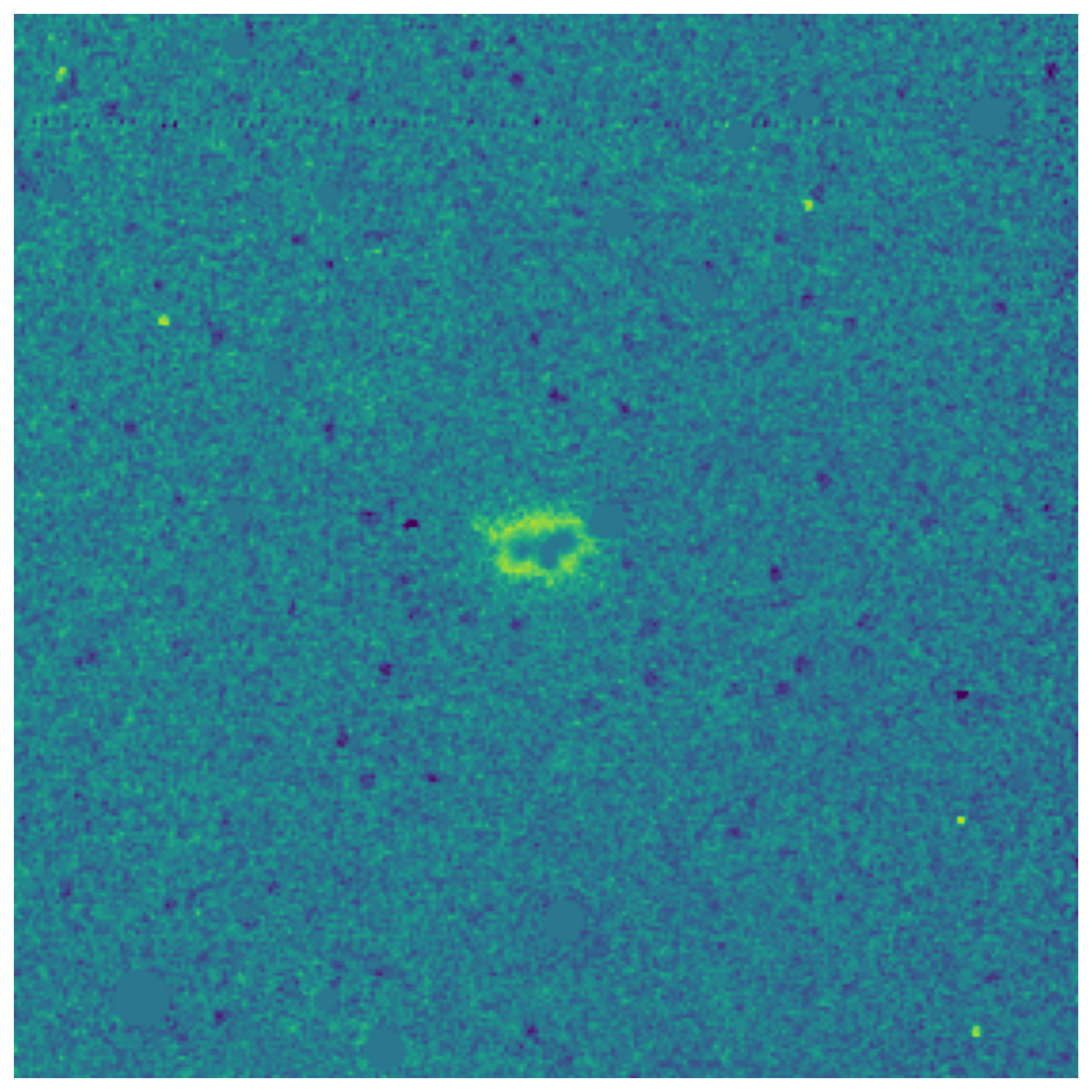}
\begin{overpic}[width=0.4\textwidth]{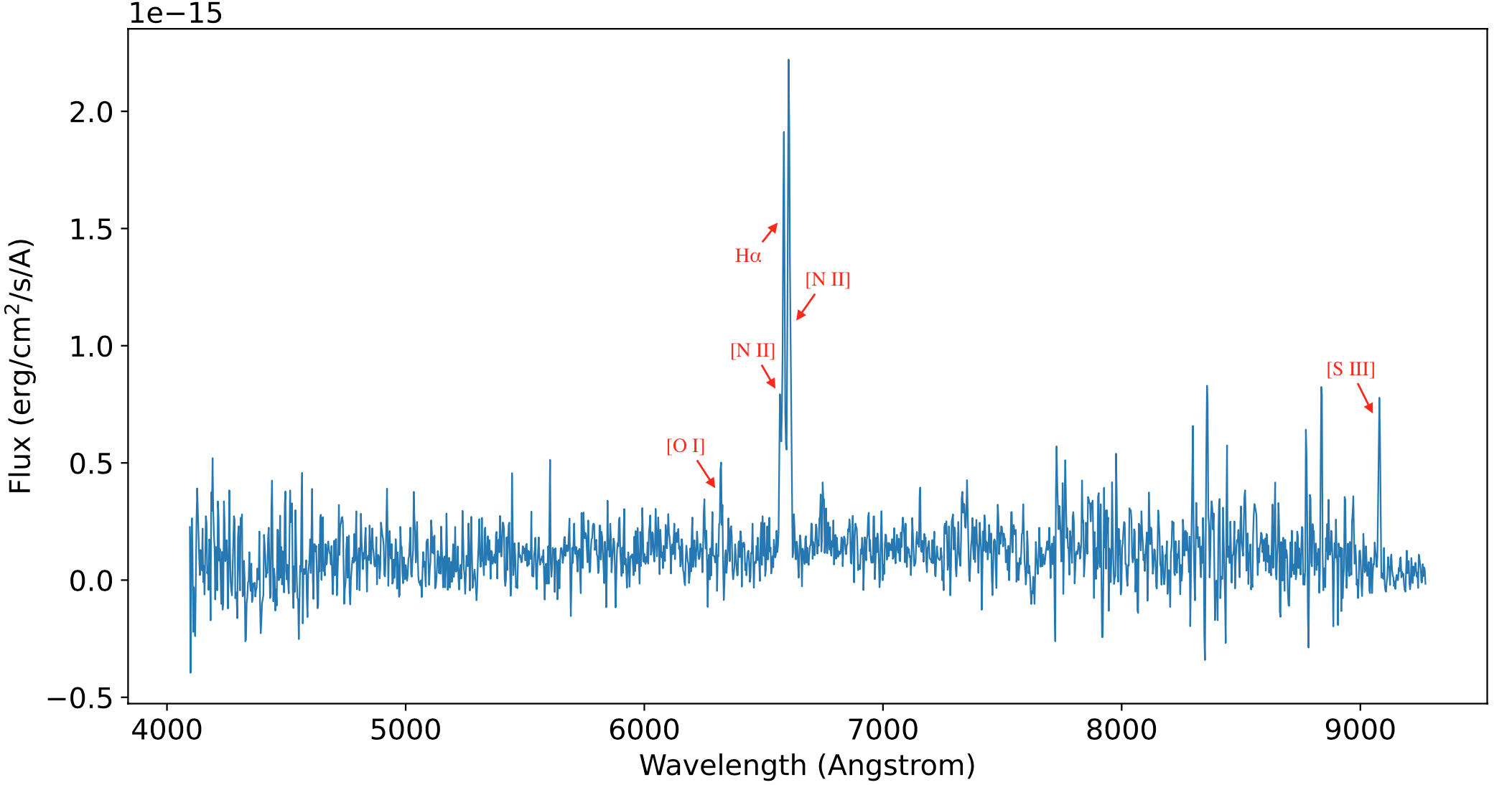}
\put(70,45){\small{YP1826-1029}}
\end{overpic}
\includegraphics[width=0.2\textwidth]{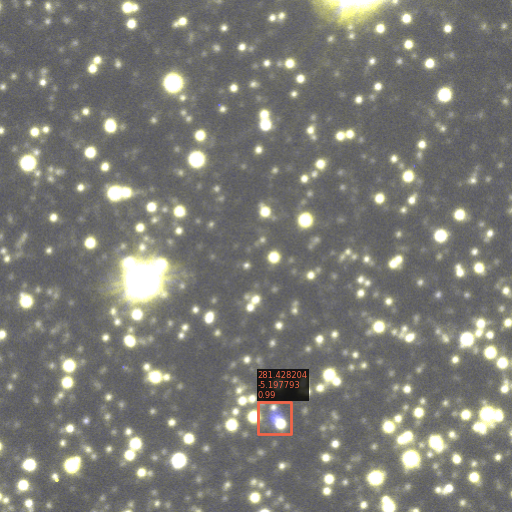}
\includegraphics[width=0.205\textwidth]{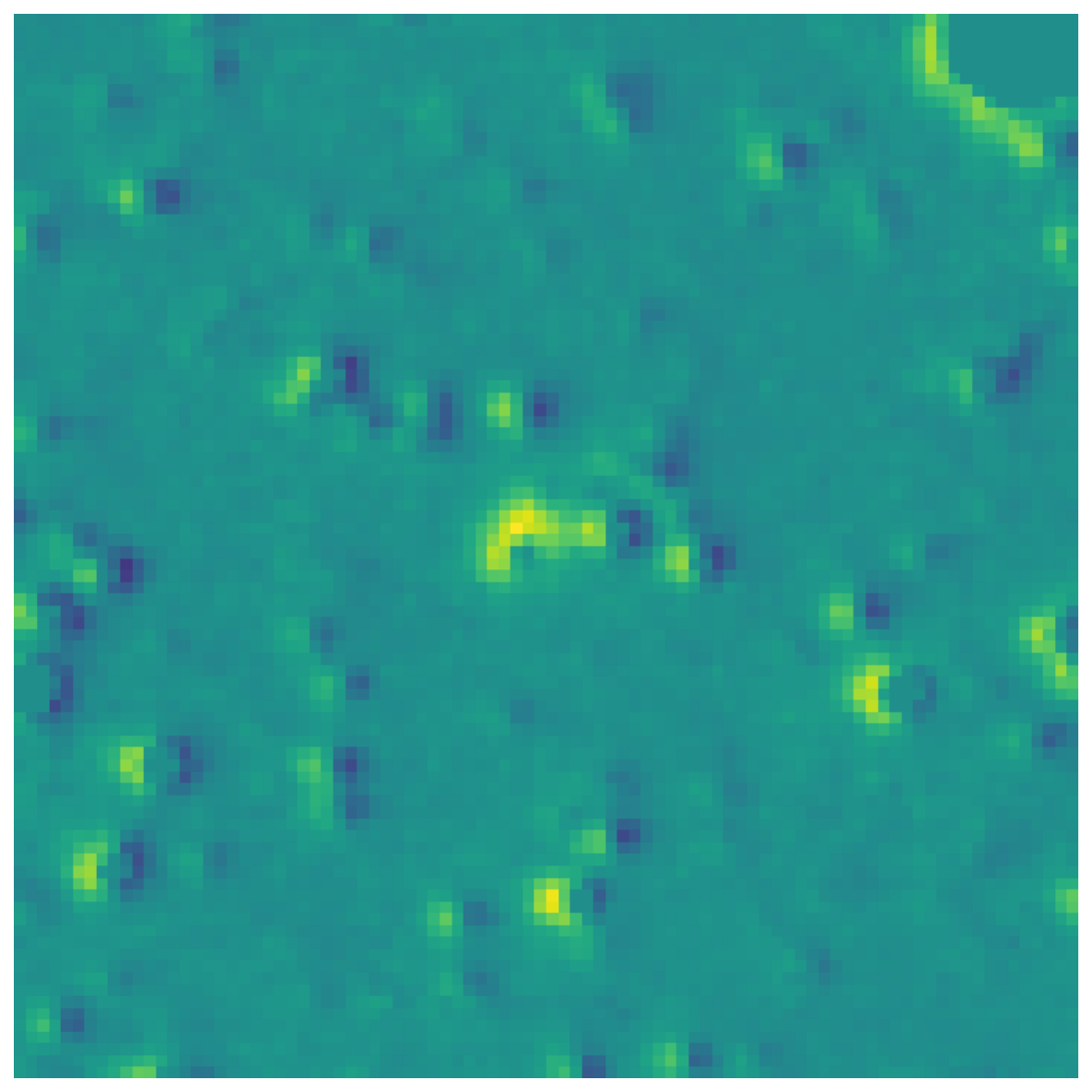}
\includegraphics[width=0.205\textwidth]{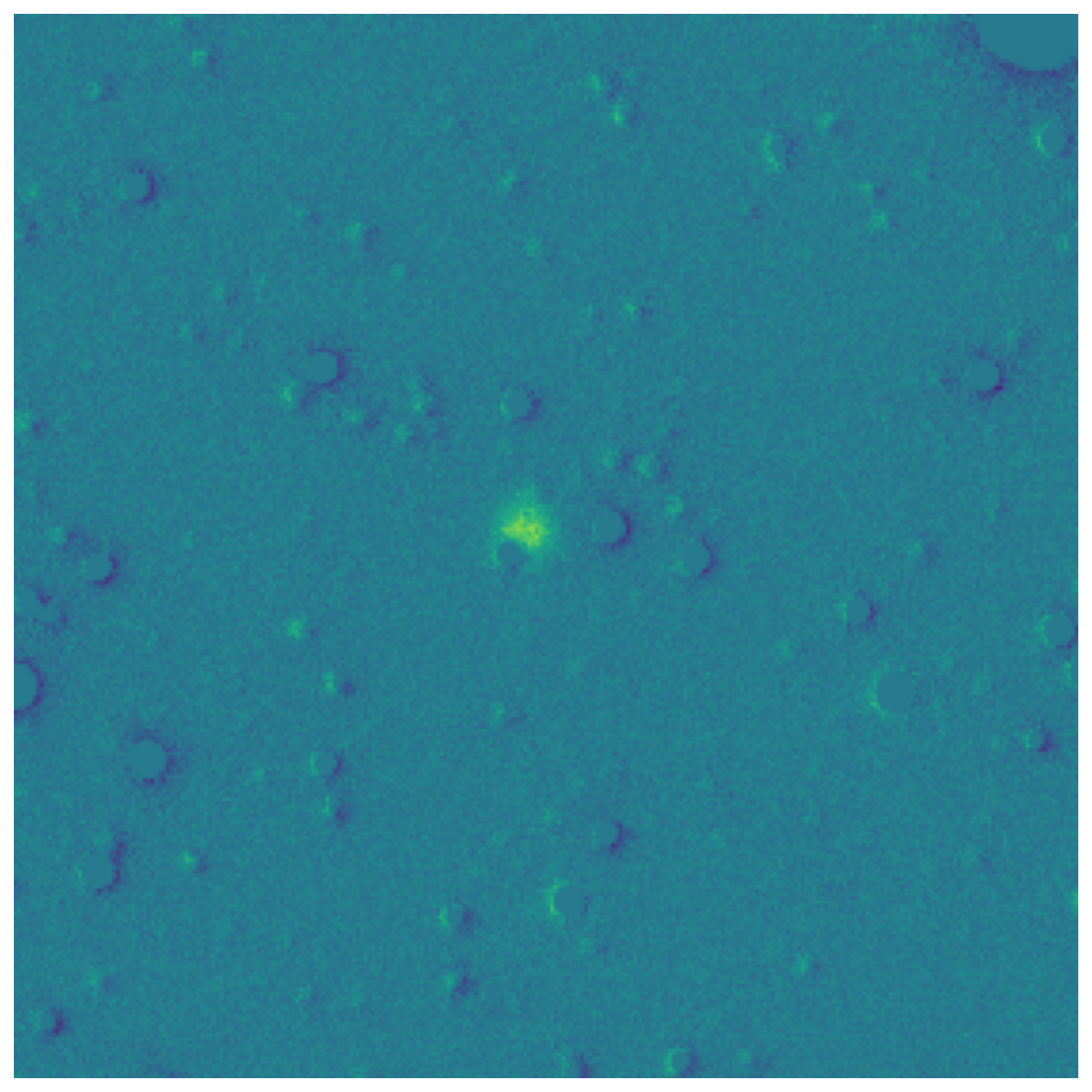}
\begin{overpic}[width=0.4\textwidth]{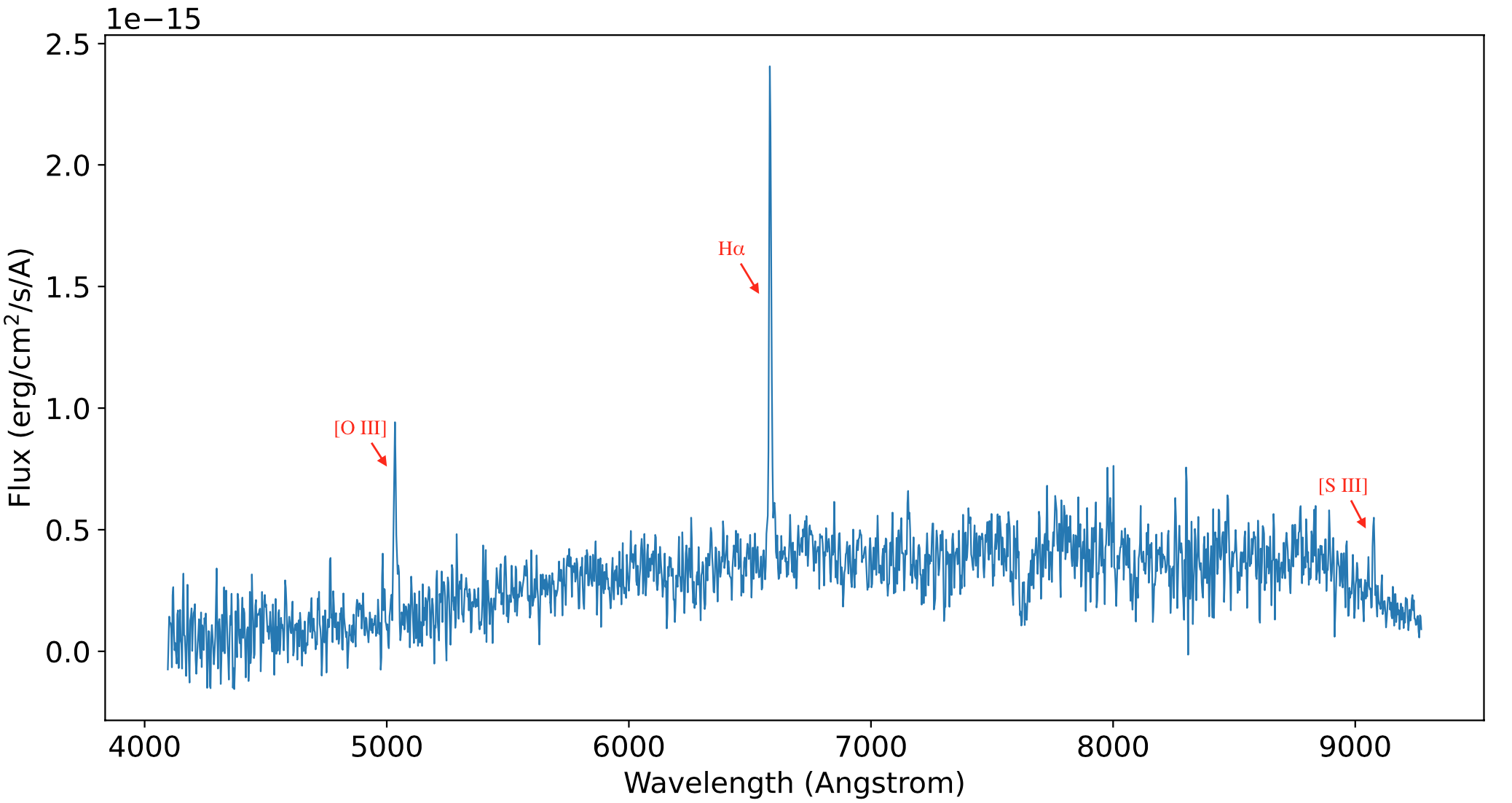}
\put(70,45){\small{YP1845-0511}}
\end{overpic}
\caption{True PNe: VPHAS+ detected PNG images labelled by red boxes, SHS H$\alpha$-Rband quotient images (H$\alpha$ image for YP1759-2036), VPHAS+ H$\alpha$-Rband quotient images, SAAO spectra with spectral lines labelled.}
\label{fig:true_pn}
\end{figure*}

\begin{figure*}[h!]
\begin{minipage}{\textwidth}
\includegraphics[width=0.2\textwidth]{fig/YP1028-5714.png}
\includegraphics[width=0.205\textwidth]{fig/YP1028-5714_shs_q.pdf}
\includegraphics[width=0.205\textwidth]{fig/YP1028-5714_vphas_q.pdf}
\begin{overpic}[width=0.4\textwidth]{fig/YP1028-5714_spec.png}
\put(70,45){\small{YP1028-5714}}
\end{overpic}
\includegraphics[width=0.2\textwidth]{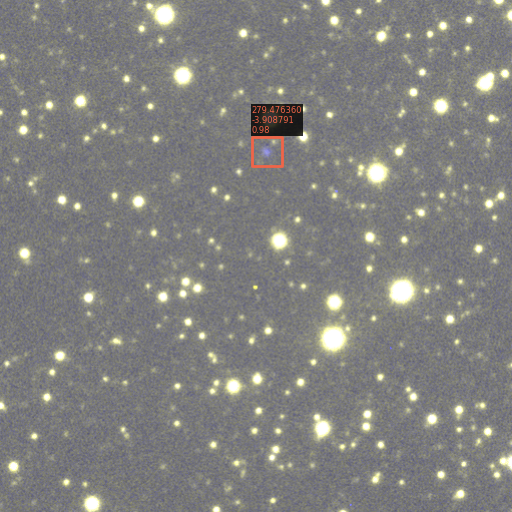}
\includegraphics[width=0.205\textwidth]{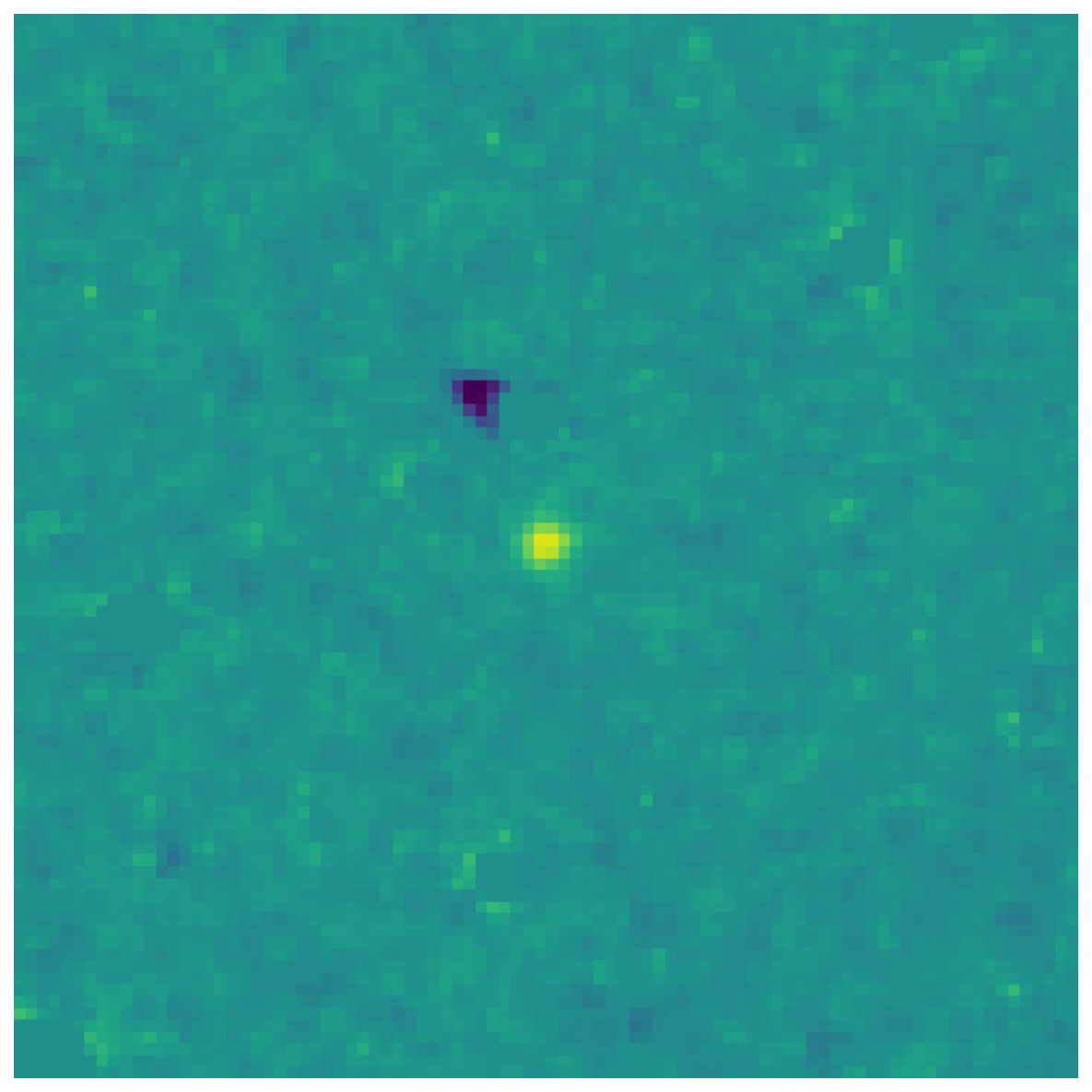}
\includegraphics[width=0.205\textwidth]{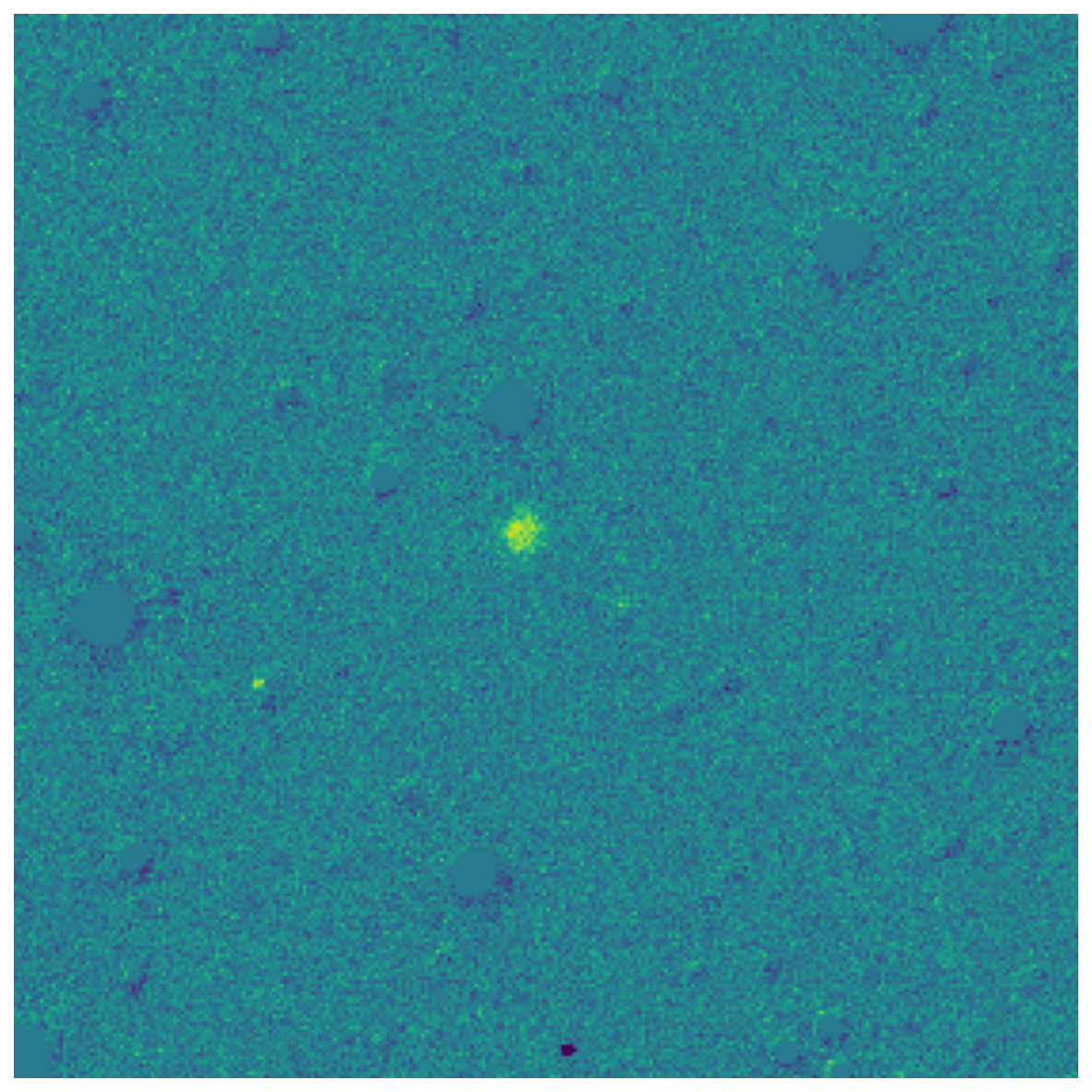}
\begin{overpic}[width=0.4\textwidth]{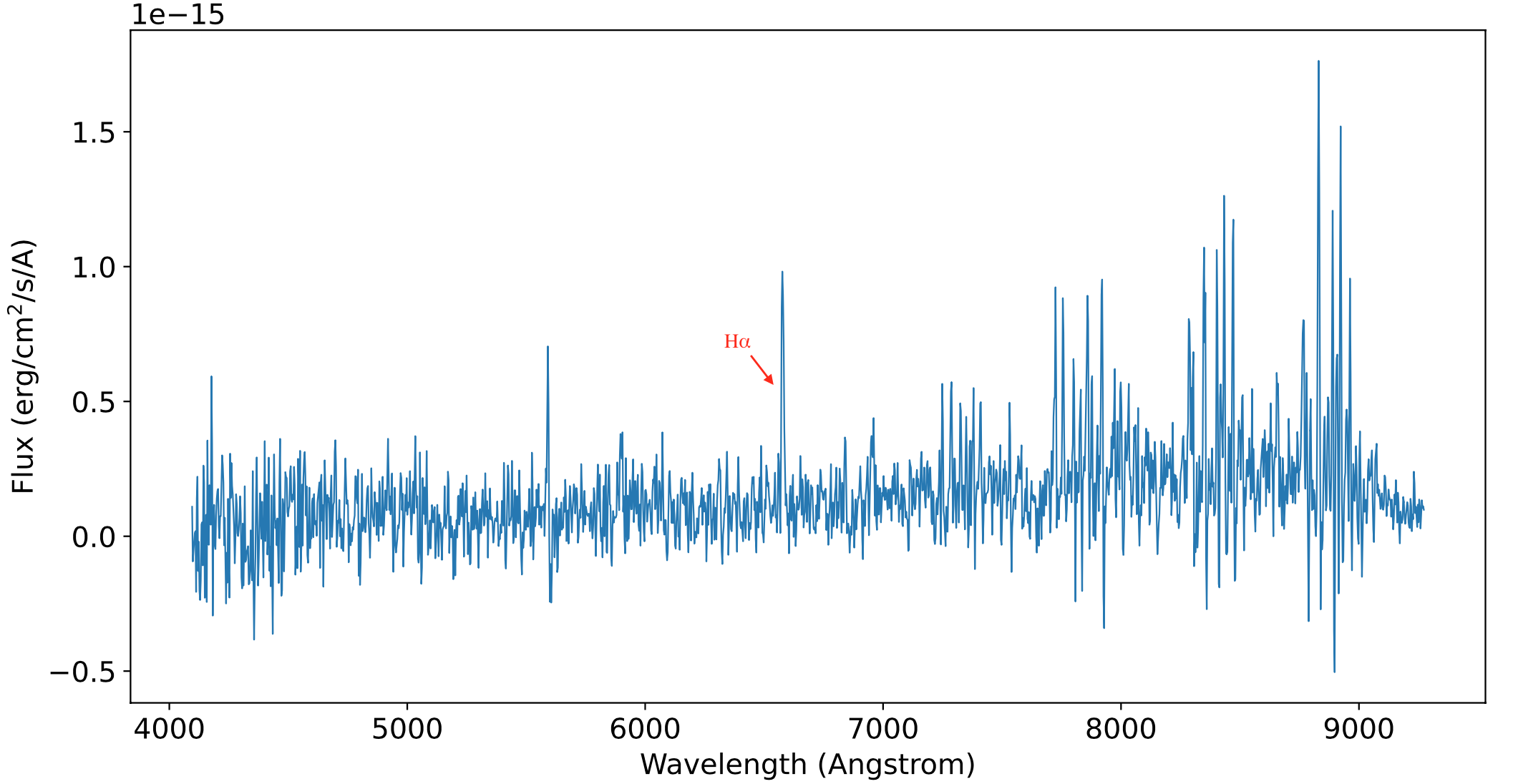}
\put(70,45){\small{YP1837-0354}}
\end{overpic}
\caption{The same as Figure \ref{fig:true_pn}, for likely PNe.}
\label{fig:likely_pn}
\end{minipage}
\begin{minipage}{\textwidth}
\includegraphics[width=0.2\textwidth]{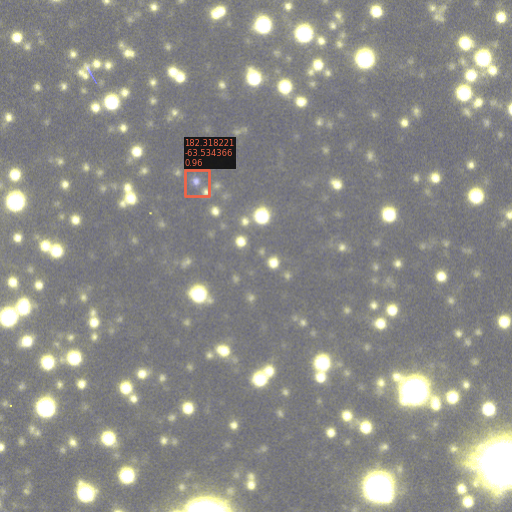}
\includegraphics[width=0.205\textwidth]{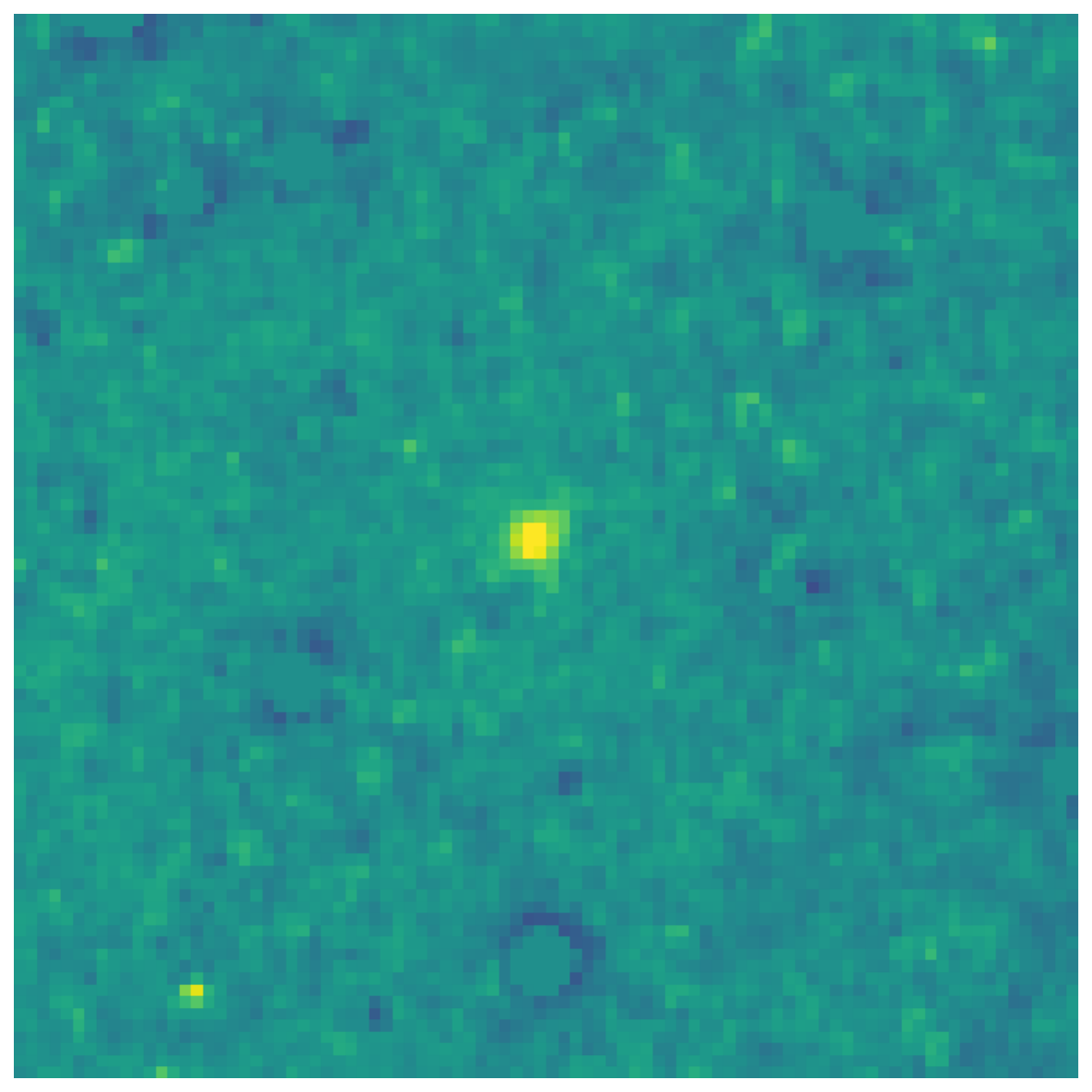}
\includegraphics[width=0.205\textwidth]{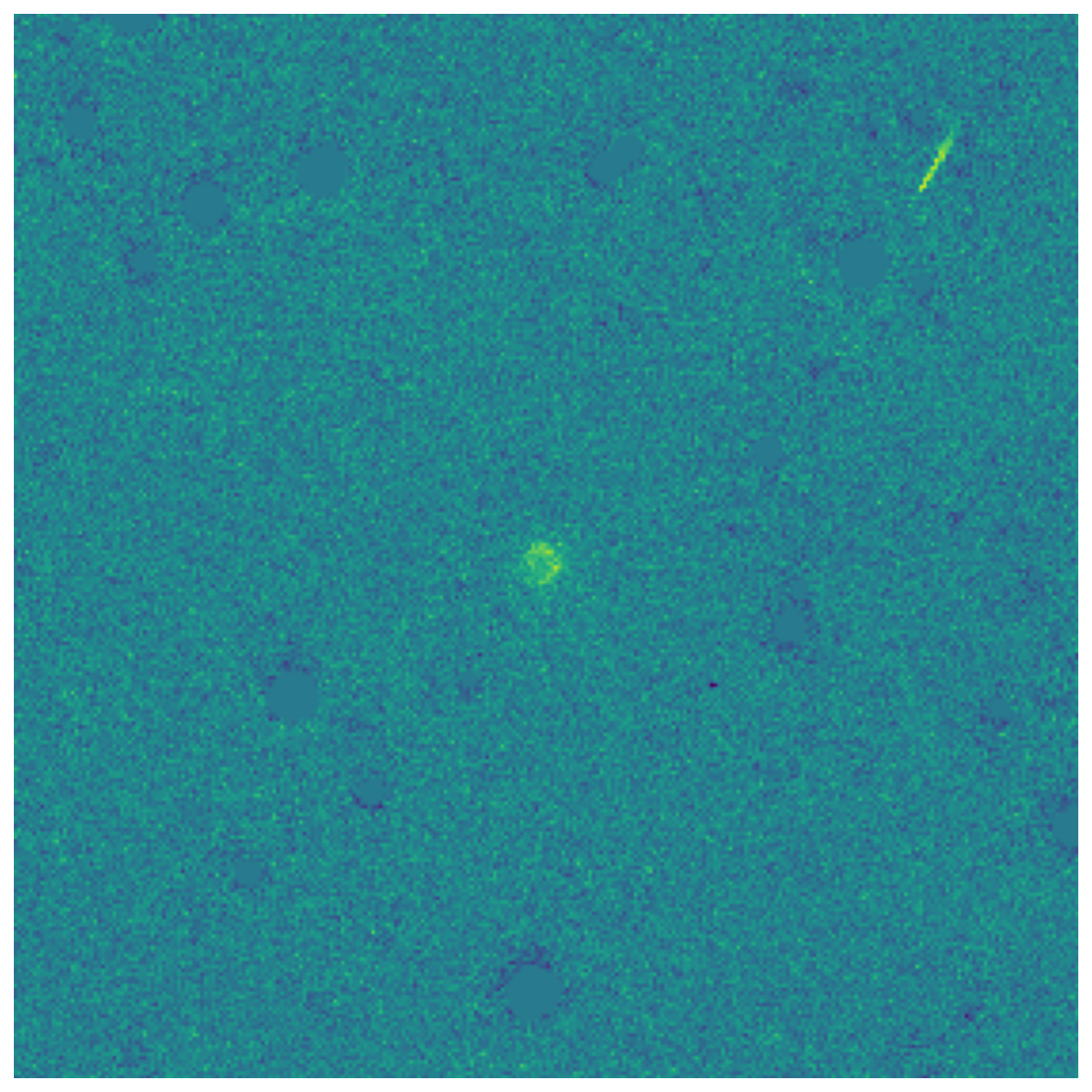}
\begin{overpic}[width=0.4\textwidth]{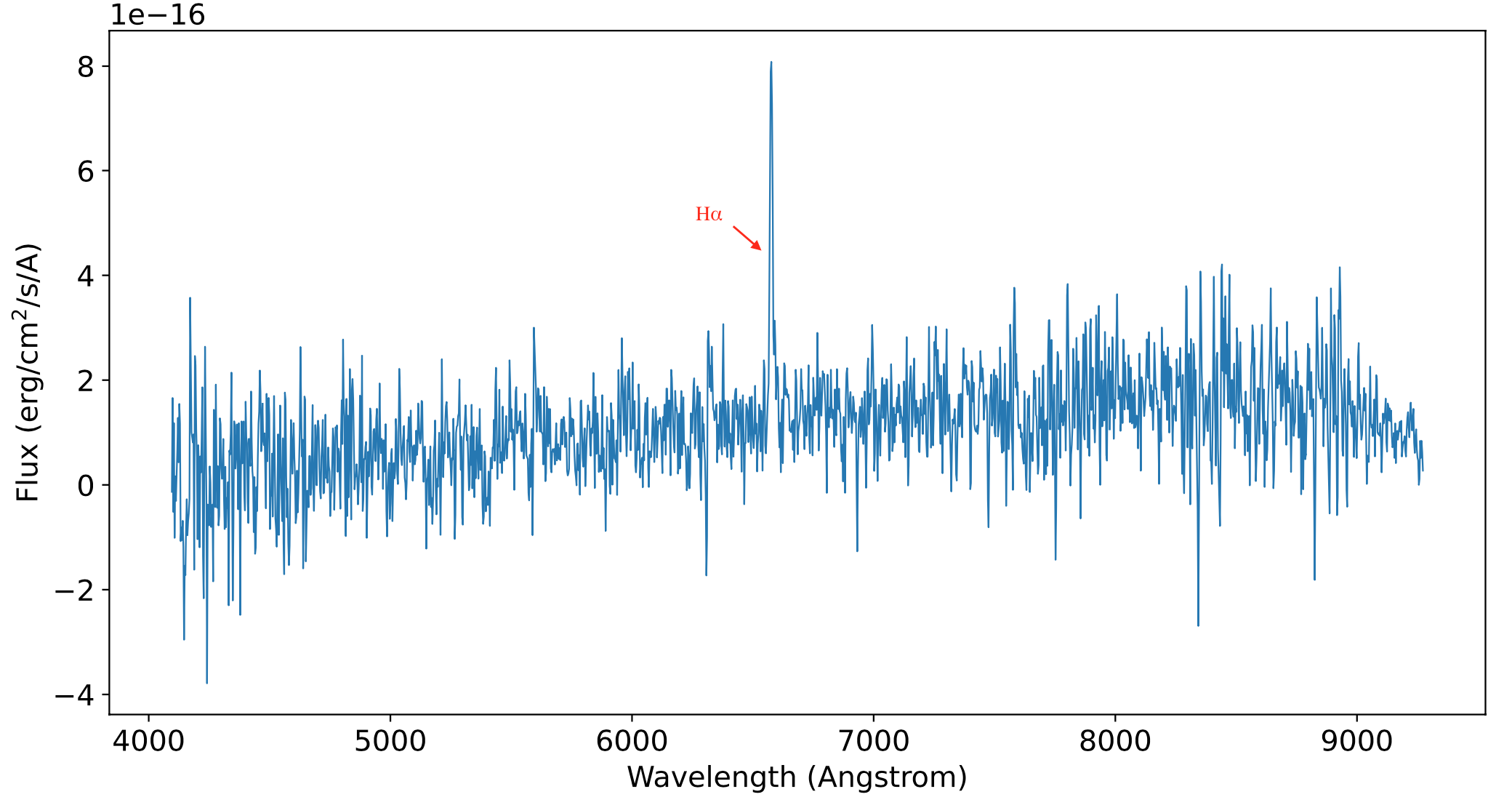}
\put(70,45){\small{YP1209-6332}}
\end{overpic}
\includegraphics[width=0.2\textwidth]{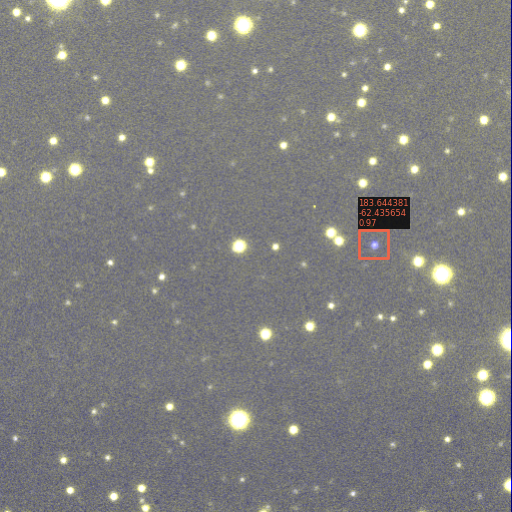}
\includegraphics[width=0.205\textwidth]{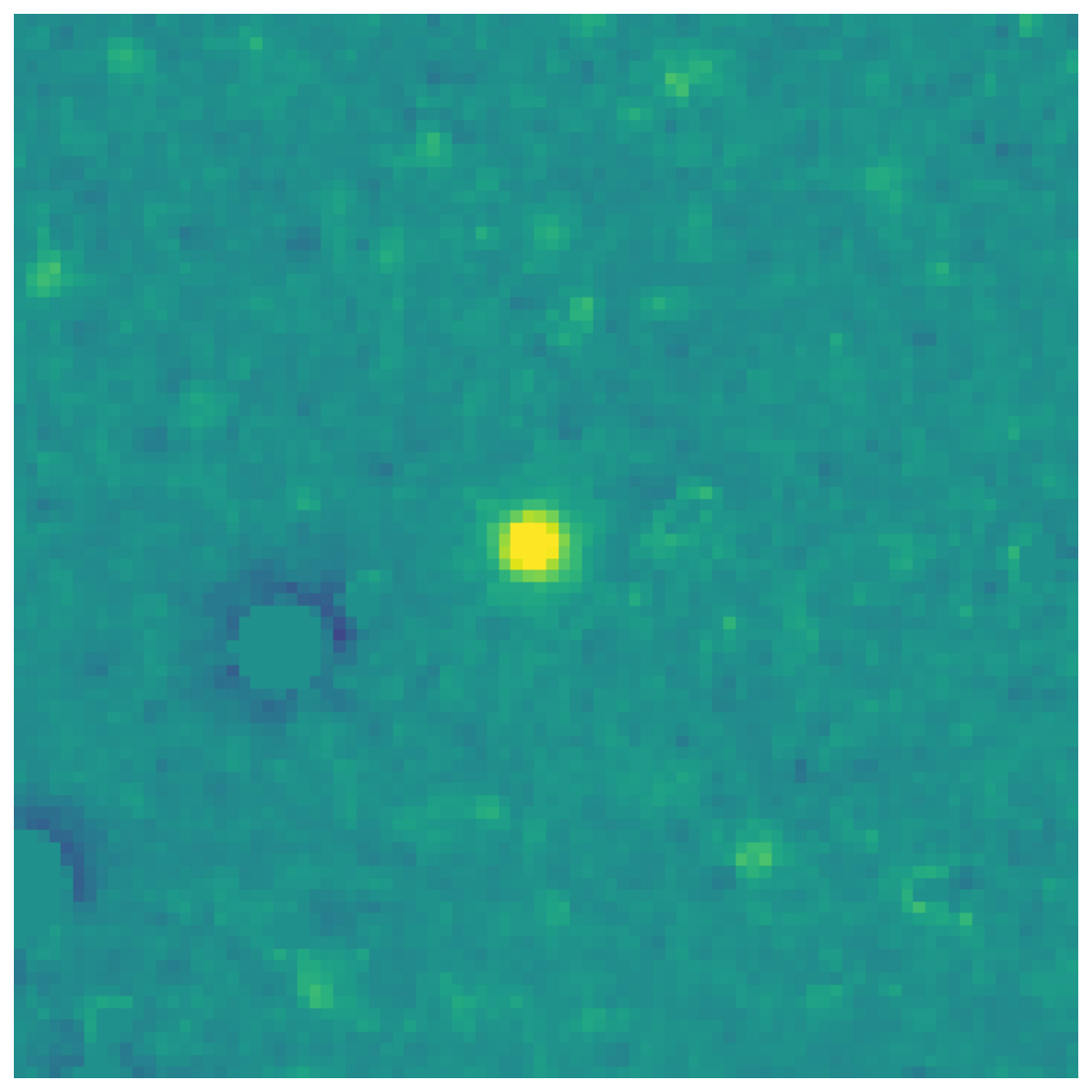}
\includegraphics[width=0.205\textwidth]{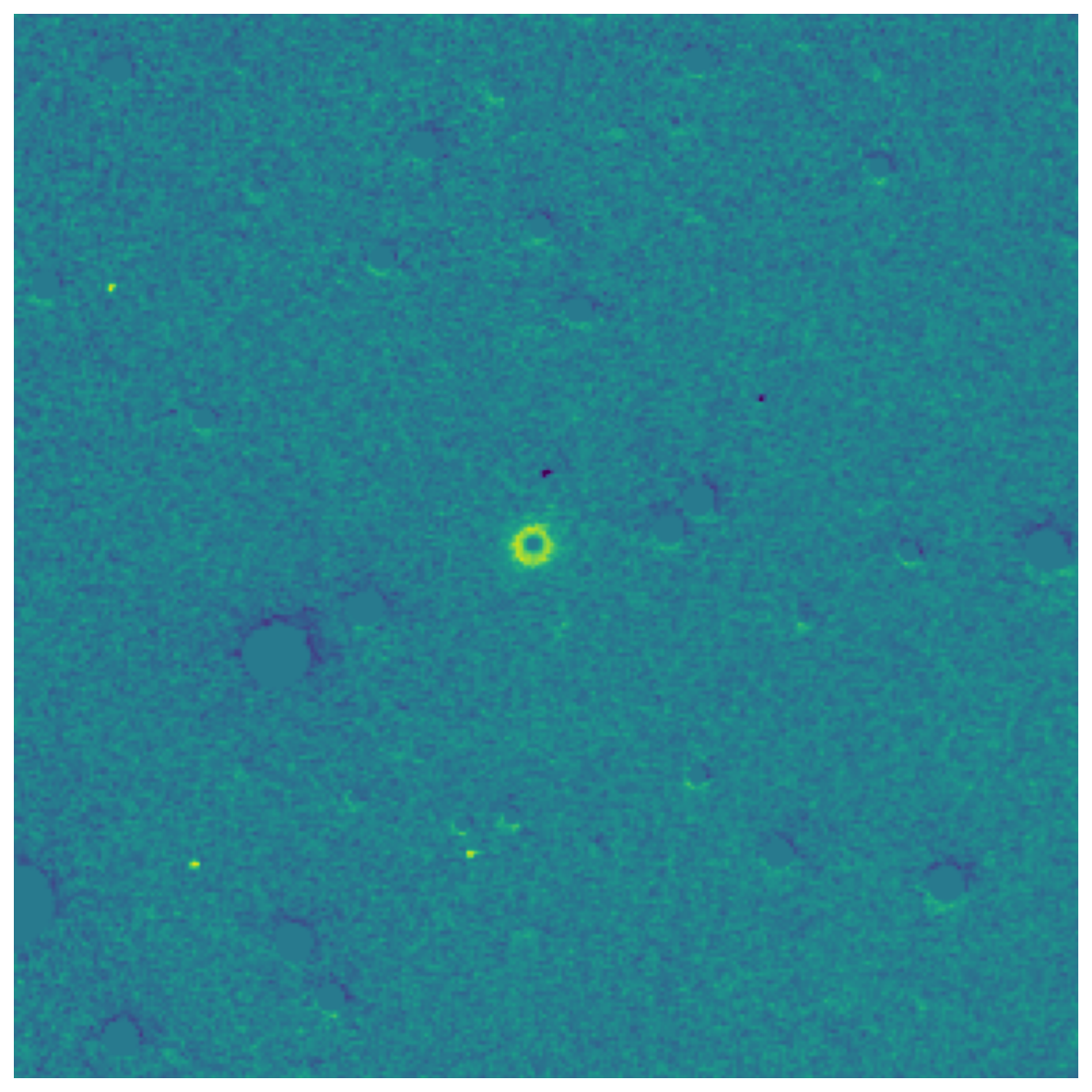}
\begin{overpic}[width=0.4\textwidth]{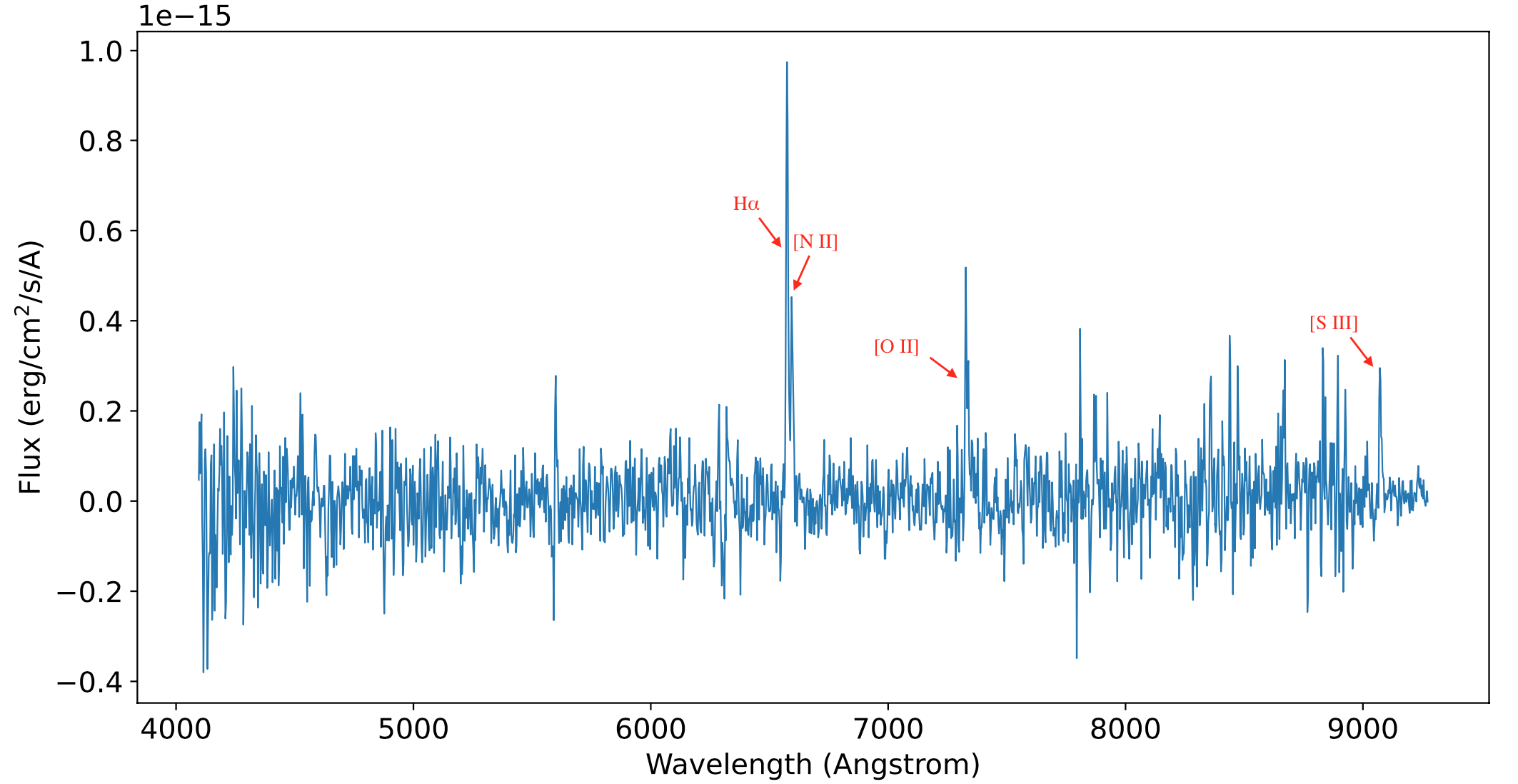}
\put(70,45){\small{YP1214-6226}}
\end{overpic}
\includegraphics[width=0.2\textwidth]{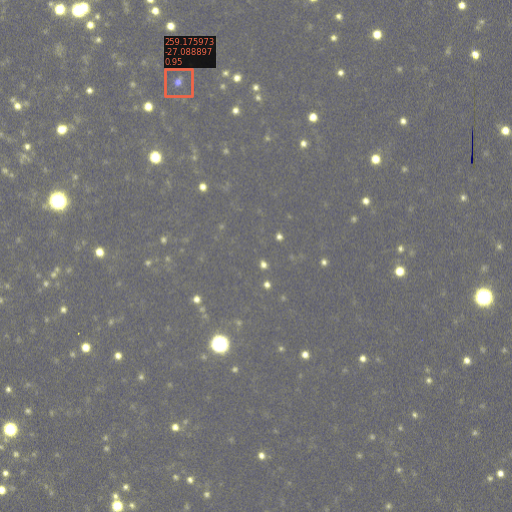}
\includegraphics[width=0.205\textwidth]{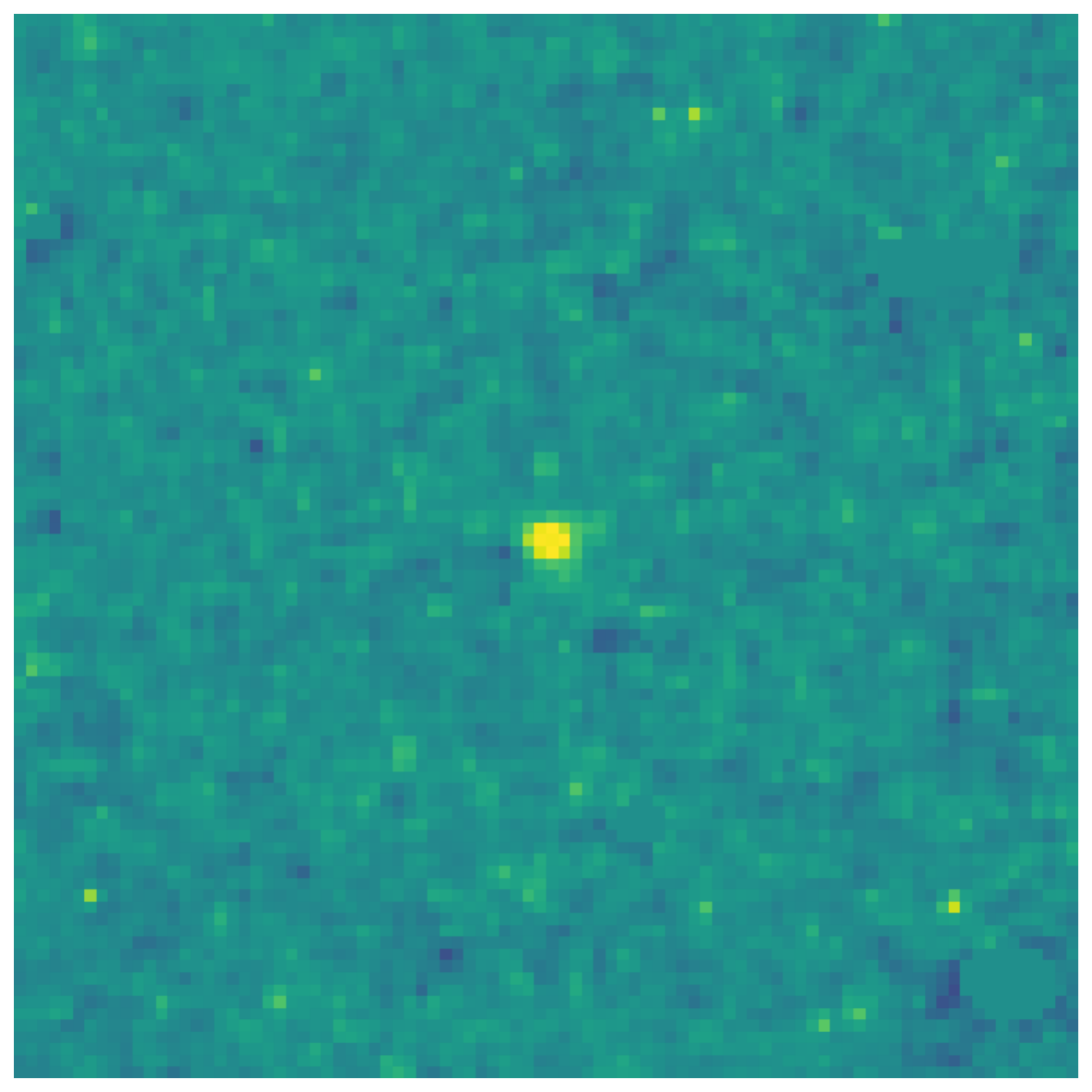}
\includegraphics[width=0.205\textwidth]{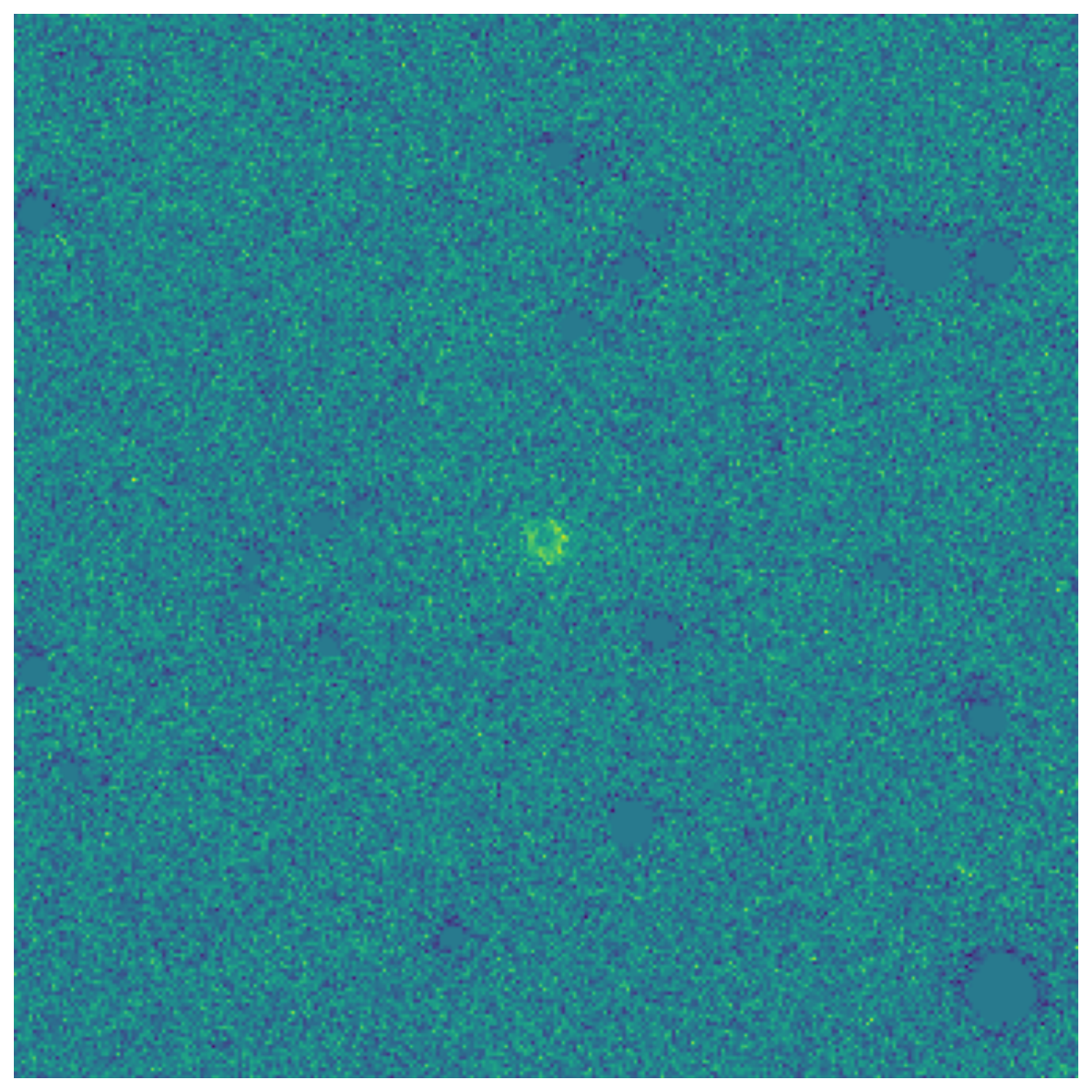}
\begin{overpic}[width=0.4\textwidth]{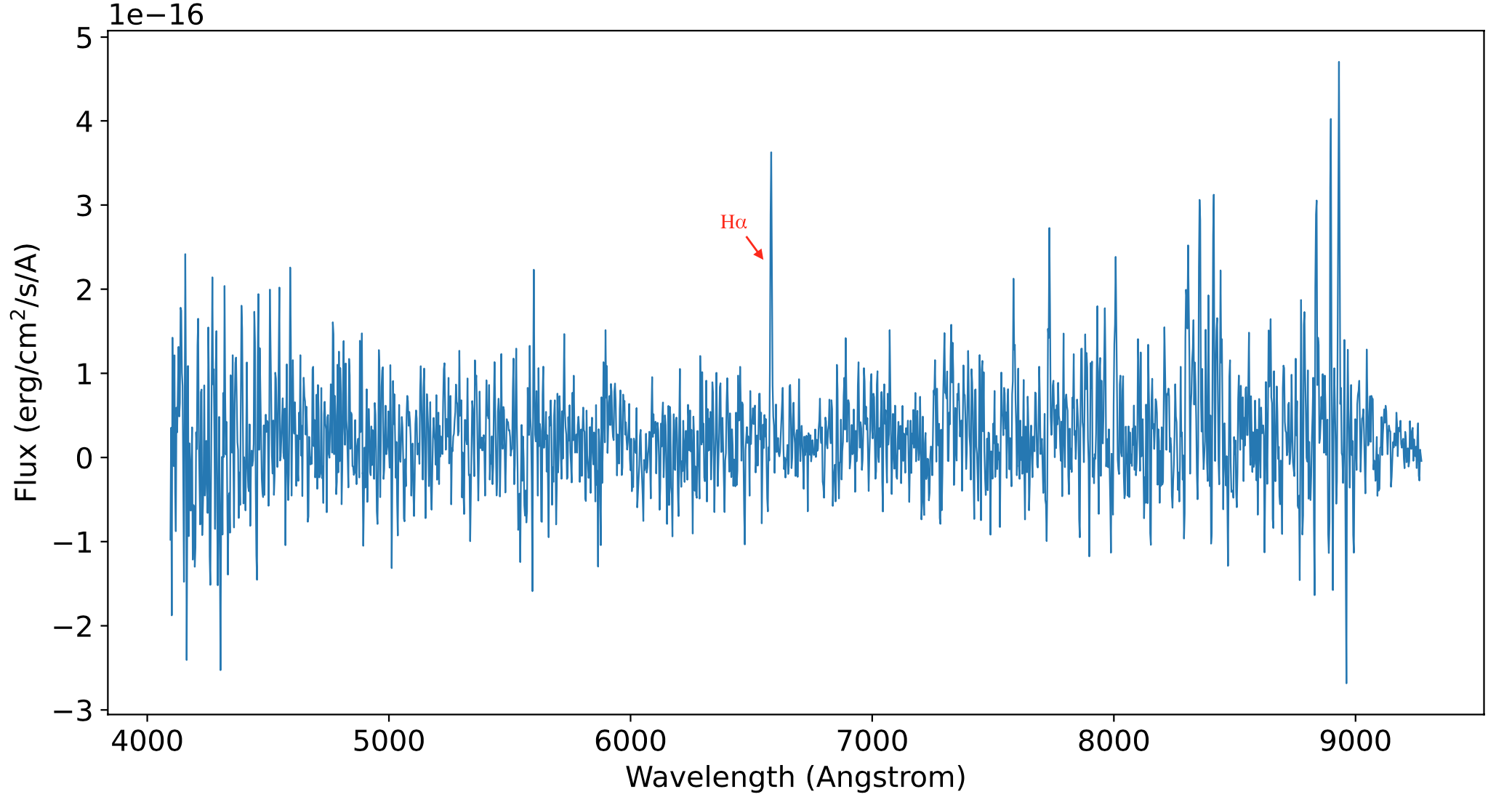}
\put(70,45){\small{YP1716-2705}}
\end{overpic}
\includegraphics[width=0.2\textwidth]{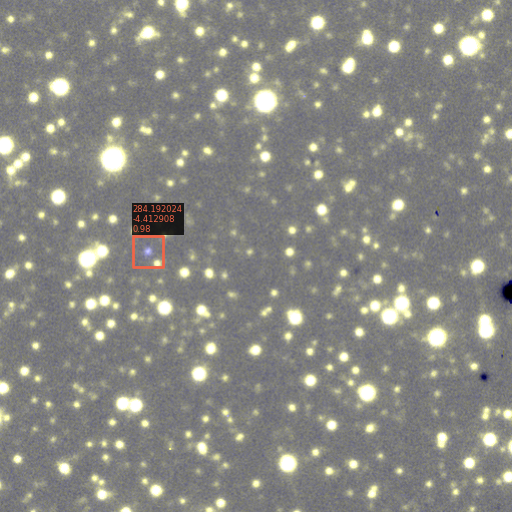}
\includegraphics[width=0.205\textwidth]{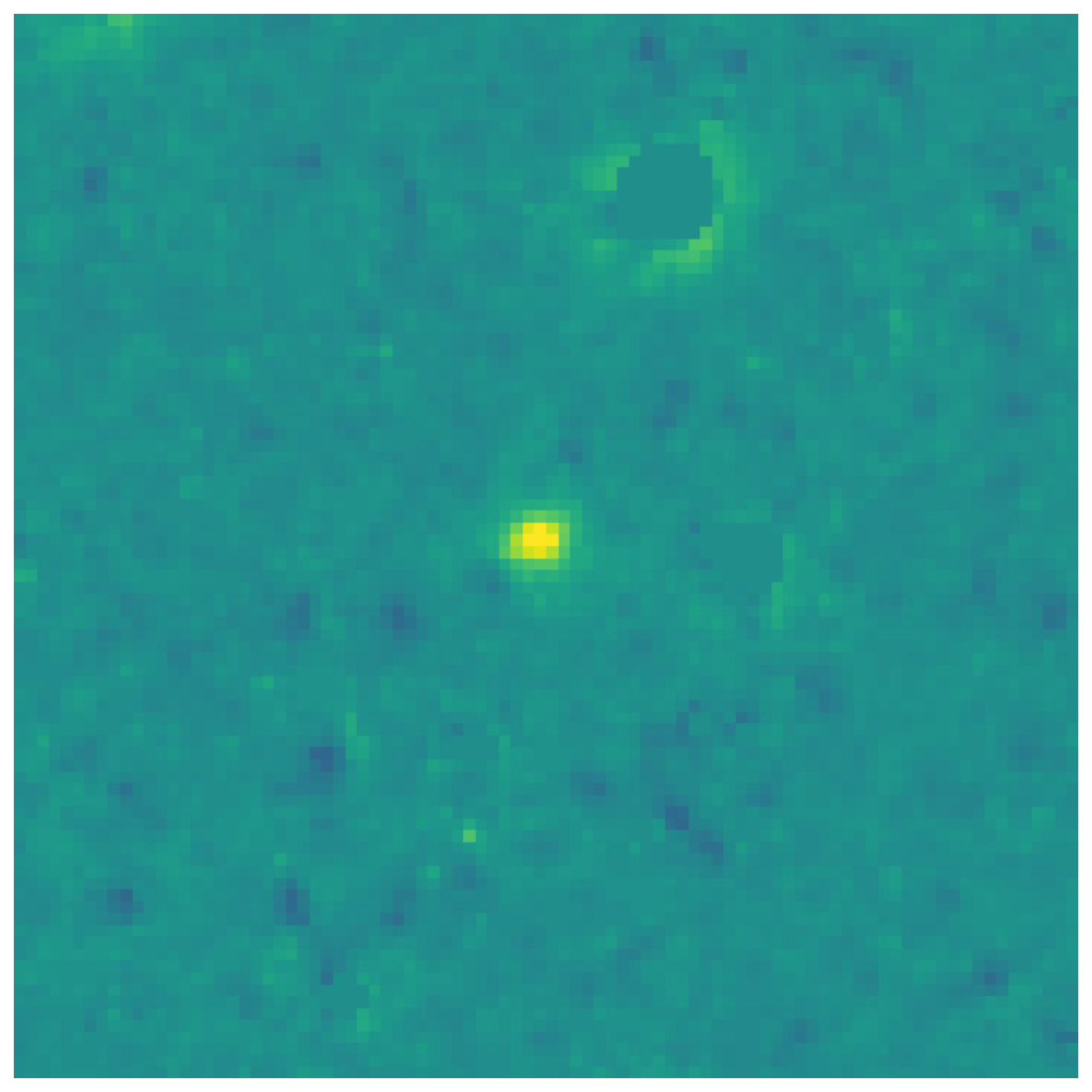}
\includegraphics[width=0.205\textwidth]{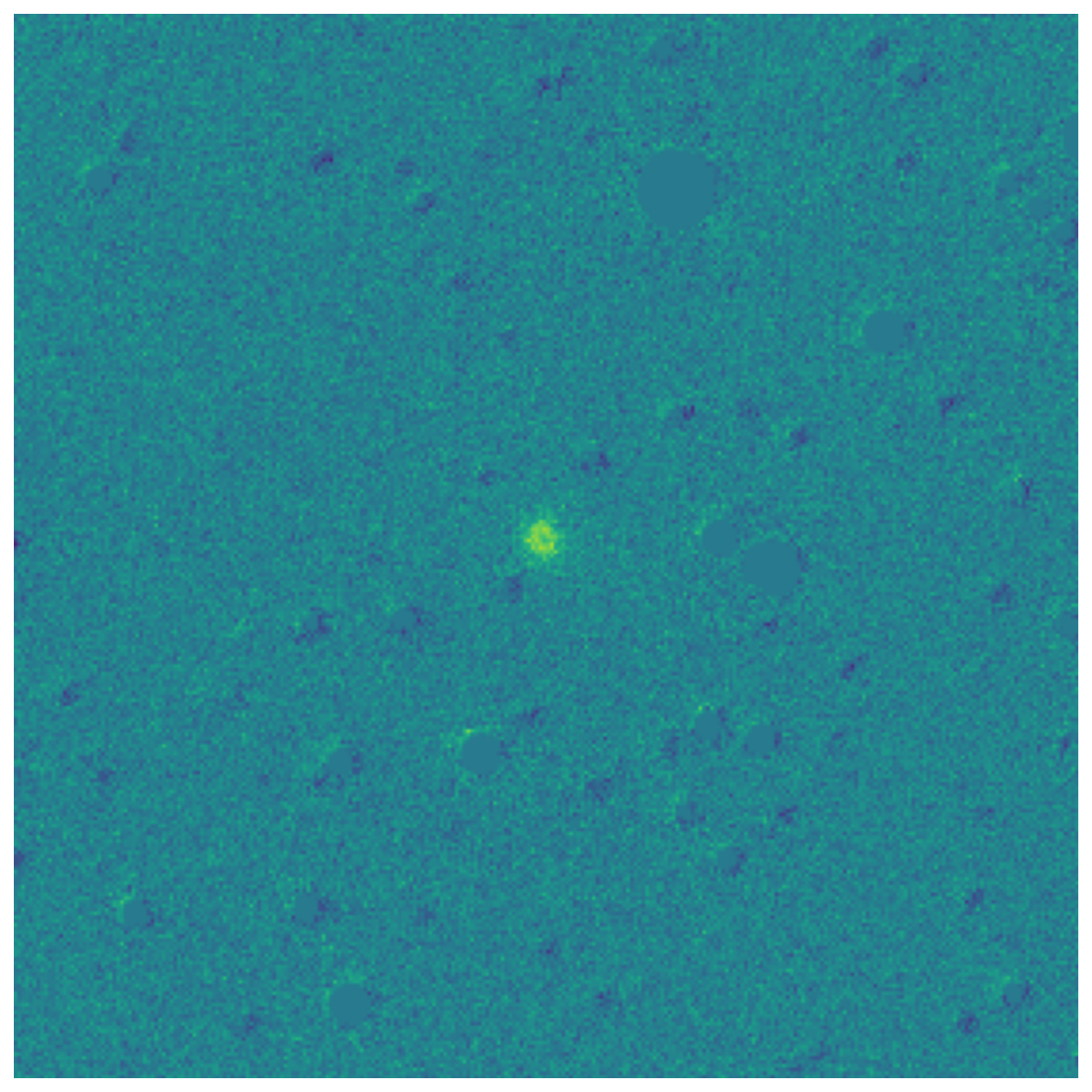}
\begin{overpic}[width=0.4\textwidth]{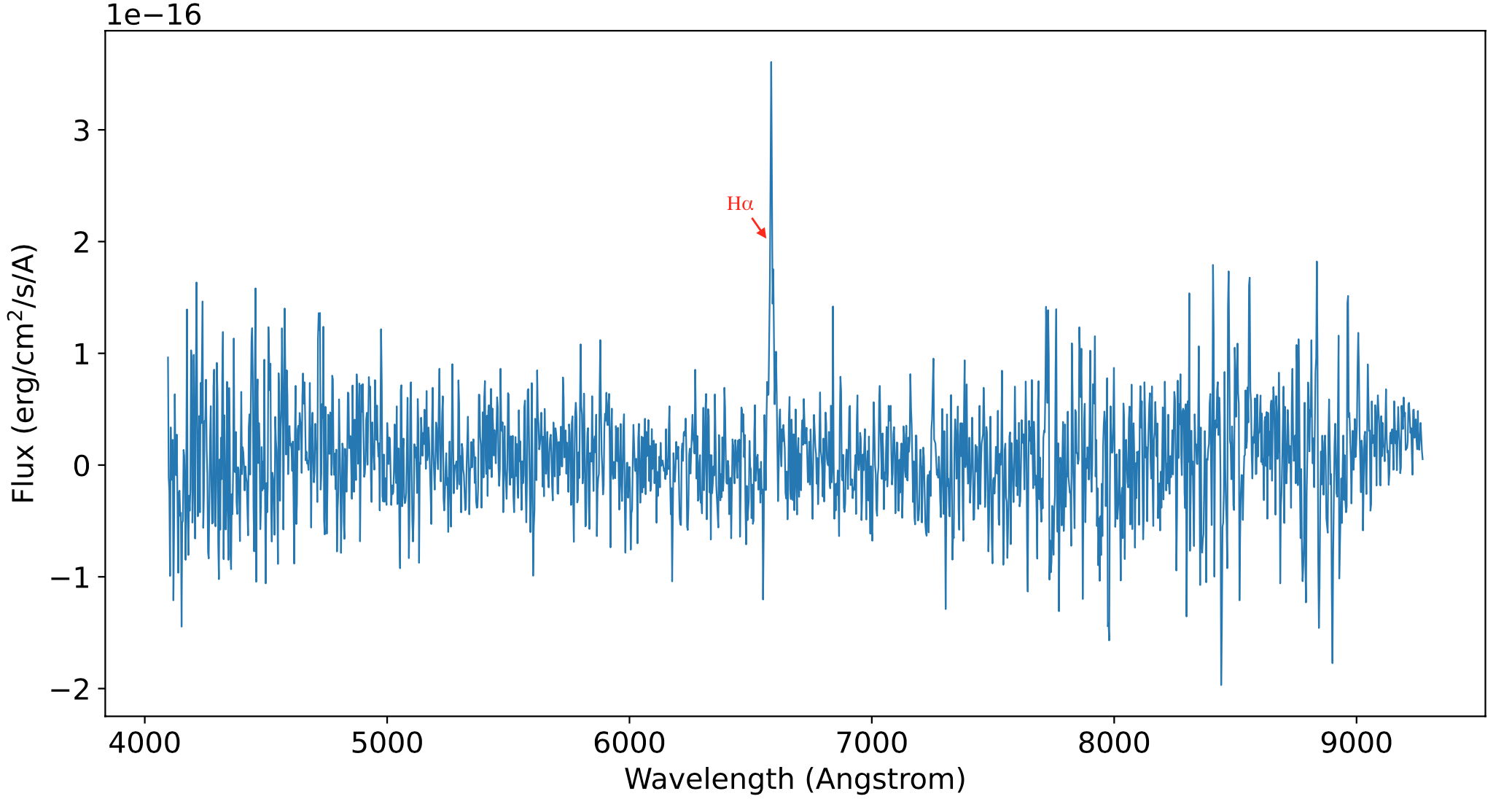}
\put(70,45){\small{YP1856-0424}}
\end{overpic}
\caption{The same as Figure \ref{fig:true_pn}, for possible PNe.}
\label{fig:possible_pn}
\end{minipage}
\end{figure*}

\begin{figure*}
\begin{minipage}{\textwidth}
\includegraphics[width=0.2\textwidth]{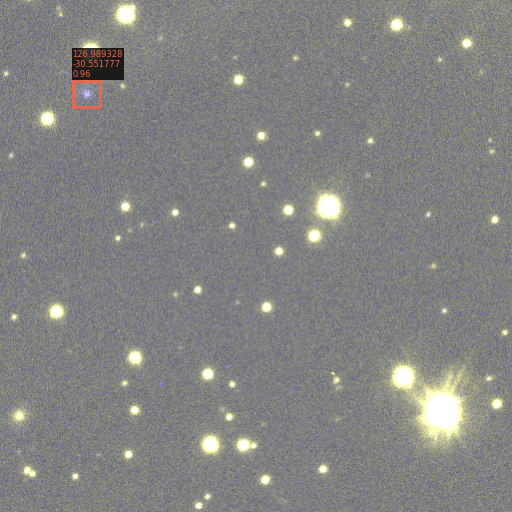}
\includegraphics[width=0.205\textwidth]{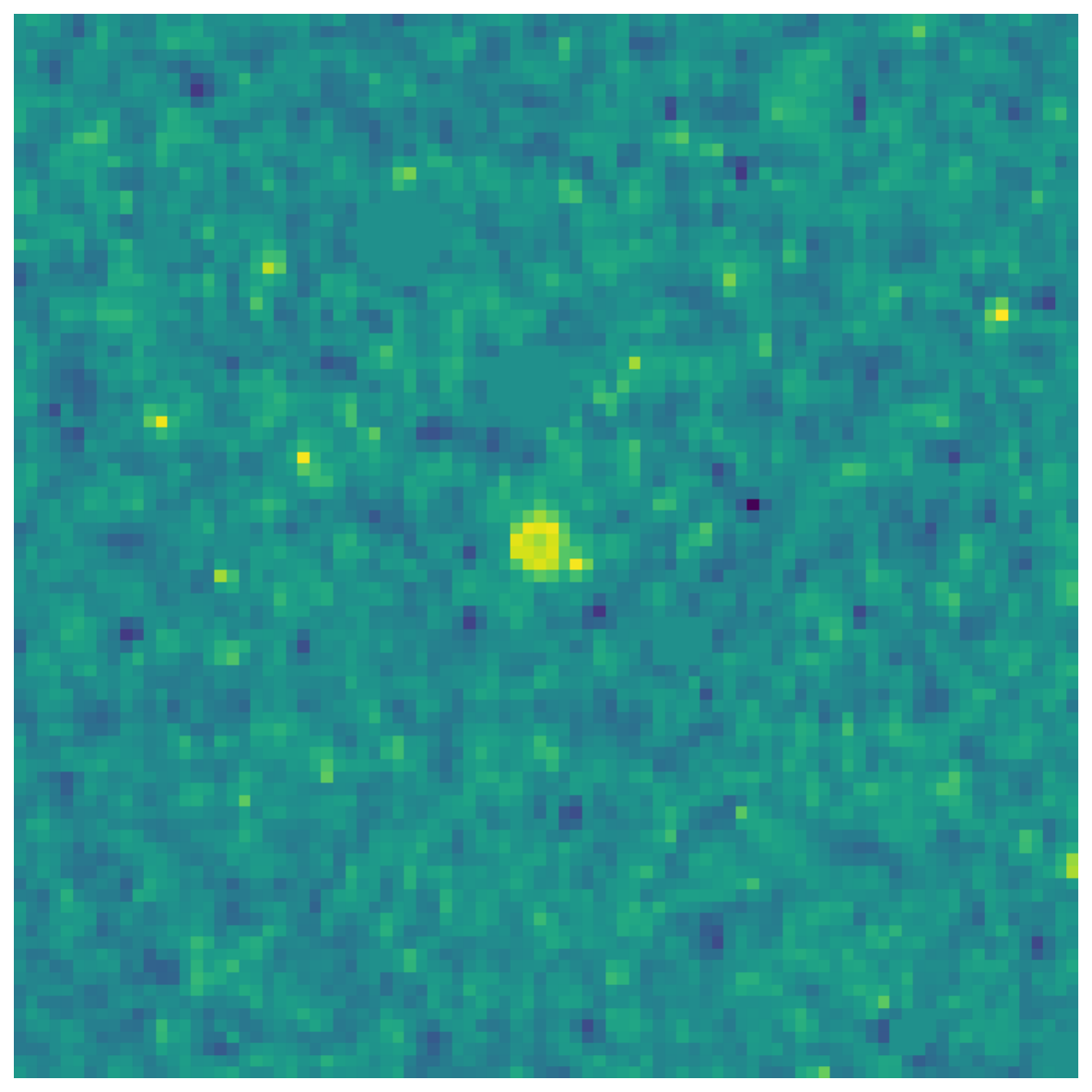}
\includegraphics[width=0.205\textwidth]{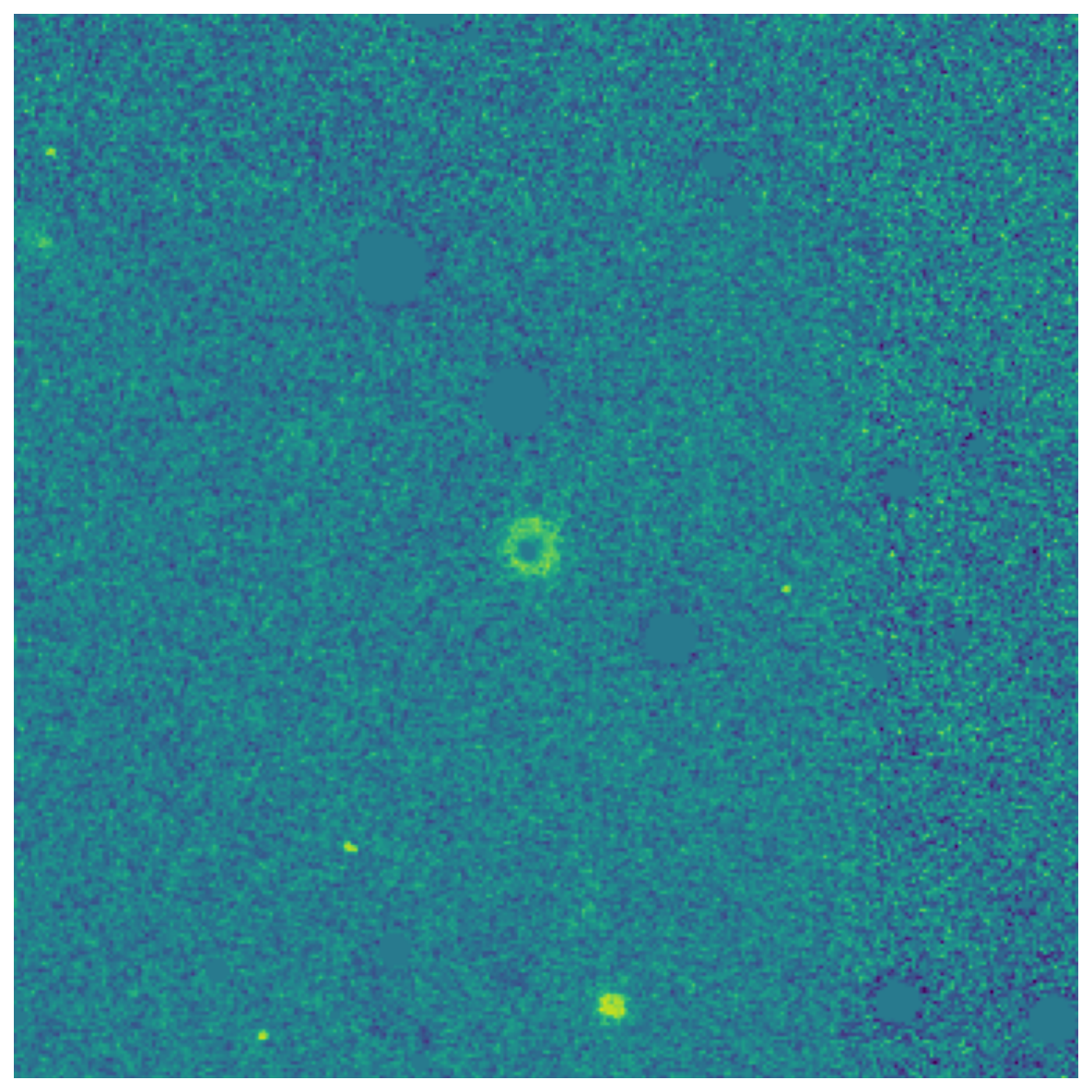}
\begin{overpic}[width=0.4\textwidth]{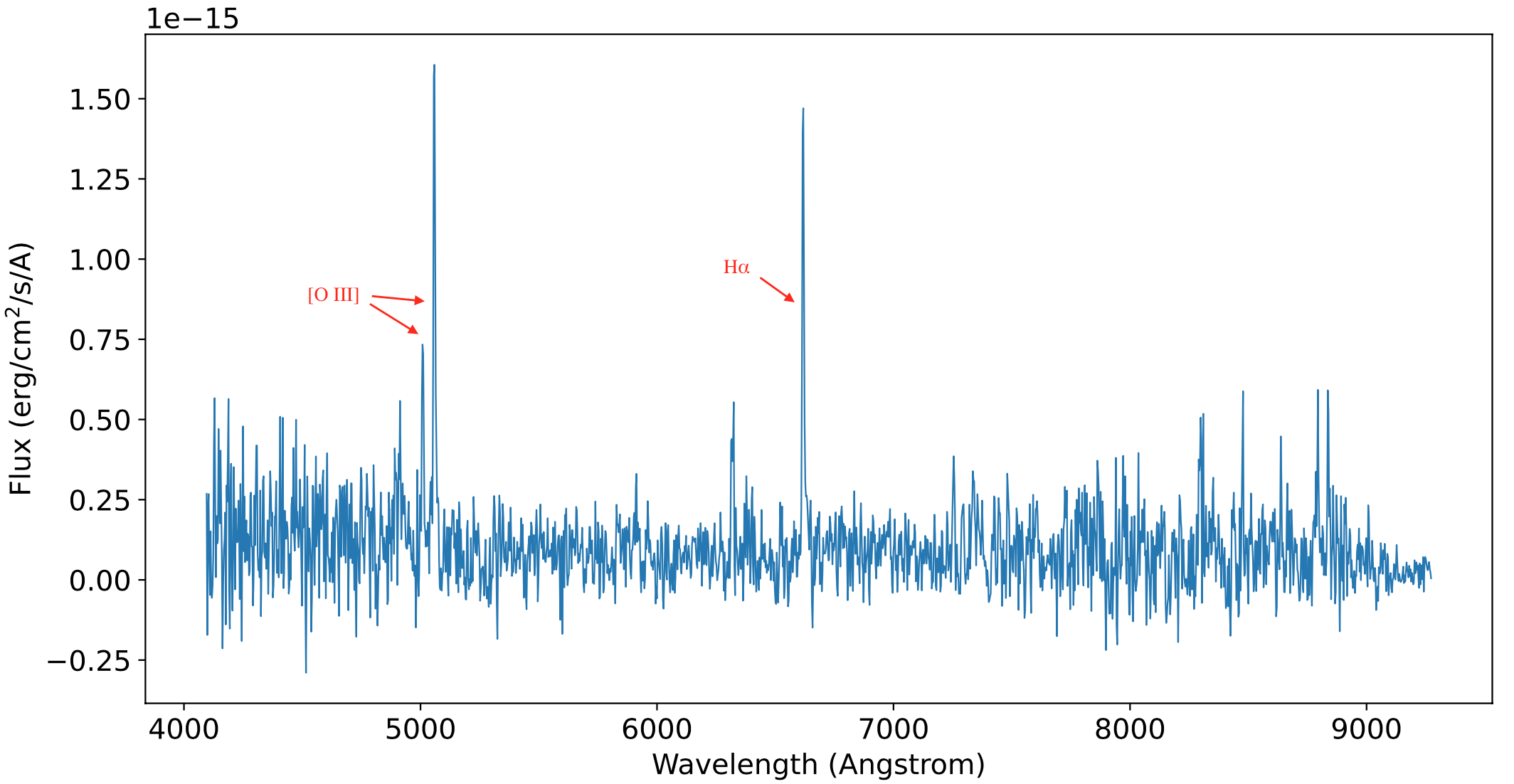}
\put(70,45){\small{YP0827-3033}}
\end{overpic}
\includegraphics[width=0.2\textwidth]{fig/YP1151-6247.png}
\includegraphics[width=0.205\textwidth]{fig/YP1151-6247_shs_q.pdf}
\includegraphics[width=0.205\textwidth]{fig/YP1151-6247_vphas_q.pdf}
\begin{overpic}[width=0.4\textwidth]{fig/YP1151-6247_spec.png}
\put(70,45){\small{YP1151-6247}}
\end{overpic}
\includegraphics[width=0.2\textwidth]{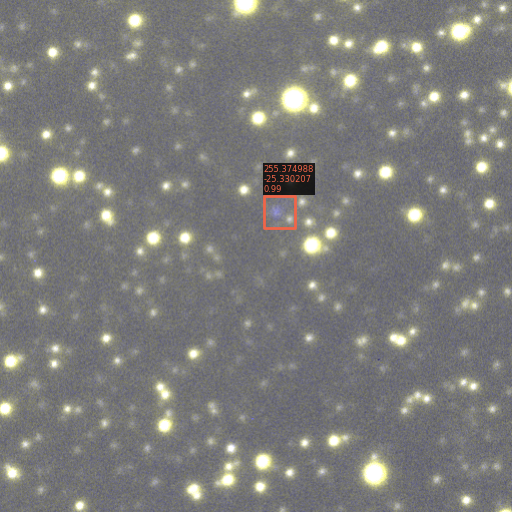}
\hspace*{1.49in}
\includegraphics[width=0.205\textwidth]{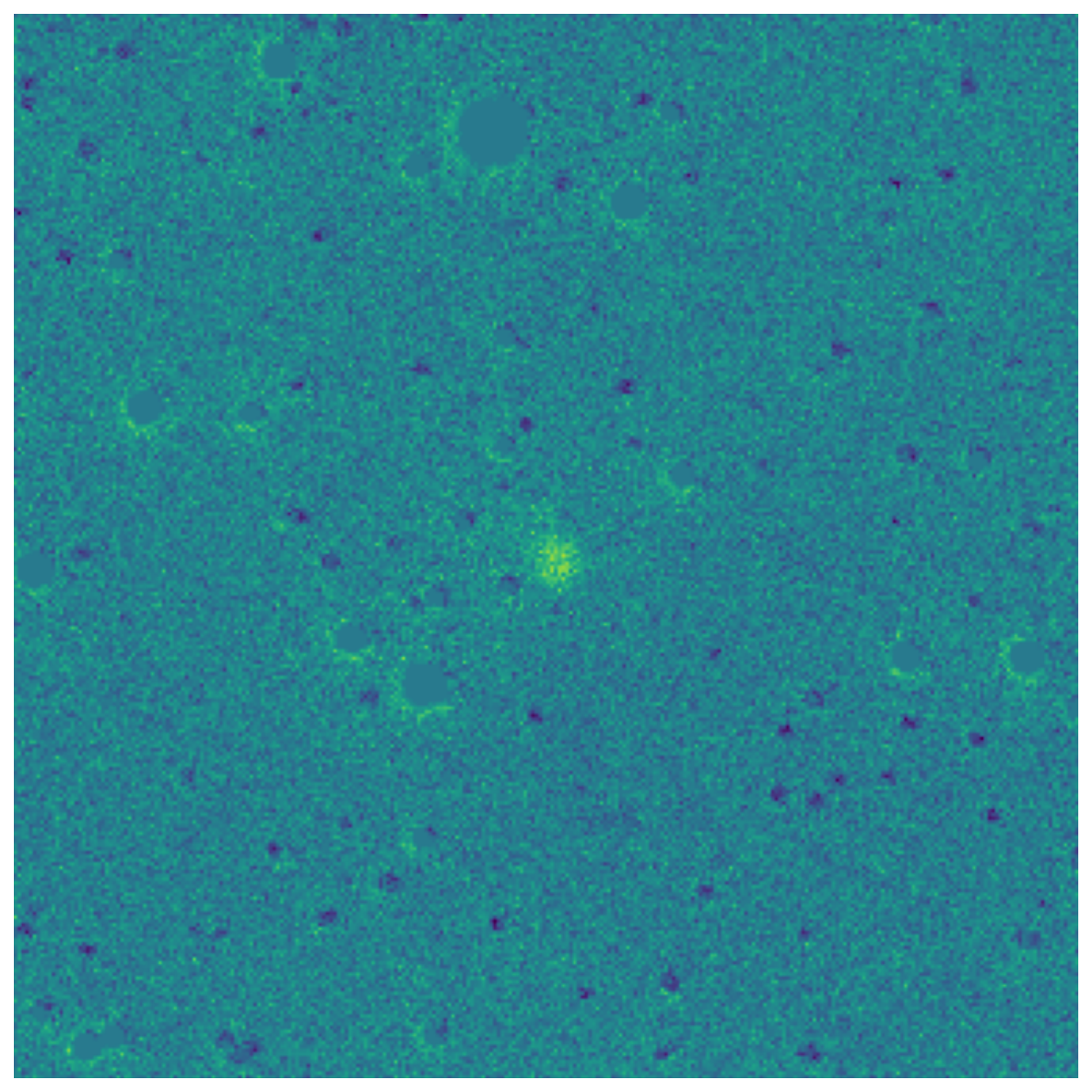}
\begin{overpic}[width=0.4\textwidth]{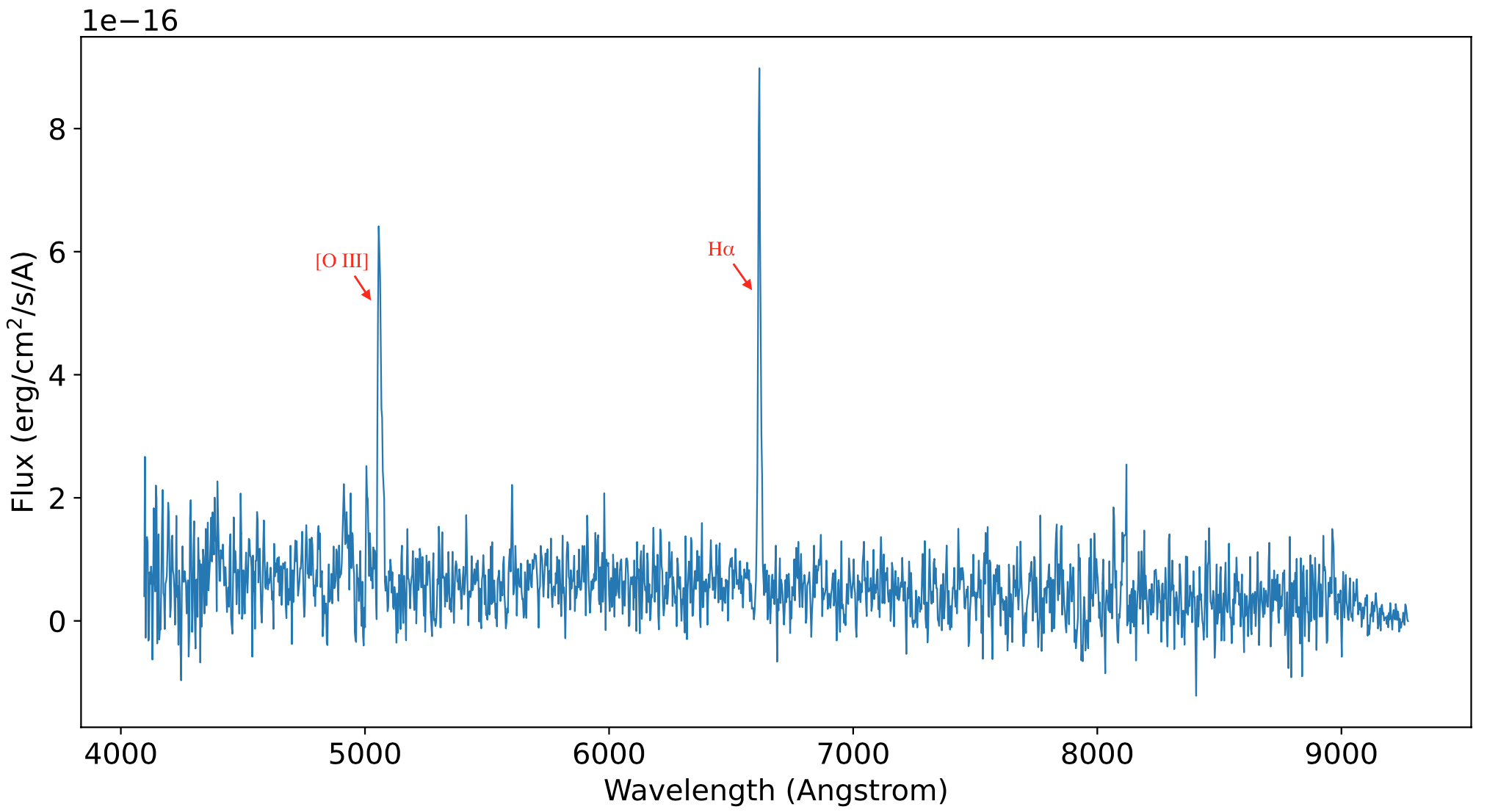}
\put(70,45){\small{YP1701-2519}}
\end{overpic}
\caption{The same as Figure \ref{fig:true_pn}, for emission-line galaxies.}
\label{fig:em_gal}
\end{minipage}
\begin{minipage}{\textwidth}
\includegraphics[width=0.2\textwidth]{fig/YP0900-4457.png}
\includegraphics[width=0.205\textwidth]{fig/YP0900-4457_shs_q.pdf}
\includegraphics[width=0.205\textwidth]{fig/YP0900-4457_vphas_q.pdf}
\begin{overpic}[width=0.4\textwidth]{fig/YP0900-4457a_spec.png}
\put(70,45){\small{YP0900-4457}}
\put(70,40){\small{Star}}
\end{overpic}
\hspace*{4.47in}
\begin{overpic}[width=0.4\textwidth]{fig/YP0900-4457b_spec.png}
\put(70,45){\small{YP0900-4457}}
\put(70,40){\small{Outflow}}
\end{overpic}
\includegraphics[width=0.2\textwidth]{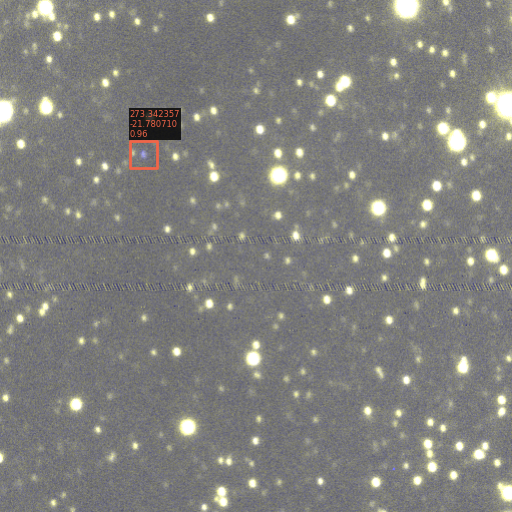}
\includegraphics[width=0.205\textwidth]{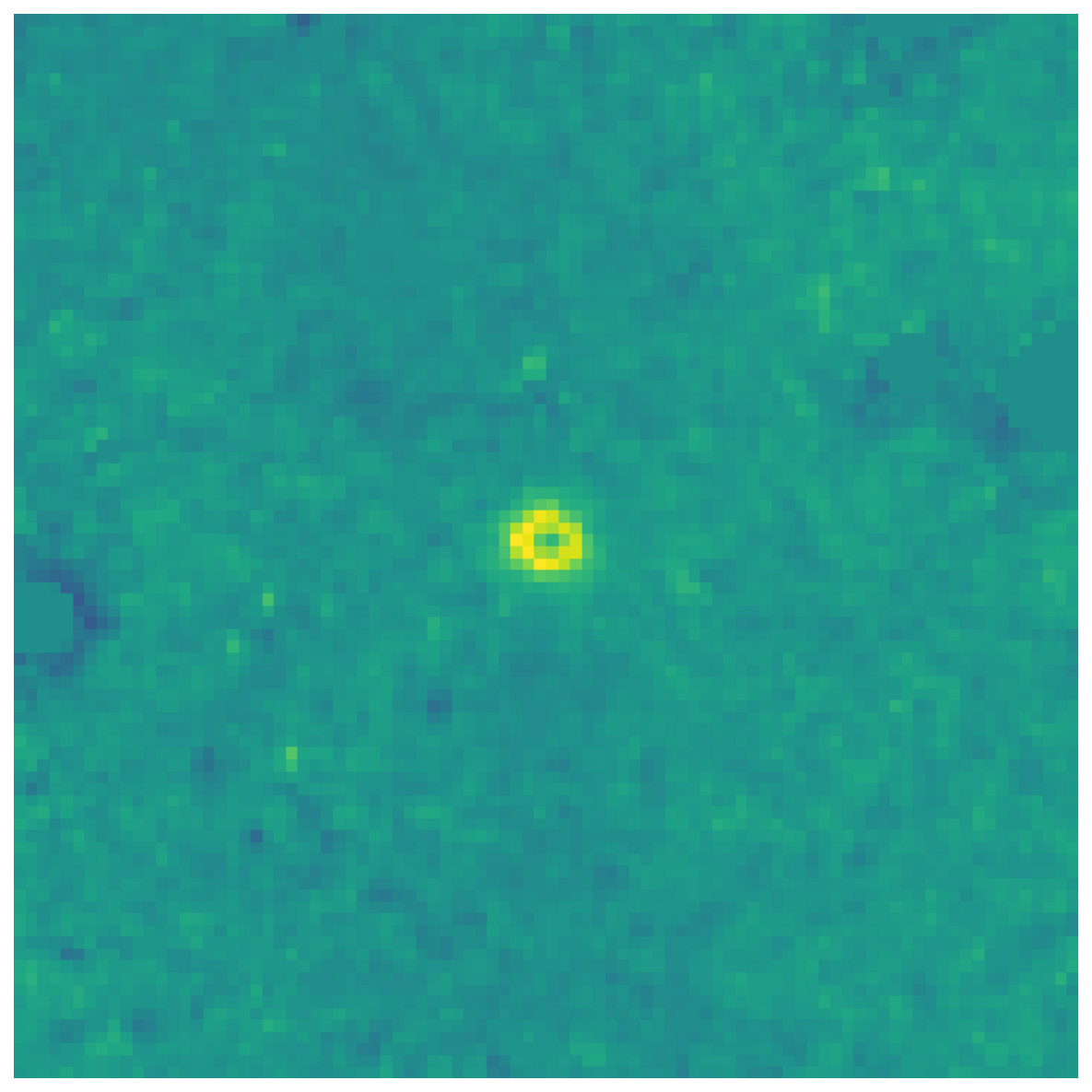}
\includegraphics[width=0.205\textwidth]{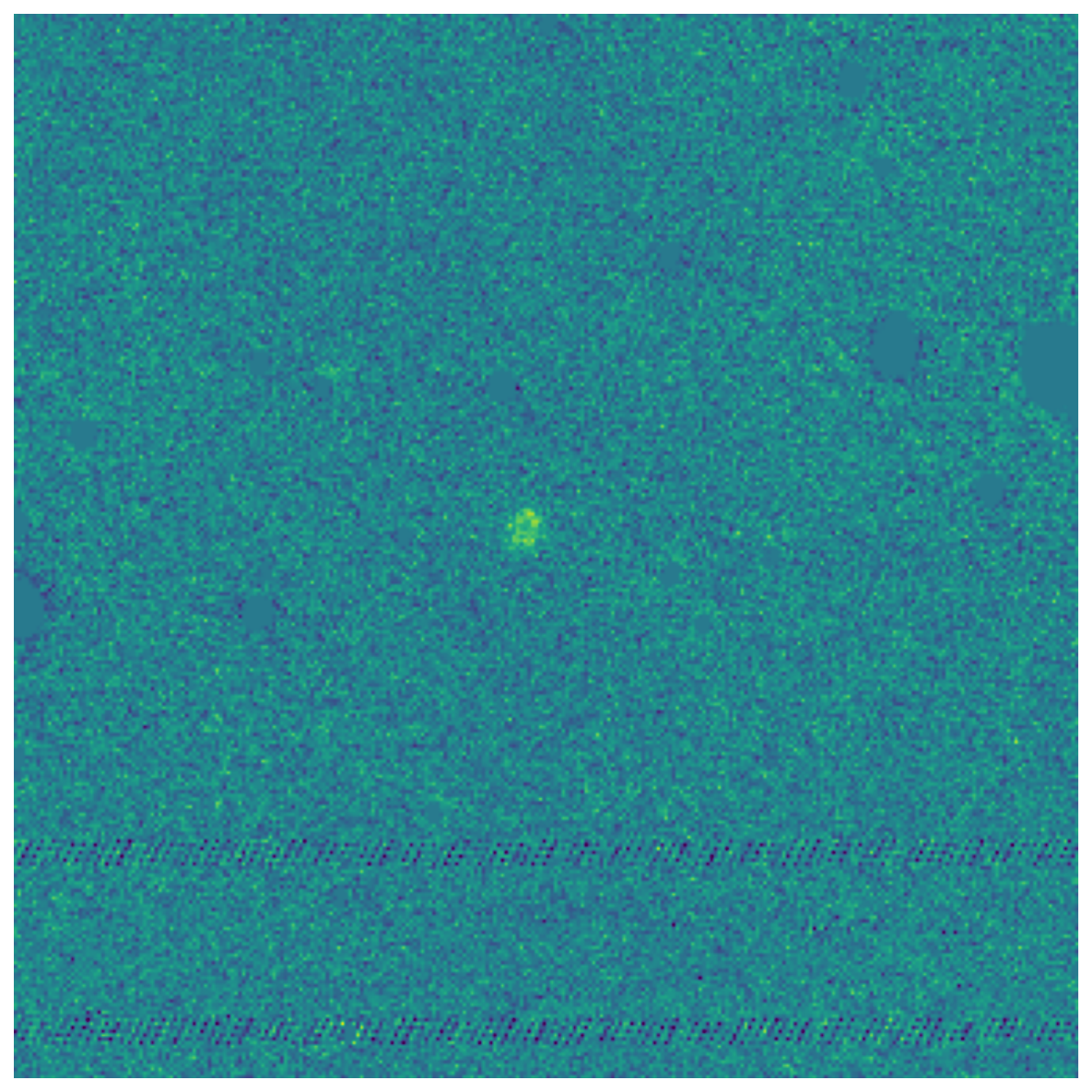}
\begin{overpic}[width=0.4\textwidth]{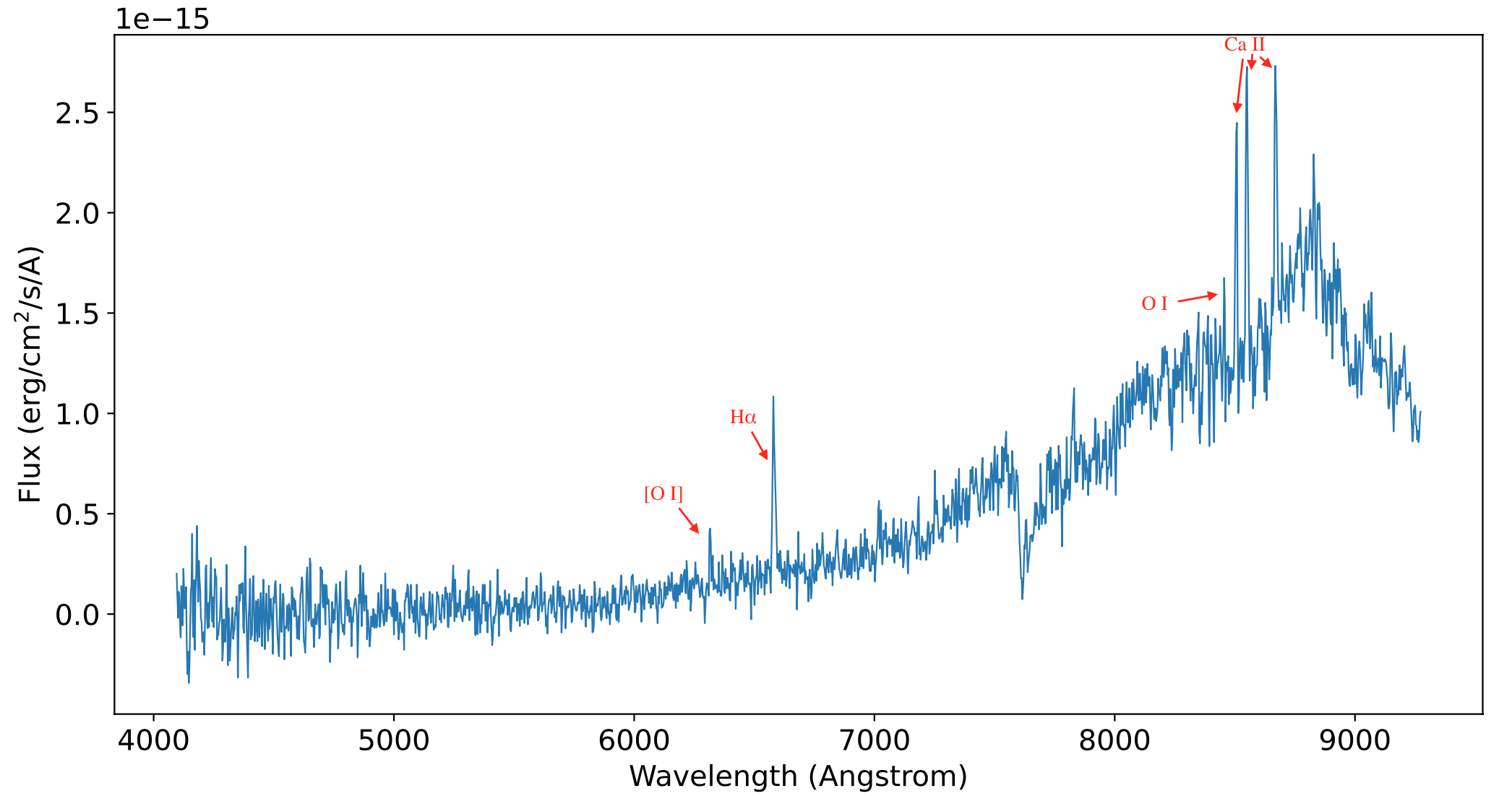}
\put(70,45){\small{YP1813-2146}}
\end{overpic}
\caption{The same as Figure \ref{fig:true_pn}, for emission-line stars.}
\label{fig:em_star}
\end{minipage}
\end{figure*}

\begin{figure*}
\begin{minipage}{\textwidth}
\includegraphics[width=0.2\textwidth]{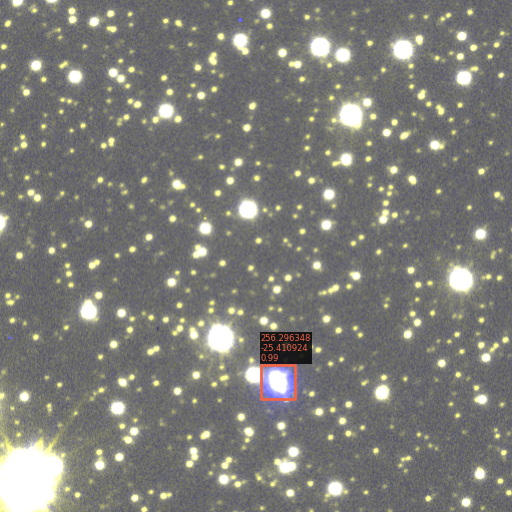}
\hspace*{1.49in}
\includegraphics[width=0.205\textwidth]{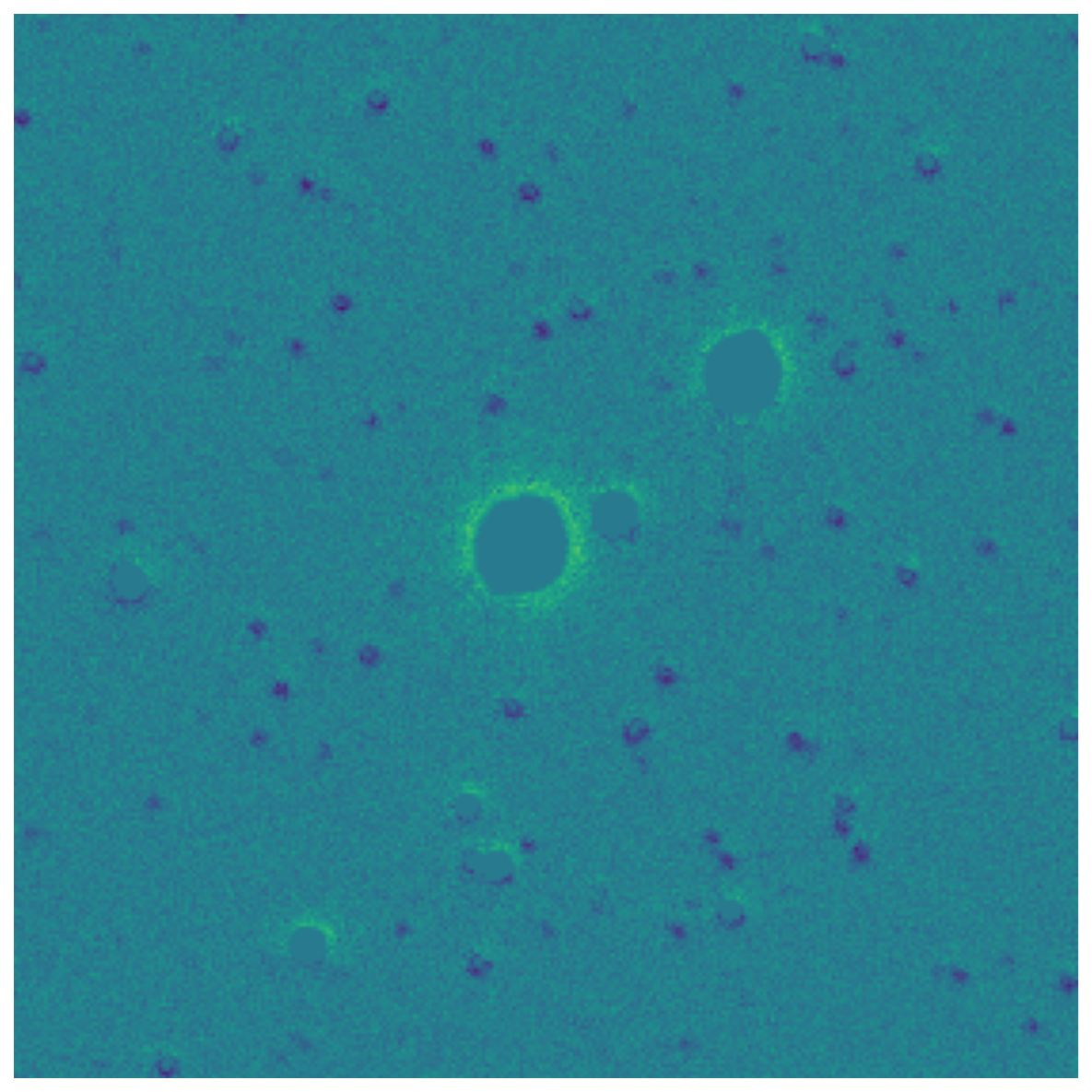}
\begin{overpic}[width=0.4\textwidth]{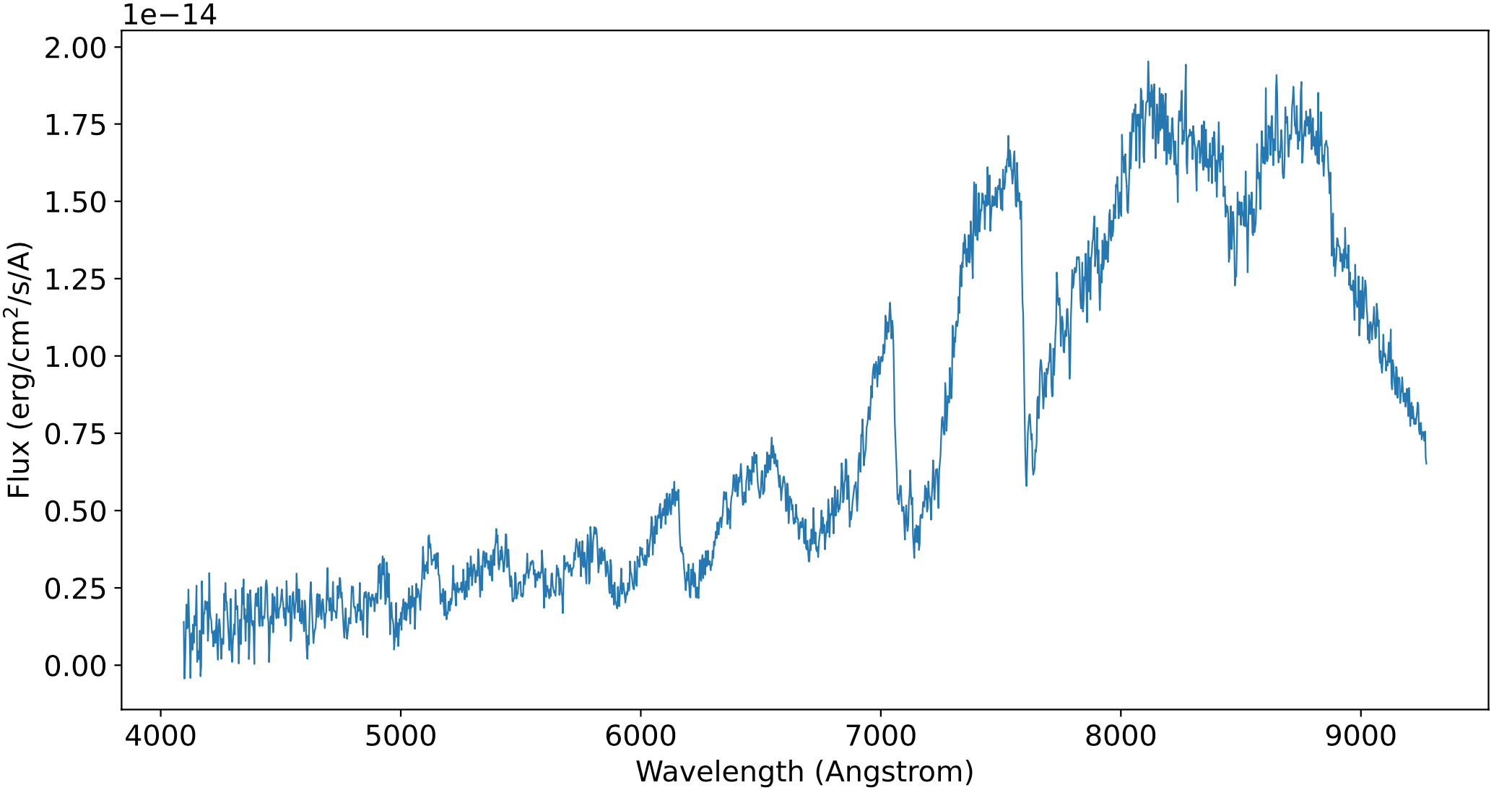}
\put(70,45){\small{YP1705-2524}}
\end{overpic}
\includegraphics[width=0.2\textwidth]{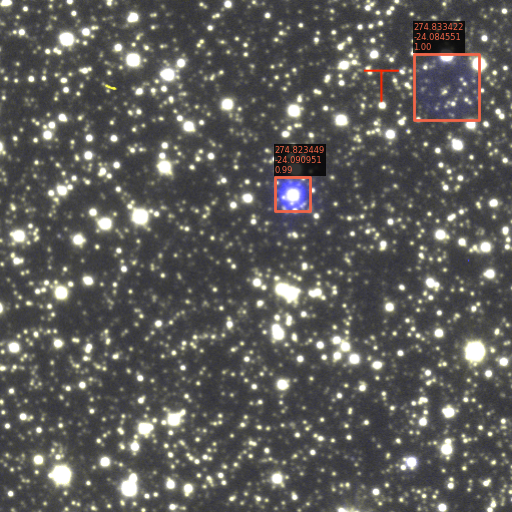}
\includegraphics[width=0.205\textwidth]{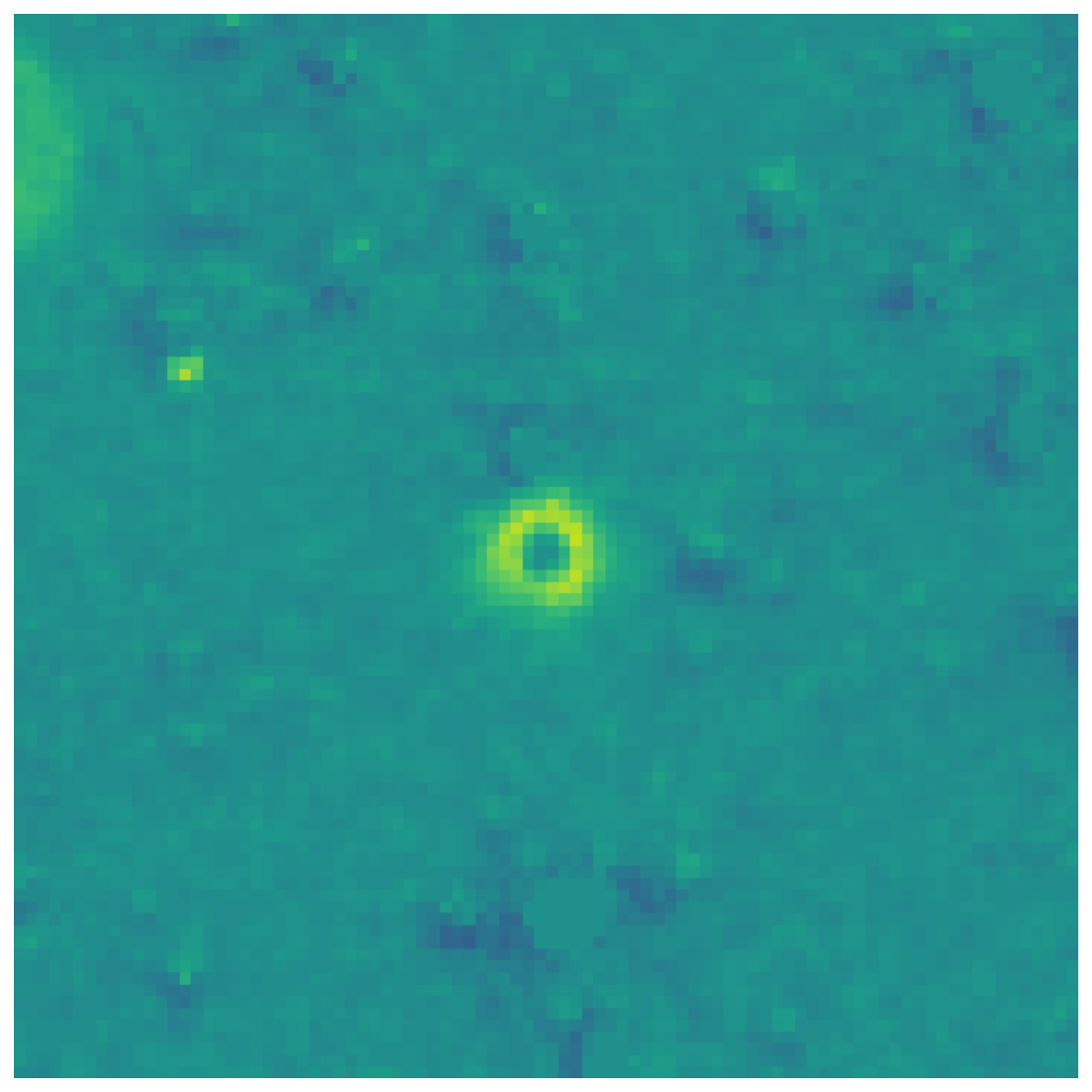}
\includegraphics[width=0.205\textwidth]{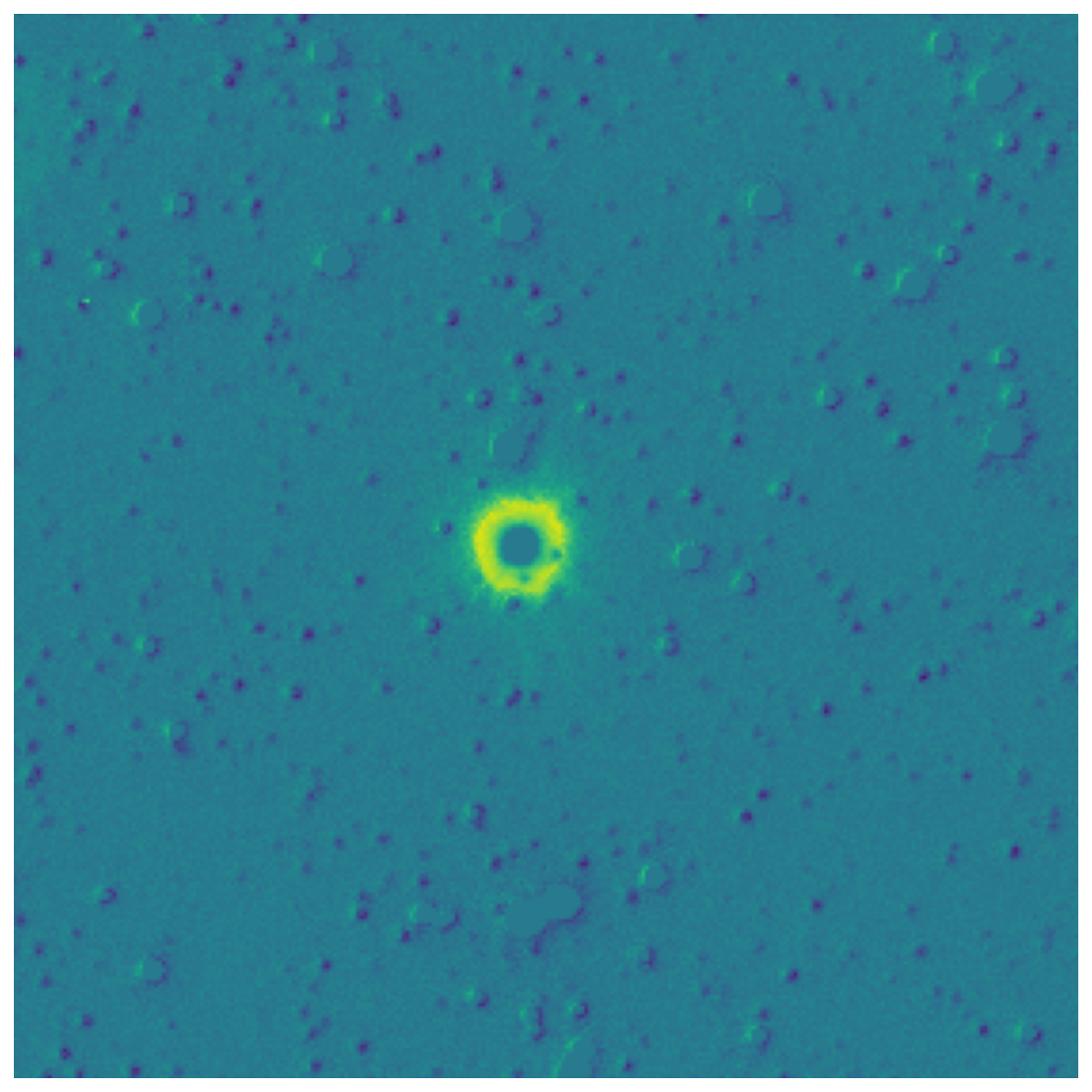}
\begin{overpic}[width=0.4\textwidth]{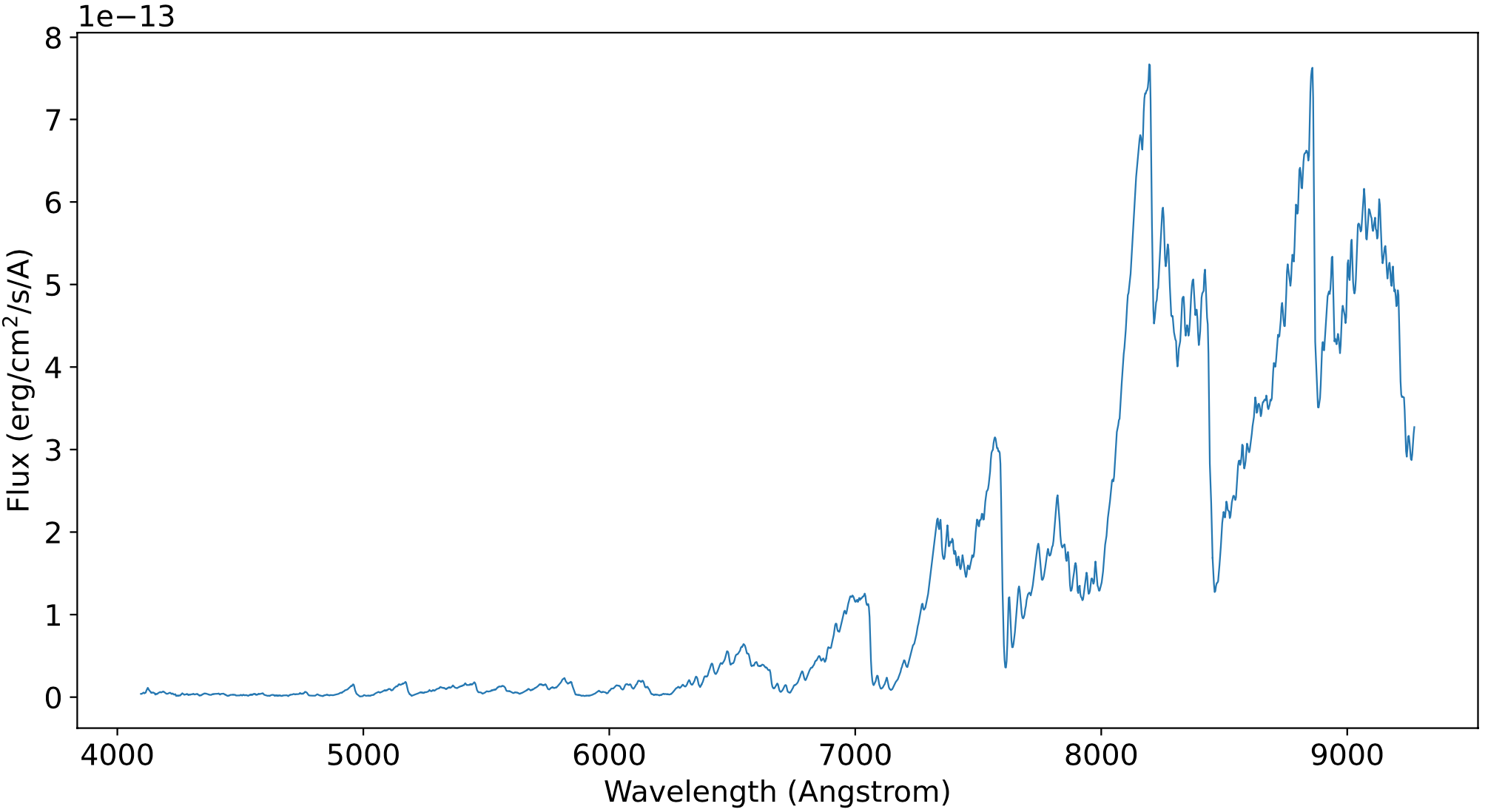}
\put(70,45){\small{YP1819-2405}}
\end{overpic}
\caption{The same as Figure \ref{fig:true_pn}, for late-type stars.}
\label{fig:lt_star}
\end{minipage}
\begin{minipage}{\textwidth}
\includegraphics[width=0.2\textwidth]{fig/YP1441-6239.png}
\includegraphics[width=0.205\textwidth]{fig/YP1441-6239_shs_q.pdf}
\includegraphics[width=0.205\textwidth]{fig/YP1441-6239_vphas_q.pdf}
\begin{overpic}[width=0.4\textwidth]{fig/YP1441-6239_spec.png}
\put(70,45){\small{YP1441-6239}}
\end{overpic}
\includegraphics[width=0.2\textwidth]{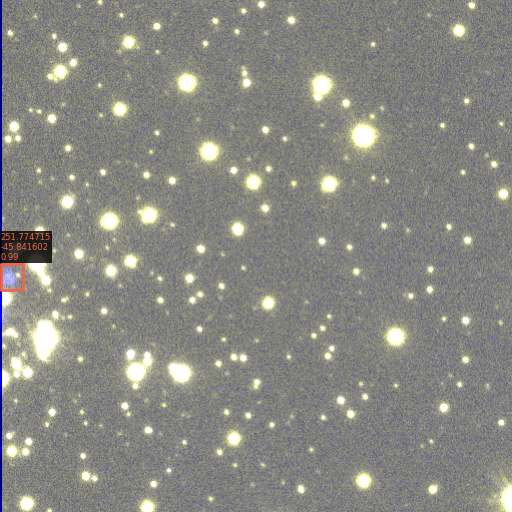}
\includegraphics[width=0.205\textwidth]{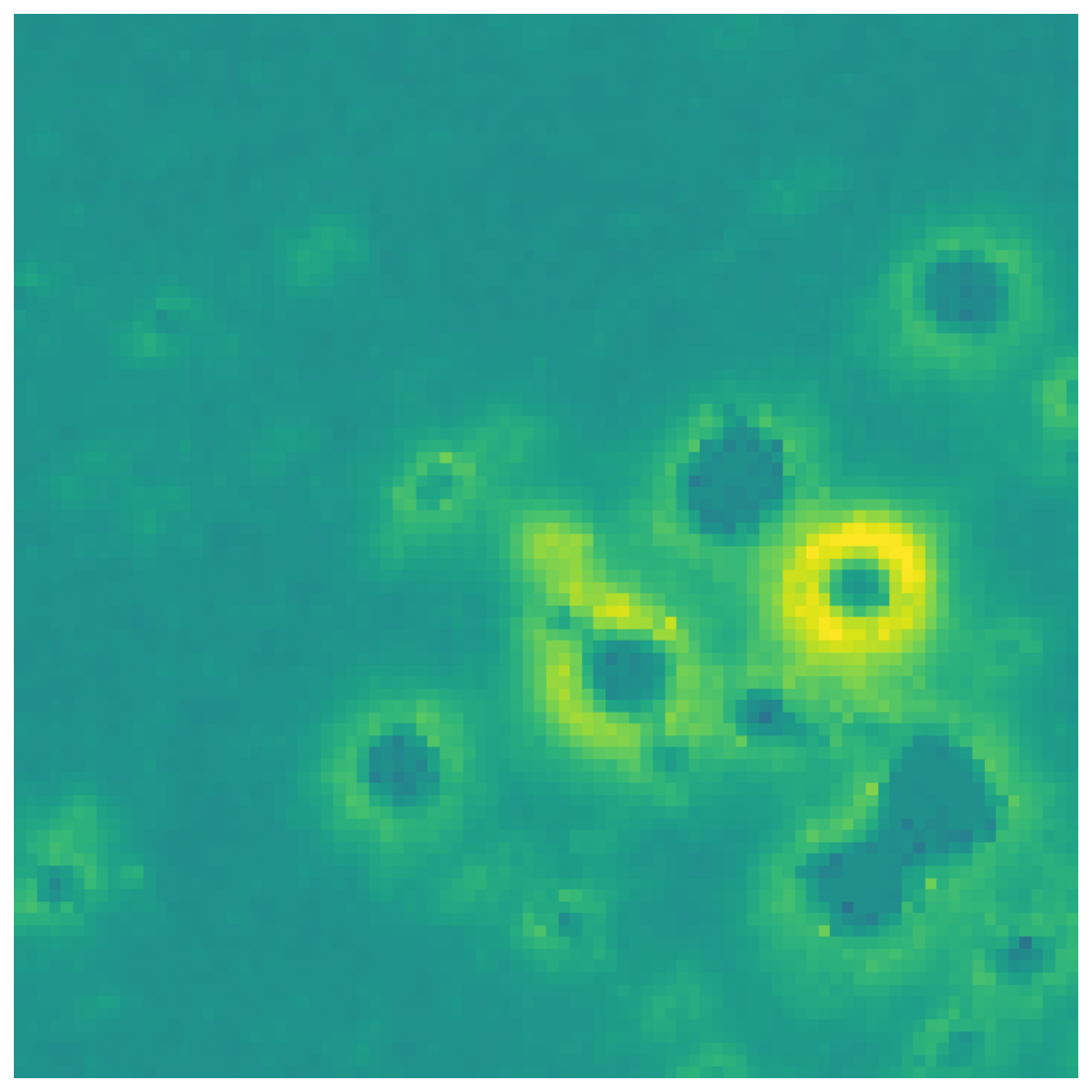}
\includegraphics[width=0.205\textwidth]{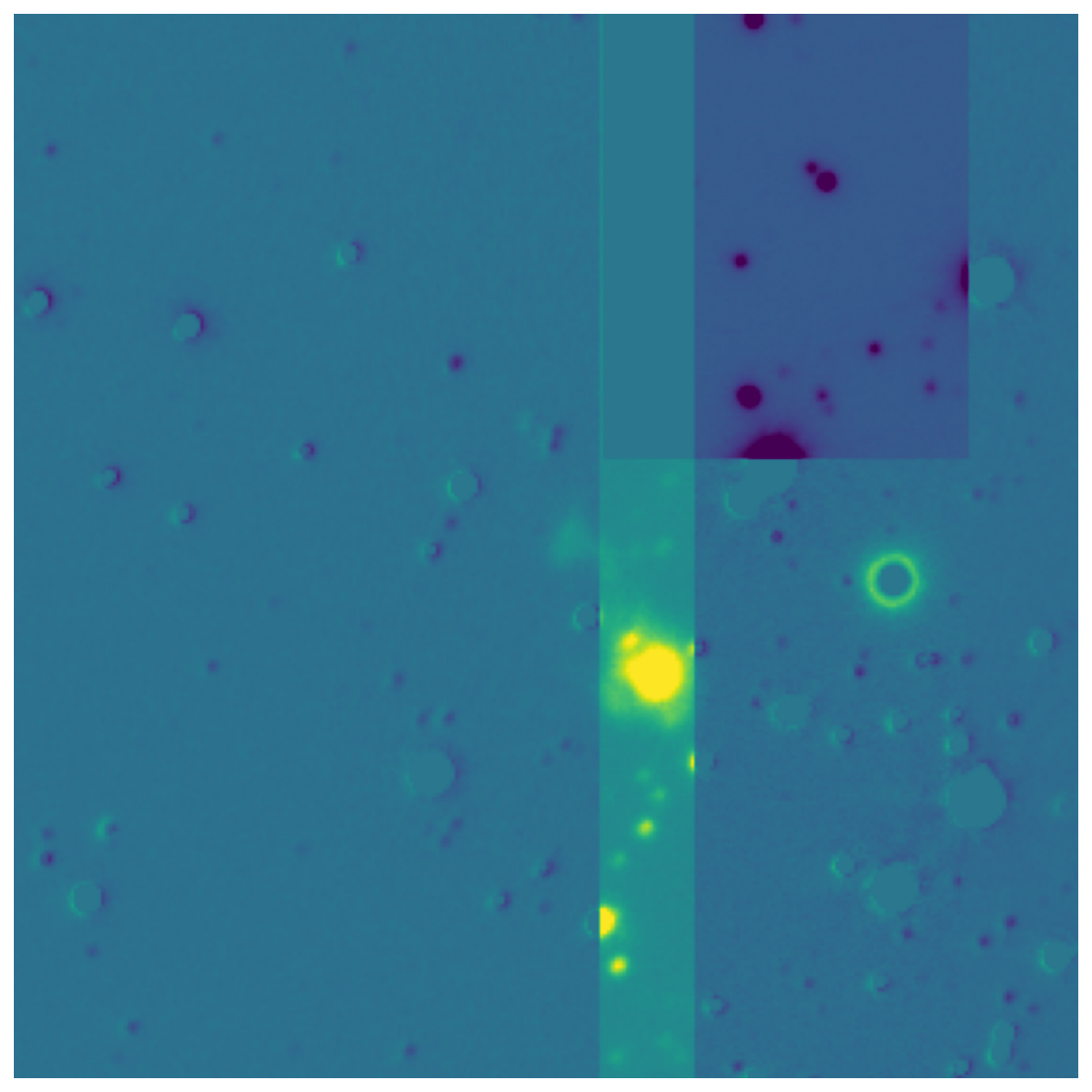}
\begin{overpic}[width=0.4\textwidth]{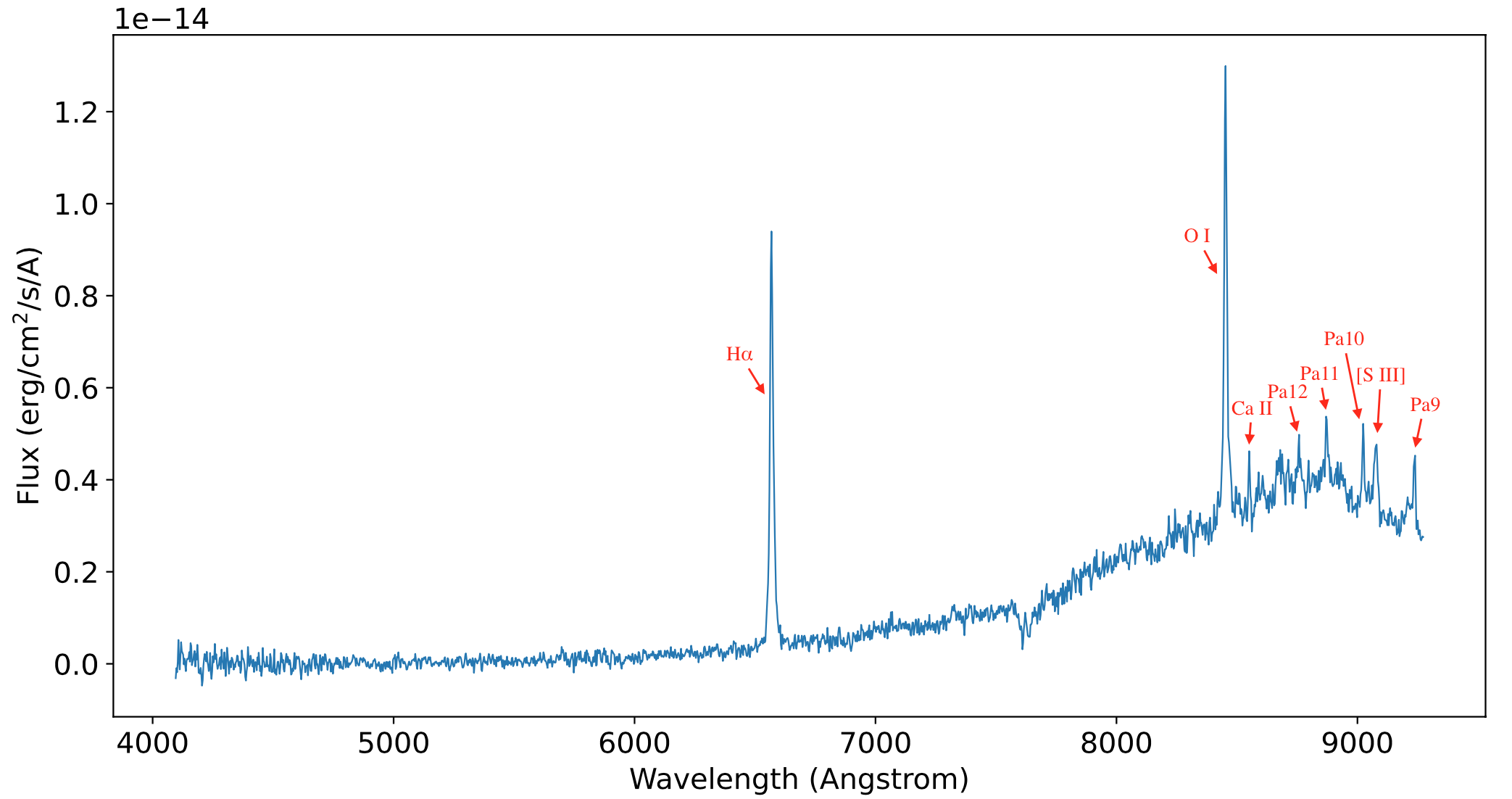}
\put(70,45){\small{YP1647-4550}}
\end{overpic}
\caption{The same as Figure \ref{fig:true_pn}, for a possible SNR fragment and a nebula in cluster.}
\label{fig:others}
\end{minipage}
\end{figure*}
\end{appendix}

\end{document}